\documentclass[11pt,a4paper]{article}
\usepackage{jheppub}

\usepackage{amsmath}
\usepackage{graphicx}
\usepackage[normalem]{ulem}
\usepackage{pst-node}
\usepackage{lscape}
\usepackage{amsfonts}
\usepackage{amsbsy}
\usepackage{graphics}
\usepackage{lscape}
\usepackage{mathrsfs}
\usepackage{dsfont}
\usepackage{subcaption}
\usepackage{color}
\usepackage{appendix}

\allowdisplaybreaks

\renewcommand{\d}{\mathrm{d}}

\newcommand{\e}{\mathrm{e}}

\newcommand{\Tr}{\mathrm{Tr}}

\newcommand{\bea}{\begin{eqnarray}}
\newcommand{\eea}{\end{eqnarray}}

\def\slash#1{\setbox0=\hbox{$#1$}  
   \dimen0=\wd0     
   \setbox1=\hbox{/} \dimen1=\wd1  
   \ifdim\dimen0>\dimen1   
      \rlap{\hbox to \dimen0{\hfil/\hfil}} 
      #1     
   \else     
      \rlap{\hbox to \dimen1{\hfil$#1$\hfil}} 
      /      
   \fi}      %

\newcommand{\sumint}{\sum_{X}\hspace{-0.5cm}\int}

\title{Polarized hyperon production in single-inclusive electron-positron annihilation at next-to-leading order}
\author[a]{Leonard Gamberg,}
\author[b,c,d]{Zhong-Bo Kang,}
\author[e,f,a]{Daniel Pitonyak,}
\author[g]{Marc Schlegel,}
\author[d]{Shinsuke Yoshida}

\affiliation[a]{Division of Science, Penn State University Berks, Reading, Pennsylvania 19610, USA}
\affiliation[b]{Department of Physics and Astronomy, University of California, Los Angeles, CA 90095, USA}
\affiliation[c]{Mani L. Bhaumik Institute for Theoretical Physics, University of California, Los Angeles, CA 90095, USA}
\affiliation[d]{Theoretical Division, Los Alamos National Laboratory, Los Alamos, NM 87545, USA}
\affiliation[e]{Department of Physics, Lebanon Valley College, Annville, PA 17003, USA}
\affiliation[f]{Department of Physics, Old Dominion University, Norfolk, VA 23529, USA}
\affiliation[g]{Department of Physics, New Mexico State University, Las Cruces, NM 88003, USA}

\emailAdd{lpg10@psu.edu}
\emailAdd{zkang@physics.ucla.edu}
\emailAdd{pitonyak@lvc.edu}
\emailAdd{schlegel@nmsu.edu}
\emailAdd{syoshida@lanl.gov}

\abstract{We study the production of polarized  $\Lambda$-hyperons in electron-positron annihilation. We are particularly interested in the transverse-spin dependence of the cross section for unpolarized incident electron-positron pairs. At high energies this process may be described in the collinear twist-3 framework, where the hadronization transition of partons into a transversely polarized $\Lambda$-hyperon can be written in terms of collinear twist-3 fragmentation matrix elements. We calculate the hard partonic cross sections and interference terms in perturbative QCD to next-to-leading order accuracy. We find that the QCD equation of motion plays a crucial role in our analysis. As a byproduct, assuming the validity of QCD factorization for twist-3 observables at next-to-leading order, we derive the evolution equation for the relevant twist-3 fragmentation matrix element.}

\begin{document}
\maketitle
\flushbottom

\section{Introduction}
A proper understanding of polarization effects for $\Lambda$-hyperons produced in high-energy reactions is a longstanding challenge in hadronic physics. In fact, surprisingly large polarizations were found in early experiments at Fermi-Lab (along with follow-up measurements) in $pA\to \Lambda X$ fixed target processes already 40 years ago  \cite{Bunce:1976yb,Schachinger:1978qs,Heller:1983ia,Lundberg:1989hw,Yuldashev:1990az,Ramberg:1994tk}. Other fixed target measurements of this reaction were reported by the NA48 Collaboration \cite{Fanti:1998px} and the HERA-B Collaboration \cite{Abt:2006da}. At CERN, $\Lambda$ polarization was also measured in $pp$ collisions at moderate center-of-mass (c.m.)~energy again close to 40 years ago \cite{Erhan:1979xm}. Interestingly, the polarization of $\Lambda$-hyperons was investigated just recently at the LHC by the ATLAS Collaboration \cite{ATLAS:2014ona}. Although only a tiny polarization, essentially consistent with zero, was found in the ATLAS measurements in the mid-rapidity region, this experimental pursuit shows that the polarization of $\Lambda$-hyperons can be studied at the highest LHC energies and could be larger in different kinematical regions at forward rapidities.

Theoretically, the hadronization of partons into hadrons in high-energy processes is described in terms of non-perturbative matrix elements of certain QCD operators, which can be extracted from fits to experimental data. However, this would be a very difficult task to do on the basis of data taken from $pp$ or $pA$ reactions alone. One reason is that these processes are mediated purely by the strong force, and therefore the analytical description is complicated due to many competing effects that enter the QCD factorization formulas for spin observables in $pp$ or $pA$ reactions. This is comparable to the extraction of parton distribution functions (PDFs) -- one would not want to rely on data only from $pp$ reactions in order to extract PDFs.

The situation becomes simpler for processes that involve electromagnetic interactions, such as semi-inclusive deep-inelastic electron-nucleon scattering (SIDIS). Here, polarized $\Lambda$'s may be produced in $ep\to e\Lambda X$ or in the equivalent quasi-real photo-production processes. Experimental studies of these reactions have been performed by the HERMES Collaboration \cite{Airapetian:2006ee,Airapetian:2007mx,Airapetian:2014tyc}, as well as in neutrino-nucleon scattering by the NOMAD Collaboration \cite{Astier:2000ax,Astier:2001ve}.

The process of SIDIS at HERMES kinematics is subject to transverse-momentum dependent (TMD) factorization. Here, intrinsic parton transverse momenta are explicitly taken into account in the corresponding fragmentation functions (FFs). Studies of these TMD FFs responsible for $\Lambda$ polarization within the TMD factorization framework have been presented in Refs.~\cite{Anselmino:2000vs,Anselmino:2001js,Boer:2010ya}. For more general information on the current theoretical and experimental status of FFs, we refer the reader to the recent review of Ref.~\cite{Metz:2016swz}.

Perhaps the cleanest possible process both experimentally and theoretically to get access to polarized $\Lambda$ FFs is single-inclusive $\Lambda$ production in electron-positron annihilation, $e^+e^-\to \Lambda\,X$. In principle, when calculating this process in perturbative QCD to leading order, one can directly map out the dependence of the corresponding FFs on the longitudinal momentum fraction $z$ of the the fragmenting parton momentum carried by the hadron. In this sense, single-inclusive annihilation plays the same role for FFs as inclusive DIS does for PDFs.

Data on polarized $\Lambda$ fragmentation in this reaction has been provided by the OPAL Collaboration \cite{Ackerstaff:1997nh} at LEP. This measurement was performed on the $Z$-pole, i.e., at a c.m. energy equal to the mass of the $Z$-boson. While a substantial  {\it longitudinal} polarization of the $\Lambda$'s was detected by OPAL, the {\it transverse} polarization was found to be zero within error bars. Interestingly, Belle has measured recently the production of unpolarized $\Lambda$'s~\cite{Niiyama:2017wpp} in $e^+e^-$ annihilation. In addition, Belle data~\cite{Abdesselam:2016nym,Guan:2018ckx} on the transverse $\Lambda$ polarization show a significant non-zero effect in this process.

In this paper we  (re-)investigate the process $e^+e^-\to \Lambda^{\uparrow}X$ from the point of view of perturbative QCD and calculate the hard scattering factors to next-to-leading order (NLO) accuracy. This calculation is particularly challenging for transverse spin observables because they are suppressed in this process by a factor of $1/Q$ compared to the unpolarized production rate, where $Q=\sqrt{s}$ is the hard scale of the process and $\sqrt{s}$ the c.m. energy of the incident leptons. As a result, the theoretical description is more involved and is beyond a simple partonic picture that may be used to understand unpolarized observables. 

A suitable framework to describe transverse spin observables in single-inclusive processes is the so-called {\it collinear twist-3 formalism}~\cite{Qiu:1991wg,Qiu:1998ia,Kouvaris:2006zy,Eguchi:2006qz,Eguchi:2006mc,Koike:2009ge,Beppu:2010qn,Koike:2011nx,Kang:2010zzb,Metz:2012ct,Kanazawa:2013uia,Beppu:2013uda} (see Ref.~\cite{Pitonyak:2016hqh} for a recent review), where one deals with collinear three-parton PDFs and FFs. In this framework, calculations at LO for various hyperon production processes have been performed in Refs.~\cite{Kanazawa:2000cx,Zhou:2008fb,Kanazawa:2015jxa,Koike:2015zya,Koike:2017fxr}, where, in particular, analyses of fragmentation effects involving transversely polarized $\Lambda$'s were pioneered in Refs.~\cite{Kanazawa:2015jxa,Koike:2017fxr}.

Our motivation for this work is twofold: 1) Only very limited NLO calculations within the collinear twist-3 framework exist in the literature \cite{Vogelsang:2009pj,Kang:2012ns,Dai:2014ala,Yoshida:2016tfh,Chen:2016dnp,Chen:2017lvx}.
These studies mostly focused on NLO corrections for so-called pole contributions of three-parton correlations in the nucleon that are relevant for naive time-reversal odd (T-odd) observables like single-spin asymmetries.
By contrast, pole contributions do not exist for fragmentation correlators~\cite{Meissner:2008yf,Gamberg:2010uw,Boer:2010ya} and therefore the calculation is different from a technical standpoint (see, e.g., Refs.~\cite{Metz:2012ct,Kanazawa:2013uia}).  We expect this feature to persist in NLO calculations for fragmentation processes. (We note that observables involving nucleon non-pole three-parton correlators do exist for T-even processes~\cite{Koike:2008du,Zhou:2009jm,Liang:2012rb,Metz:2012fq,Kanazawa:2014tda,Koike:2016ura}.) In order to fully understand the NLO dynamics for fragmentation, we choose to study the simplest process available, $e^+e^-\to \Lambda^{\uparrow}X$. 2) If a future global NLO QCD analysis of the available polarized $\Lambda$ data involving data sets from different experiments is to be performed, a NLO calculation for this process will be needed.

The rest of the paper is organized as follows. In section \ref{Corr} we will define all of the relevant soft fragmentation matrix elements. In section \ref{LO} we calculate the spin-dependent cross section to LO. In section \ref{NLO} we extend the calculation to NLO accuracy, which then allows us to discuss evolution equations in section \ref{Evol}. We conclude in section \ref{Concl} and give an outlook for future work.

\section{Twist-3 fragmentation correlators\label{Corr}}

In this section we will introduce and review all of the fragmentation matrix elements that are needed for a
factorized perturbative QCD (pQCD) twist-3 description of the spin-dependent cross section for $e^+e^-\to \Lambda\,X$.
This section is to be a self-contained reference for the reader, with the main calculations for the observable given in sections~\ref{LO} and~\ref{NLO}. In the following we denote the four-momentum of the $\Lambda$-hyperon that is produced in a fragmentation process as $P_h^\mu$. We will neglect hadron masses as is typical in a pQCD calculation, and we thus consider $P_h^\mu$ to be a light-like vector, $P_h^2=0$. We introduce an adjoint light-like vector $n^\mu$ with $n^2=0$ and $P_h\cdot n=1$. Note that these two conditions do not completely fix the choice of $n$~\cite{Kanazawa:2015ajw}. However, both $P_h$ and $n$ are needed to define what is meant by the term {\it transverse}. If we define the projector 
\bea
g_T^{\mu\nu}\equiv g^{\mu \nu}-P_h^\mu n^\nu-P_h^\nu n^\mu,\label{eq:transverseProj}
\eea
then the {\it transverse} part of a four-vector $a^\mu$ is defined as $a_T^\mu=g^{\mu \nu}_T\,a_\nu$. In order to discuss the spin-dependent fragmentation correlators we also need to introduce a four-spin vector $S_h^\mu$. In the rest frame of the hadron the zeroth component of $S_h^\mu$ vanishes, while the spatial components indicate the polarization of the hadron in the rest frame. The normalization of $S_h^\mu$ is then chosen to be $S_h^2=-1$, and we also have $P_h\cdot S_h=0$.

\subsection{Two-parton correlations}

\begin{figure}
\centering
\begin{subfigure}[b]{0.4\textwidth}
\includegraphics[width=\textwidth]{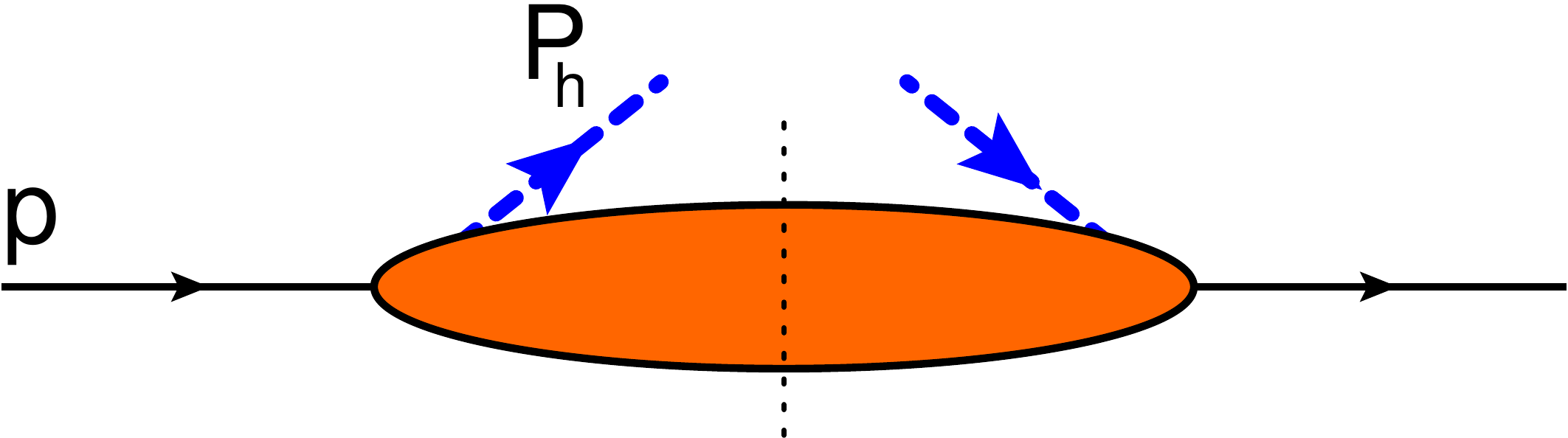}
\caption{Quark fragmentation}
\label{fig:CorrQQ}
\end{subfigure}
\hskip 0.1in
\begin{subfigure}[b]{0.4\textwidth}
\includegraphics[width=\textwidth]{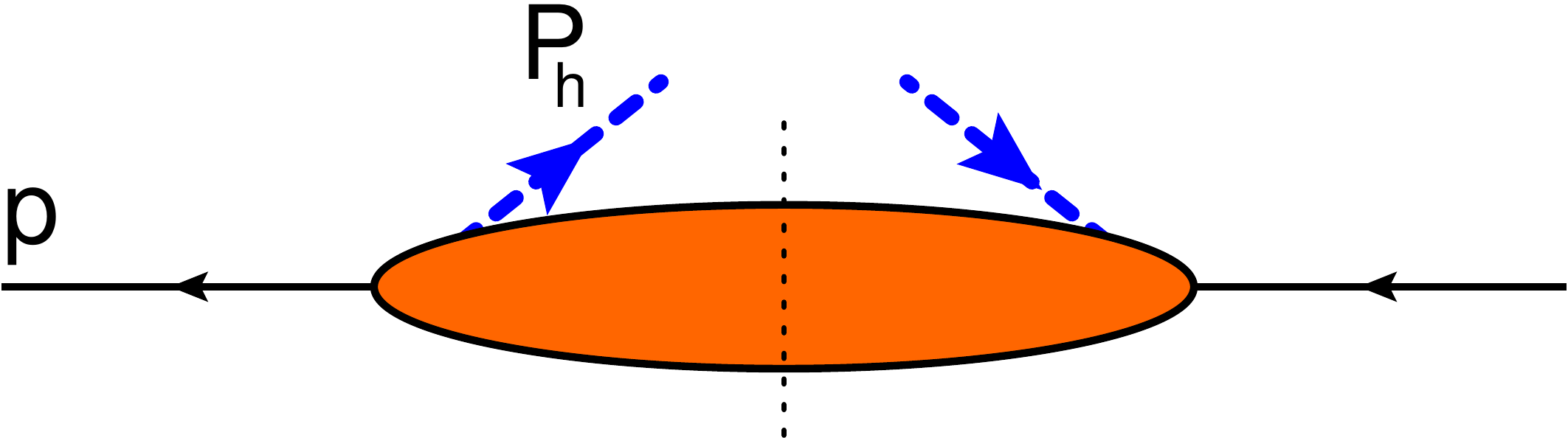}
\caption{Antiquark fragmentation}
\label{fig:CorrQbQb}
\end{subfigure}
\hskip 0.1in
\\[0.3cm]
\begin{subfigure}[b]{0.4\textwidth}
\includegraphics[width=\textwidth]{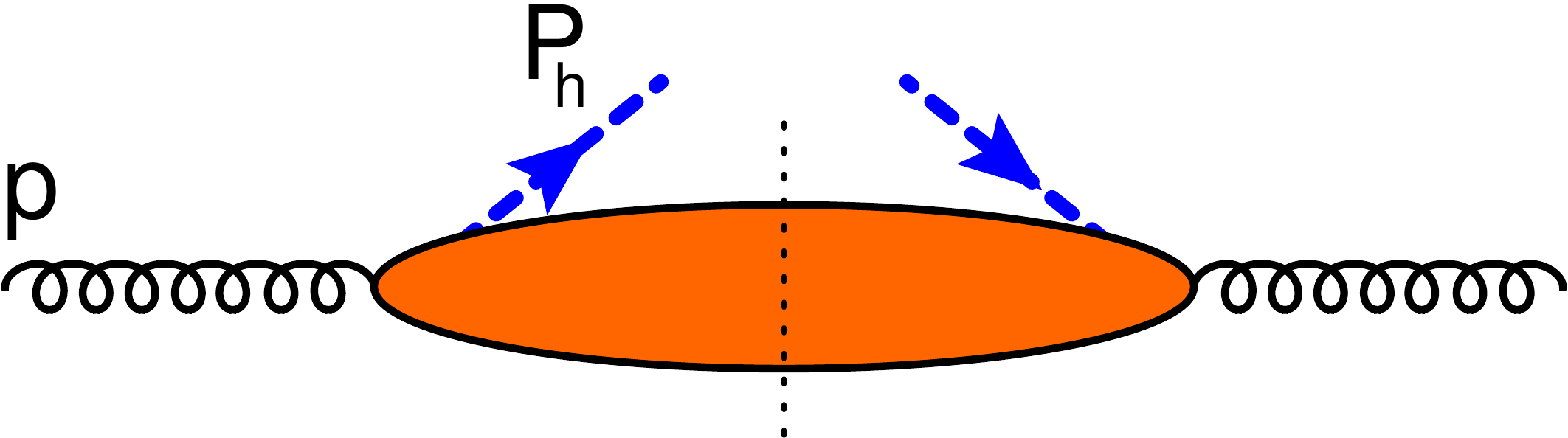}
\caption{Gluon fragmentation}
\label{fig:CorrGG}
\end{subfigure}

\caption{Diagrammatic representation of two-parton fragmentation correlations.}
\label{fig:2PartonFF}
\end{figure}

Based on a partonic interpretation of the fragmentation process~\cite{Ji:1993vw,Chen:1994ar}, a matrix element that describes the hadronization of a parton into a jet of hadrons may be written as $\langle X | \,\phi(0)\,|0\rangle$, where $\phi$ stands for a generic partonic field (quark, anti-quark, or gluon) and $|X\rangle$ is an arbitrary hadronic multi-particle state which forms an (unobserved) jet. If one of the hadrons of the jet is detected and its four-momentum $P_h$ and four-spin $S_h$ are measured, we may write instead $\langle P_h S_h;X | \,\phi(0)\,|0\rangle$. In order to implement the soft fragmentation process into a pQCD formula one can view  the   ``square'' as a cut forward  transition amplitude and sum over all possible unobserved hadron states. In this way fragmentation correlators are defined. 

\subsubsection{Intrinsic twist-3}
In single-inclusive high-energy processes, kinematical approximations are applied on the parton momenta in the factorization of the cross section into
a hard partonic cross section and the non-perturbative correlation functions.
For fragmentation, one assumes that the jet of hadrons that is produced by a highly-energetic parton moves into the same direction as the parton. To be precise, one approximates  the momentum $P_h$ of the detected hadron to be \emph{collinear} to the initial parton's momentum $p$. Since the parton ``decays" into many particles, the detected hadron only carries a fraction $z$ of the initial parton momentum. Hence, the kinematical approximation on the parton momentum reads
\bea
p^\mu &\simeq & \frac{1}{z}P_h^\mu.\label{eq:collkinAppr}
\eea

The correlator that describes the fragmentation of a quark of flavor $q$ into a hadron with momentum $P_h$ and spin $S_h$ is  represented in fig.~\ref{fig:CorrQQ}. It  can be expressed in terms of fragmentation functions based on constraints of hermiticity and parity~\cite{Ji:1993vw,Mulders:1995dh,Boer:1997mf},
\begin{eqnarray}
\Delta_{ij}^{q}(z) & = & \frac{1}{N_{c}}\sumint\int_{-\infty}^{\infty}\tfrac{\d\lambda}{2\pi}\,\mathrm{e}^{-i\frac{\lambda}{z}}\langle0|[\infty ;0]\,q_{i}(0)\,|P_{h}S_{h};X\rangle\langle P_{h}S_{h};X|\,\bar{q}_{j}(\lambda n)\,[\lambda;\infty ]|0\rangle\label{eq:Deltaq} \\
 & = & \frac{z^{2\varepsilon}}{z}\Bigg(\slash P_{h}\,D_{1}^{q}(z)-S_{hL}\,\slash P_{h}\gamma_{5}\,G_{1}^{q}(z)-\tfrac{1}{2}[\slash P_{h},\slash S_{h}]\gamma_{5}\,H_{1}^{q}(z)\nonumber \\
 &  & \hspace{1cm}-M_{h}\epsilon^{P_{h}n\alpha S_{h}}\gamma_{\alpha}\,D_{T}^{q}(z)-M_{h}\slash S_{hT}\gamma_{5}\,G_{T}^{q}(z)+M_{h}\,E^{q}(z)\nonumber \\
 &  &\hspace{1cm} -M_{h}\,S_{hL}\,i\gamma_{5}\,E_{L}^{q}(z)+M_{h}\tfrac{i}{2}[\slash P_{h},\slash n]\,H^{q}(z)+M_{h}\,S_{hL}\,\tfrac{1}{2}[\slash P_{h},\slash n]\gamma_{5}\,H_{L}^{q}(z)\Bigg).\nonumber
\end{eqnarray}
The definition of the correlator $\Delta^q(z)$ includes the quark field operator $q(x)$ as well as collinear Wilson lines $[a;b]$ of gluon fields $A^\mu(x)$ that run along the light-like vector $n$,
\bea
[a;b]\equiv \mathcal{P}\mathrm{e}^{-ig\mu^{\varepsilon}\int_a^b \d t\,(n\cdot A)(t n)}.\label{eq:Wilson}
\eea
The Wilson line renders the correlator $\Delta^q(z)$ color gauge invariant. Since it is a collinear Wilson line, it reduces to unity in the so-called light-cone ($n\cdot A=0$) gauge of the gluon fields $A^\mu$. The Wilson line may be in the fundamental representation (for quark/anti-quark FFs) or in the adjoint representation (for gluons). The number of colors in eq. (\ref{eq:Deltaq}) is denoted by $N_c$ (=3 in QCD).
The second line in eq.~(\ref{eq:Deltaq}) is a well-known parameterization of the collinear correlator $\Delta^q(z)$, and we rely on the notation established in ref.~\cite{Kanazawa:2015ajw}. (Note that $S_{hL}\equiv M_h (n\cdot S_h)$.) The first three functions in this parameterization, $D_1$, $G_1$ and $H_1$, are twist-2 FFs and describe the fragmentation of unpolarized quarks, longitudinally polarized quarks, and transversely polarized quarks. The structures proportional to the the hadron mass $M_h$ are \emph{intrinsic} twist-3 fragmentation correlation functions~\cite{Kanazawa:2015ajw}. We note that the whole purpose of the appearance of the hadron mass $M_h$ in parameterizations like eq.~(\ref{eq:Deltaq}) (and subsequent parameterizations below) is to match mass dimensions. Other scales may be possible as well, resulting in a redefinition of the twist-3 fragmentation function. In this paper we will focus on the chiral-even functions $D_T$ and $G_T$ only.
To treat the ultraviolet (UV) and infrared (IR) divergences that enter the factorized definitions of the fragmentation functions~\cite{Collins:1981uw,Collins:2011zzd}, 
we defined the parameterization in arbitrary $d=4-2\varepsilon$ dimensions.

A correlator for intrinsic anti-quark fragmentation may be pictorially represented as in fig.~\ref{fig:CorrQbQb} and defined likewise,
\begin{eqnarray}
\Delta_{ij}^{\bar{q}}(z) & = & \frac{1}{N_{c}}\sumint\int_{-\infty}^{\infty}\tfrac{\d\lambda}{2\pi}\,\mathrm{e}^{-i\frac{\lambda}{z}}\langle0|[\infty ;0]\,\bar{q}_{j}(0)\,|P_{h}S_{h};X\rangle\langle P_{h}S_{h};X|\,q_{i}(\lambda n)\,[\lambda;\infty ]|0\rangle.\label{eq:Deltaqb} 
\end{eqnarray}
The parameterization for $\Delta^{\bar{q}}(z)$ is the same as the one for $\Delta^q(z)$, with the obvious replacement of the flavor index $q$ by the anti-flavor index $\bar{q}$. In addition, the $E$ and $G$ anti-quark FFs acquire a different sign, i.e., ($D_1^q,\,H_1^q,\,D_T^q,\,H^q,\,H_L^q)\to (+D_1^{\bar{q}},\,+H_1^{\bar{q}},\,+D_T^{\bar{q}},\,+H^{\bar{q}},\,+H_L^{\bar{q}}$), respectively, and ($G_1^q,\,G_T^q,\,E^q,\,E_L^q)\to (-G_1^{\bar{q}},\,-G_T^{\bar{q}},\,-E^{\bar{q}},\,-E_L^{\bar{q}}$), respectively.

The correlator for intrinsic gluon fragmentation is shown as a diagram in fig.~\ref{fig:CorrGG}. Mathematically, it can be written as
\begin{eqnarray}
\Delta^{g;\mu\nu}(z) & = & \frac{1}{N_{c}^{2}-1}\sumint\int_{-\infty}^{\infty}\tfrac{\d\lambda}{2\pi}\,\mathrm{e}^{-i\frac{\lambda}{z}}\langle0|F^{n\mu}(0)\,[0\,;\,\infty]|P_{h}S_{h};X\rangle\langle P_{h}S_{h};X|[\infty\,;\,\lambda]\,F^{n\nu}(\lambda n)|0\rangle\nonumber \\
 & = & \frac{z^{2\varepsilon}}{z^2}\Big(-g_{T}^{\mu\nu}\,D_{1}^{g}(z)-S_L\,i\epsilon^{P_{h}n\mu\nu}\,G_{1}^{g}(z)\nonumber\\
 & &\hspace{1cm}-M_h \,n^{\lbrace \mu} \epsilon^{\nu \rbrace P_h n S_{hT}}\,D_T^g(z)-i\,M_h \,n^{[\mu} \epsilon^{\nu ]P_h n S_{hT}}\,G_T^g(z)\Big).\label{eq:DeltagColl}
\end{eqnarray}
The matrix elements in the first line include the gluonic field-strength tensor $F^{\mu \nu}$. The symbols $\lbrace \mu\,\, \nu\rbrace$ and $[\mu\,\, \nu]$ that appear in the parameterization indicate symmetrization and antisymmetrization in the indices $\mu$ and $\nu$. As before, the FFs $D_1^g$ and $G_1^g$ are  twist-2 objects that describe the fragmentation of unpolarized and polarized gluons. The structures proportional to the hadron mass $M_h$ are intrinsic twist-3 gluon fragmentation correlation functions.

\subsubsection{Kinematical twist-3}

A different kind of two-parton fragmentation correlator is specific to twist-3 observables and takes the transverse motion of the fragmenting partons into account. Such contributions are called {\it kinematical} twist-3~\cite{Kanazawa:2015ajw}. Instead of the approximation in eq.~(\ref{eq:collkinAppr}), one adds a transverse parton momentum $p_T$ that is considered to be a small deviation from the otherwise collinear motion of the jet hadrons ``in" the parton,
\bea
p^\mu &\simeq & \frac{1}{z}P_h^\mu + p_T^\mu\,.\label{eq:pTAppr}
\eea
In fact, in practice one performs a Taylor expansion of the perturbative hard scattering subprocess with respect to $p_T$ to first order. This expansion is often called the {\it collinear expansion} (see, e.g., Ref.~\cite{Kouvaris:2006zy}). While the zeroth order constitutes the twist-2 contributions, the first order in this expansion yields the kinematical twist-3 contributions. Since single-particle inclusive processes are not directly sensitive to this transverse motion, the $p_T$-dependence will ultimately be integrated out. This leaves us with collinear matrix elements.

The kinematical twist-3 fragmentation correlations for quarks and anti-quarks are written in terms of $p_T$-dependent gauge-invariant matrix elements,\\

\bea
\Delta_{ij}^{q}(z,p_T) & = & \frac{1}{N_{c}}\sumint\int_{-\infty}^{\infty}\tfrac{\d\lambda}{2\pi}\int \tfrac{\d^{d-2}z_T}{(2\pi)^{d-2}}\,\mathrm{e}^{-i\frac{\lambda}{z}-i p_T\cdot z_T}\nonumber \\
&&\times\, \langle0|\mathcal{W}[0_T]\,q_{i}(0)\,|P_{h}S_{h};X\rangle\langle P_{h}S_{h};X|\,\bar{q}_{j}(\lambda n+z_T)\,\mathcal{W}^\dagger[z_T]|0\rangle,\label{eq:DeltaqTMD}\\[0.3cm]
\Delta_{ij}^{\bar{q}}(z,p_T) & = & \frac{1}{N_{c}}\sumint\int_{-\infty}^{\infty}\tfrac{\d\lambda}{2\pi}\int \tfrac{\d^{d-2}z_T}{(2\pi)^{d-2}}\,\mathrm{e}^{-i\frac{\lambda}{z}-i p_T\cdot z_T}\nonumber\\
&&\times\, \langle0|\mathcal{W}[0_T]\,\bar{q}_{j}(0)\,|P_{h}S_{h};X\rangle\langle P_{h}S_{h};X|\,q_{i}(\lambda n+z_T)\,\mathcal{W}^\dagger[z_T]|0\rangle.\label{eq:DeltaqbTMD} 
\eea
The Wilson line is non-trivial for a TMD correlator, and one may assume a common ``staple-like" form~\cite{Collins:2002kn,Ji:2002aa,Belitsky:2002sm,Boer:2003cm,Collins:2004nx,Bacchetta:2006tn}. However, the full $p_T$ dependence is in fact not needed in the collinear twist-3 formalism -- it is the ``$p_T$-weighted" correlator that enters, 
\bea
\Delta_{\partial;ij}^{q;\rho}(z) & = & \int \d ^{d-2}p_T\,p_T^\rho\,\Delta^q_{ij}(z,p_T)\nonumber\\
& = & \frac{z^{2\varepsilon}}{z}\,M_h\,\Big(\epsilon^{P_{h}n\rho S_{hT}}\,\slash P_{h}\,D_{1T}^{\perp(1),q}(z)-S_{hT}^{\rho}\,\slash P_{h}\gamma_{5}\,G_{1T}^{\perp(1),q}(z)\nonumber \\
 &  &\hspace{1.5cm} +\tfrac{i}{2}\,[\slash P_{h},\gamma^{\rho}_T]\,H_{1}^{\perp(1),q}(z)+\tfrac{1}{2}\,S_L\,[\slash P_{h},\gamma^{\rho}_T]\gamma_{5}\,H_{1L}^{\perp(1),q}(z)\Big).\label{eq:Deltaqkintw3}
\eea
The parameterization in the second line is again taken from ref.~\cite{Kanazawa:2015ajw}. This correlator may also be depicted as in fig.~\ref{fig:CorrQQ} but with $p$ defined as in eq.~(\ref{eq:pTAppr}). Note that it is entirely proportional to the hadron mass $M_h$, which indicates the twist-3 nature of these correlations.

Also, the anti-quark kinematical twist-3 FFs may be derived from the TMD anti-quark correlator eq.~(\ref{eq:DeltaqbTMD}) by $\Delta^{\bar{q},\rho}_\partial(z)=\int d^{d-2}p_T\,p_T^\rho\,\Delta^{\bar{q}}(z,p_T)$. The parameterization is the same as in eq.~(\ref{eq:Deltaqkintw3}) but with a different sign for $G_{1T}^{\perp (1)}$, i.e., ($D_{1T}^{\perp (1),q},\,H_1^{\perp (1),q},\,H_{1L}^{\perp (1),q})\to (+D_{1T}^{\perp (1),\bar{q}},\,+H_1^{\perp (1),\bar{q}},\,+H_{1L}^{\perp (1),\bar{q}}$), respectively, and $G_{1T}^{\perp (1),q}\to - G_{1T}^{\perp (1),\bar{q}}$.

Lastly, we discuss the kinematical twist-3 contributions for gluons. The gluon TMD fragmentation correlator can be defined as

\bea
\Delta^{g,\mu\nu}(z,p_T) & = & \frac{1}{N^2_{c}-1}\sumint\int_{-\infty}^{\infty}\tfrac{\d\lambda}{2\pi}\int \tfrac{\d^{d-2}z_T}{(2\pi)^{d-2}}\,\mathrm{e}^{-i\frac{\lambda}{z}-i p_T\cdot z_T}\nonumber \\
&&\hspace{-1cm}\times\,\langle0|\mathcal{W}[0_T]\,F^{n\mu}(0)\,|P_{h}S_{h};X\rangle\langle P_{h}S_{h};X|\,F^{n\nu}(\lambda n+z_T)\,\mathcal{W}^\dagger[z_T]|0\rangle.\label{eq:DeltagTMD}
\eea

Then, the kinematical twist-3 correlator for gluons is again obtained by a $p_T$-weighting,
\begin{eqnarray}
\Delta_{\partial}^{g;\mu\nu;\rho}(z) & = & \int \d^{d-2}p_T\,p_T^\rho\,\Delta^{g,\mu \nu}(z,p_T)\nonumber\\
 & = & \frac{z^{2\varepsilon}}{z^2}M_{h}\Big(g_{T}^{\mu\nu}\,\epsilon^{P_{h}n\rho S_{h}}\,D_{1T}^{\perp(1)g}(z)+\,i\epsilon^{P_{h}n\mu\nu}\,S_{hT}^{\rho}\,G_{1T}^{\perp(1)g}(z)\nonumber \\
 &  & \hspace{1.5cm}-\tfrac{1}{2}\Big(g_{T}^{\rho\lbrace \mu}\epsilon^{\nu\rbrace P_{h}n S_{h}}+S_{hT}^{\lbrace \mu}\epsilon^{\nu\rbrace P_{h}n\rho}\Big)\,H_{1}^{(1)g}(z)\Big).\label{eq:DeltapartgColl}
\end{eqnarray}
Each of the intrinsic and kinematic twist-3 FFs depend on the momentum fraction $z$. The support of these functions is $z\in [0,1]$. An implicit assumption is that these FFs vanish for $z=1$.

\subsection{Dynamical twist-3}

\begin{figure}
\centering
\begin{subfigure}[b]{0.4\textwidth}
\includegraphics[width=\textwidth]{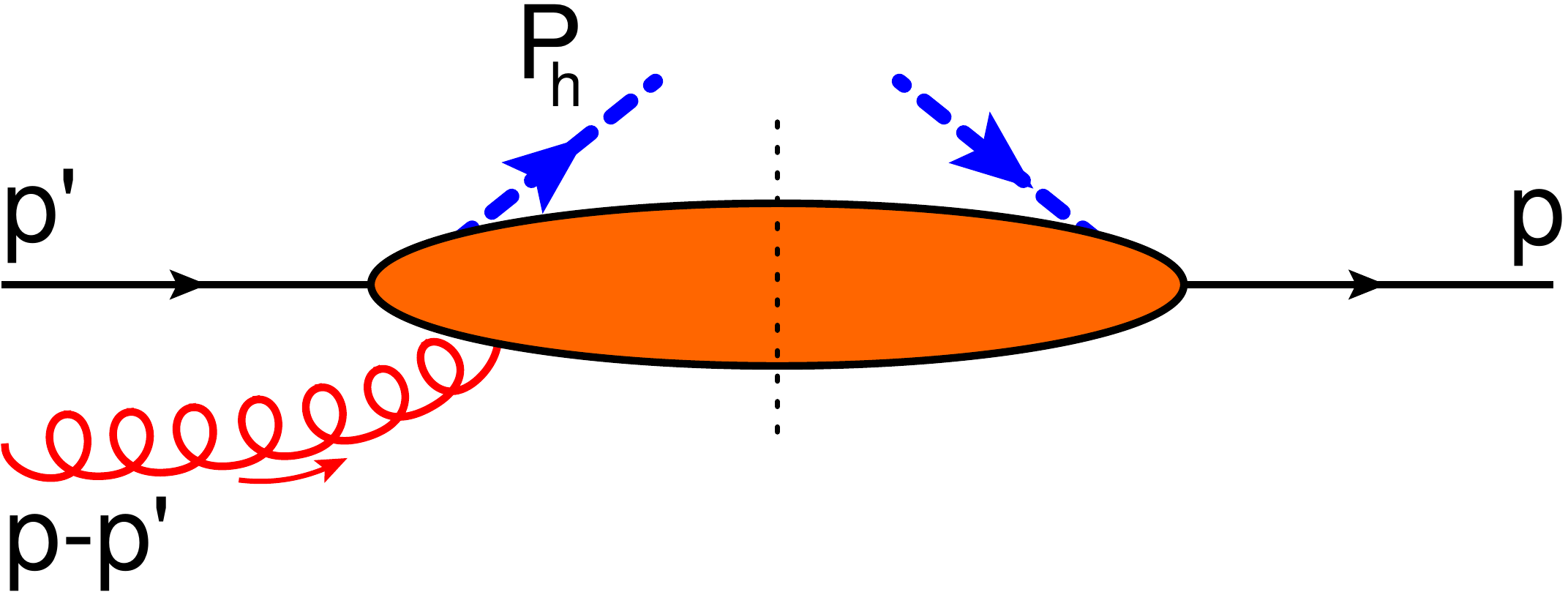}
\caption{$qg$ fragmentation}
\label{fig:CorrQGQ}
\end{subfigure}
\hskip 0.1in
\begin{subfigure}[b]{0.4\textwidth}
 \raisebox{0.15cm}{\includegraphics[width=\textwidth]{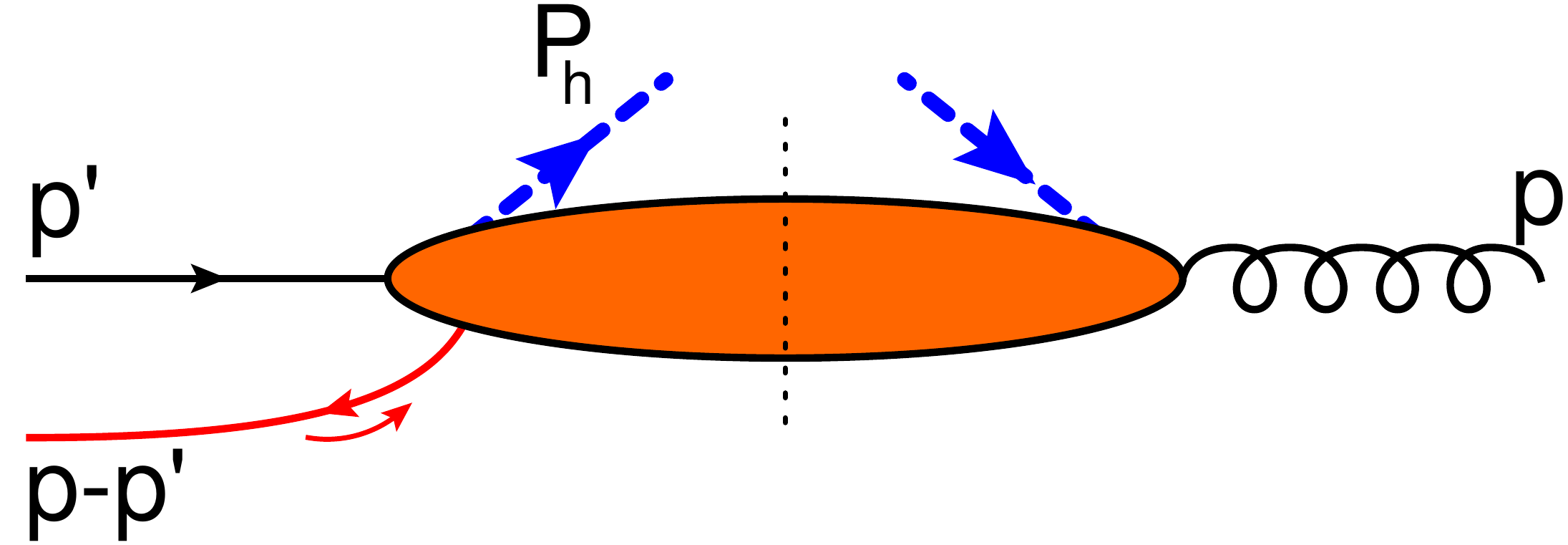}}
\caption{$q\bar{q}$ fragmentation}
\label{fig:CorrQQbG}
\end{subfigure}
\hskip 0.1in \\[0.3cm]
\begin{subfigure}[b]{0.4\textwidth}
\includegraphics[width=\textwidth]{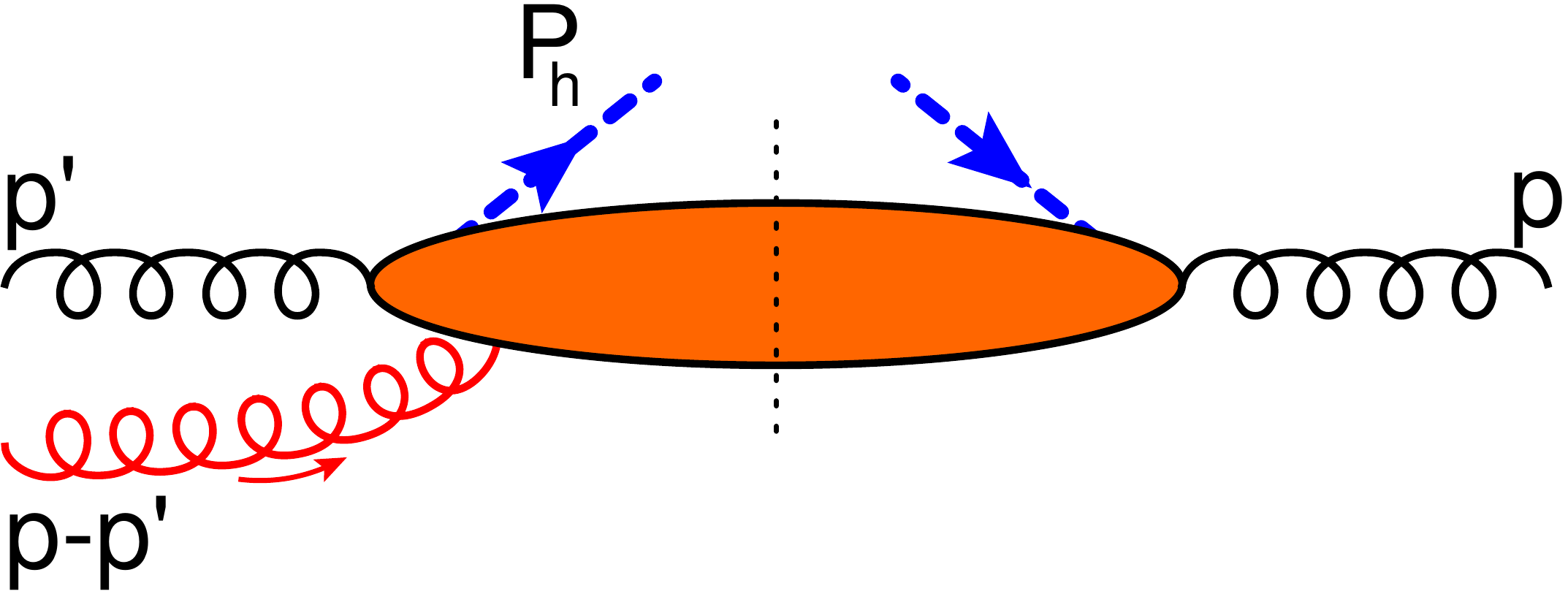}
\caption{$gg$ fragmentation}
\label{fig:CorrGGG}
\end{subfigure}

\caption{Diagrammatic representation of three-parton fragmentation correlations.}
\label{fig:3PartonFF}
\end{figure}

Matrix elements involving three partonic fields are called {\it dynamical twist-3} FFs. In general, such structures are generated through an interference of two amplitudes:~one that is a coherent fragmentation of two partons into a hadron, and another that is the ordinary one-parton fragmentation. The relevant matrix elements are depicted in fig.~\ref{fig:3PartonFF}. 

Since more than one parton is responsible for the hadronization into the observed hadron, naturally the matrix element will depend on more than one momentum. To ensure momentum conservation, the momenta in the two-parton amplitude are $p'$, $(p-p')$ while the momentum in the one-parton amplitude is $p$. As for the other twist-3 effects, only collinear matrix elements are needed for single-inclusive processes. Hence, we approximate, analogous to eq.~(\ref{eq:collkinAppr}), both partons to move collinearly in the same direction as the observed hadron,
\begin{equation}
p^\mu \simeq  \frac{1}{z}P_h^\mu \,,\,\,\,\,\,\,\,  p^{\prime \mu}\simeq \frac{1}{z^\prime}P_h^\mu.\label{eq:collkinApprdyn}
\end{equation}

For later convenience, we may rewrite the second momentum fraction $z^\prime$ as $z^\prime=z/\beta$. The collinear dynamical twist-3 matrix elements then depend on the light-cone momentum fraction $z$ and the parameter $\beta$. It is a well-known property that so-called soft-pole fragmentation matrix elements vanish \cite{Meissner:2008yf}. In other words, if $D(z,\beta)$ is a generic dynamical twist-3 fragmentation function, then $D(z,\beta=1)=0$ and $D(z,\beta=0)=0$.  Therefore, the support properties for $D(z,\beta)$ are $0 \le z \le 1$ and $0 < \beta < 1$. The last condition is equivalent to $z < z^\prime < \infty$. It has also been shown in ref.~\cite{Kanazawa:2015ajw} that the derivative with respect to $\beta$ vanishes for $\beta=1$, i.e., $(\partial D(z,\beta)/ \partial \beta) |_{\beta=1}=0$.  This proof can be easily modified to show $(\partial D(z,\beta)/ \partial \beta) |_{\beta=0}=0$ as well.

\subsubsection{Quark-gluon-quark correlations \label{qgqcor}}

An important class of dynamical twist-3 FFs are those involving quark-gluon correlations, cf. fig.~\ref{fig:CorrQGQ}. This means that a quark and a gluon radiated into the final state of a particular process together fragment and hadronize. Mathematically, this diagram can be expressed as follows,
\begin{eqnarray}
\Delta_{F;ij}^{qg;\rho}(z,\beta) & = & \frac{1}{N_{c}}\sumint\int_{-\infty}^{\infty}\tfrac{\d\lambda}{2\pi}\int_{-\infty}^{\infty}\tfrac{\d\mu}{2\pi}\,\e^{-i\frac{\lambda}{z}\beta}\e^{-i\frac{\mu}{z}(1-\beta)}\langle0|\,[\infty ;\,0]\,q_{i}(0)\,|P_{h}S_{h};X\rangle\nonumber \\
 &  & \hspace{3.5cm}\times\,\langle P_{h}S_{h};X|\,\bar{q}_{j}(\lambda m)\,[\lambda ;\,\mu ]\,ig\mu^{\varepsilon}\,F^{n\rho}(\mu n)\,[\mu ;\,\infty]|0\rangle\nonumber \\[0.3cm]
 & = & z^{2\varepsilon}\tfrac{M_{h}}{z}\Big(\epsilon^{P_{h}n\rho S_{h}}\slash P_{h}\,i(\hat{D}_{FT}^{qg})^{\ast}(z,\beta)+S_{hT}^{\rho}\,\slash P_{h}\gamma_{5}\,(\hat{G}_{FT}^{qg})^{\ast}(z,\beta)\nonumber \\
 &  & +\tfrac{i}{2}[\slash P_{h},\gamma^{\rho}_T]\,i(\hat{H}_{FU}^{qg})^{\ast}(z,\beta) -\tfrac{1}{2}S_{hL}[\slash P_{h},\gamma^{\rho}_T]\gamma_{5}\,(\hat{H}_{FL}^{qg})^{\ast}(z,\beta)\Big).\label{eq:DeltaqF}
\end{eqnarray}
The parameterization in the second equation is taken from ref.~\cite{Kanazawa:2015ajw}, where for later convenience we present the complex conjugated correlator. Note that each of the three-parton FFs are complex due to the lack of a time-reversal constraint~\cite{Boer:2003cm}. 

The situation for anti-quark-gluon fragmentation is handled as above for the intrinsic and kinematical twist-3 cases. The relevant matrix element reads
\bea
\Delta_{F;ij}^{\bar{q}g;\rho}(z,\beta) & = & \frac{1}{N_{c}}\sumint\int_{-\infty}^{\infty}\tfrac{\d\lambda}{2\pi}\int_{-\infty}^{\infty}\tfrac{\d\mu}{2\pi}\,\e^{-i\frac{\lambda}{z}\beta}\e^{-i\frac{\mu}{z}(1-\beta)}\langle0|\,\bar{q}_{j}(0)\,[0;\,\infty]|P_{h}S_{h};X\rangle\nonumber\\
 &  & \hspace{2cm}\times\,\langle P_{h}S_{h};X|\,[\infty;\,\mu]\,ig\mu^{\varepsilon}\,F^{n\rho}(\mu n)\,[\mu ;\,\lambda]\,q_{i}(\lambda n)\,|0\rangle.\label{eq:DeltaqbF} 
\eea
The parameterization of $\Delta_{F}^{\bar{q}g}$ is similar to the one for $\Delta_F^{qg}$ but with the obvious replacements $(\hat{D}_{FT}^{qg},\hat{G}_{FT}^{qg},\hat{H}_{FU}^{qg},\hat{H}_{FL}^{qg})\to (+ \hat{D}_{FT}^{\bar{q}g}, - \hat{G}_{FT}^{\bar{q}g}, + \hat{H}_{FU}^{\bar{q}g}, + \hat{H}_{FL}^{\bar{q}g})$, respectively.

\subsubsection{Quark-anti-quark-gluon correlations}

The situation where a quark-anti-quark fragmentation amplitude interferes with a one-gluon amplitude is represented by the diagram in fig.~\ref{fig:CorrQQbG}. Mathematically, the graph leads to a correlator $\Delta^{q\bar{q}}$, defined as
\bea
\Delta_{F;ij}^{q\bar{q};\rho}(z,\beta) & = & \frac{1}{N_{c}}\sumint\int_{-\infty}^{\infty}\tfrac{\d\lambda}{2\pi}\int_{-\infty}^{\infty}\tfrac{\d\mu}{2\pi}\,\e^{-i\frac{\lambda}{z}\beta}\e^{-i\tfrac{\mu}{z}(1-\beta)} \langle0|\,([\infty ;\,0]\,F^{n\rho}(0)\,[0;\,\infty])_{ba}\,|P_{h}S_{h};X\rangle\times\nonumber\\
 &  &\hspace{3cm} \langle P_{h}S_{h};X|\,\left(ig\mu^{\varepsilon}[\infty;\,\mu]\,q_{i}(\mu n)\right)_a\,\left(\bar{q}_{j}(\lambda n)\,[\lambda;\,\infty]\right)_b\,|0\rangle.\label{eq:DeltaqqbF}
\eea
The difference between $\Delta^{q\bar{q}}_F$ and the quark-gluon correlator~(\ref{eq:DeltaqF}) is that within $\Delta^{qg}_F$ the quark field in the one-quark fragmentation amplitude and the gluon field in the quark-gluon fragmentation amplitude exchange their role. This implies that its parameterization of $\Delta^{q\bar{q}}_F$ is completely analogous to eq.~(\ref{eq:DeltaqF}), subject to the replacements $(\hat{D}_{FT}^{qg}, \hat{G}_{FT}^{qg},\hat{H}_{FU}^{qg},\hat{H}_{FL}^{qg})\to  (\hat{D}_{FT}^{q\bar{q}}, \hat{G}_{FT}^{q\bar{q}}, \hat{H}_{FU}^{q\bar{q}}, \hat{H}_{FL}^{q\bar{q}}$), respectively.  In eq.~(\ref{eq:DeltaqqbF}) we wrote explicitly how to understand the trace of the color indices $a,\, b$ in the fundamental representation.

Another relevant correlator may be obtained from eq.~(\ref{eq:DeltaqqbF}) by exchanging the role of the quark and the anti-quark field,
\bea
\Delta_{F;ij}^{\bar{q}q;\rho}(z,\beta) & = & \frac{1}{N_{c}}\sumint\int_{-\infty}^{\infty}\tfrac{\d\lambda}{2\pi}\int_{-\infty}^{\infty}\tfrac{\d\mu}{2\pi}\,\e^{-i\frac{\lambda}{z}\beta}\e^{-i\tfrac{\mu}{z}(1-\beta)} \langle0|\,([\infty ;\,0]\,F^{n\rho}(0)\,[0;\,\infty])_{ba}\,|P_{h}S_{h};X\rangle\nonumber\\
&  &\hspace{2cm}\times\, \langle P_{h}S_{h};X|\,\left([\infty;\,\lambda]\,q_{i}(\lambda n)\right)_b\,\left(ig\mu^{\varepsilon}\bar{q}_{j}(\mu n)\,[\mu;\,\infty]\right)_a|0\rangle.\label{eq:DeltaqbqF}
\eea
The parameterization of this object is again analogous to eq.~(\ref{eq:DeltaqF}), but with different labels (and, analogously, signs) for the FFs, i.e., we may write $(\hat{D}_{FT}^{qg}, \hat{G}_{FT}^{qg},\hat{H}_{FU}^{qg}, \hat{H}_{FL}^{qg})\to (\hat{D}_{FT}^{\bar{q}q},- \hat{G}_{FT}^{\bar{q}q},\hat{H}_{FU}^{\bar{q}q},\hat{H}_{FL}^{\bar{q}q})$, respectively.  By comparing the two correlators (\ref{eq:DeltaqqbF}) and (\ref{eq:DeltaqbqF}), we find the following symmetry relations,
 \bea
\hat{D}_{FT}^{q\bar{q}}(z,\beta)&=&\hat{D}_{FT}^{\bar{q}q}(z,1-\beta)\,,\nonumber\\[0.1cm]
\hat{G}_{FT}^{q\bar{q}}(z,\beta)&=&-\hat{G}_{FT}^{\bar{q}q}(z,1-\beta)\,,\nonumber\\[0.1cm]
\hat{H}_{FU}^{q\bar{q}}(z,\beta)&=&\hat{H}_{FU}^{\bar{q}q}(z,1-\beta)\,,\nonumber\\[0.1cm]
\hat{H}_{FL}^{q\bar{q}}(z,\beta)&=&\hat{H}_{FL}^{\bar{q}q}(z,1-\beta)\,.\label{eq:QQbsym}
\eea
In addition, integration of the correlators (\ref{eq:DeltaqqbF}) and (\ref{eq:DeltaqbqF}) over $\beta$ leads to the same result. In particular, this means that
\bea
\int_0^1\d\beta\,\hat{D}_{FT}^{q\bar{q}}(z,\beta)=\int_0^1\d\beta\,\hat{D}_{FT}^{\bar{q}q}(z,\beta)\,.\label{eq:symDFTQQb}
\eea

\subsubsection{Tri-gluon correlations}
The third species of three-parton twist-3 fragmentation is represented by the diagram in fig.~\ref{fig:CorrGGG} where a two-gluon fragmentation amplitude interferes with a one-gluon amplitude. This diagram leads to a formula for the antisymmetric {\it tri-gluon correlator},
\begin{eqnarray}
  \Delta_{F}^{gg;\mu\nu\rho}(z,\beta)& = & \frac{1}{N_{c}^{2}-1}\sumint\int_{-\infty}^{\infty}\frac{\d\lambda}{2\pi}\int_{-\infty}^{\infty}\frac{\d\mu}{2\pi}\,\mathrm{e}^{-i\frac{\lambda}{z}\beta}\e^{-i\frac{\mu}{z}(1-\beta)}\,if^{\alpha\beta\gamma}\times\nonumber \\
 &  & \hspace{1cm}\langle0|\,F^{n\mu,\alpha}(0)\,|P_{h}S_{h};X\rangle  \langle P_{h}S_{h};X|\,F^{n\nu,\beta}(\lambda n)ig\mu^{\varepsilon}\,F^{n\rho,\gamma}(\mu n)\,|0\rangle\nonumber \\[0.3cm]
&=&-z^{2\varepsilon}\frac{M_h}{z^2}\Big[g_T^{\mu \nu}i\epsilon^{P_h n\rho S_h} \hat{N}_2^\ast(z,\beta)-g_T^{\mu\rho}i\epsilon^{P_h n \nu S_h} \hat{N}_2^\ast(z,1-\beta)\nonumber\\
&&\hspace{2cm}-g_T^{\nu\rho}i\epsilon^{P_h n \mu S_h}\hat{N}_1^\ast(z,\beta)\Big]\,.\label{eq:DeltagggF}
\end{eqnarray}
The matrix elements are understood to carry appropriate Wilson lines accompanying the field-strength tensors $F^{n\alpha}$ in eq.~(\ref{eq:DeltagggF})~\cite{Beppu:2010qn}, but for brevity we omitted the explicit notation of gauge links. The parameterization in (\ref{eq:DeltagggF}) is similar to ref.~\cite{Beppu:2010qn} for tri-gluon distributions in the nucleon. However, the permutation symmetry of the gluon fields for tri-gluon fragmentation (now that $|P_{h}S_{h};X\rangle  \langle P_{h}S_{h};X|$ is in between the fields of the matrix element) is such that there are two independent FFs $\hat{N}_1(z,\beta)$, $\hat{N}_2(z,\beta)$ instead of one like on the PDF side. Note that the antisymmetric SU($N_c$) structure constant $f^{\alpha\beta\gamma}$ appears in the definition of $\Delta_F^{gg}$. In principle, one may also define a similar symmetric tri-gluon correlator which involves the symmetric structure constant $d^{\alpha\beta\gamma}$. However, such a matrix element will not appear in the single-inclusive spin-dependent $e^+e^-$ cross section, and for that reason we do not further elaborate on the symmetric correlator in this paper. Note, however, that the symmetric tri-gluon correlator may contribute in $pp$-collisions~\cite{Koike:2011mb,Koike:2011nx,Beppu:2013uda}.

There is a symmetry relation for the correlator $\Delta^{gg}_F$. We could write the gluon bilinear in the second matrix element of eq.~(\ref{eq:DeltagggF}) as a time-ordered bilinear. In ref.~\cite{Jaffe:1983hp} arguments are given that the time-ordering is irrelevant. This allows us to re-order the gluonic fields in the second matrix element of (\ref{eq:DeltagggF}). A subsequent relabeling of the integration variables $\mu\leftrightarrow \lambda$ leads to the relation,
\bea
\Delta_F^{gg;\mu\nu\rho}(z,\beta)=-\Delta_F^{gg;\mu\rho\nu}(z,1-\beta).\label{eq:relationDeltagggF}
\eea
Note that the sign in eq.~(\ref{eq:relationDeltagggF}) originates from an exchange of color indices in the antisymmetric structure constant $f^{\alpha\beta\gamma}$. The symmetry relation (\ref{eq:relationDeltagggF}) translates into a relation directly for the function $\hat{N}_1$,
\bea
\hat{N}_1(z,\beta)=-\hat{N}_1(z,1-\beta)\,.\label{eq:SymN1}
\eea
There is no symmetry constraint for the other function $\hat{N}_2$. This means means that $\hat{N}_2$ is the sum of a symmetric and antisymmetric part, $\hat{N}_2^{s/a}(z,\beta)\equiv (\hat{N}_2(z,\beta)\pm \hat{N}_2(z,1-\beta))/2$, respectively.

\subsection{Equation of motion relations}

The aforementioned various twist-3 fragmentation matrix elements are not completely independent of each other. In fact, one may derive constraints by means of the QCD-equation of motion (EOM) for Heisenberg field operators within matrix elements.

The QCD EOM for quark fields $q(x)$ reads
\bea
i\slash D(x)\,q(x)-m_q\,q(x) &=& 0\,,\label{eq:EoMQ}
\eea
where $D^\mu_{ab}(x)\equiv\delta_{ab}\partial^\mu-igA^\mu_{ab}(x)$ is the well-known covariant derivative in the fundamental representation. The application of this equation on matrix elements like $\Delta^q$, $\Delta_\partial^q$ and $\Delta_F^{qg}$ leads to the following EOM relations (EoMRs) (cf. ref.~\cite{Kanazawa:2015ajw}),
\bea
\frac{D_T^q(z)}{z}&=&-D_{1T}^{\perp (1),q}(z)+\int_0^1\d\beta\,\frac{\Im[\hat{D}_{FT}^{qg}-\hat{G}_{FT}^{qg}](z,\beta)}{1-\beta}\,,\label{eq:EoMQD}\\
\frac{G_T^q(z)}{z}&=&G_{1T}^{\perp (1),q}(z)-\int_0^1\d\beta\,\frac{\Re[\hat{D}_{FT}^{qg}-\hat{G}_{FT}^{qg}](z,\beta)}{1-\beta}\,,\label{eq:EoMQG}
\eea
where $\Im$ ($\Re$) indicates the imaginary (real) part of the functions.  We note that one can also derive constraints for the chiral-odd functions $H$ and $E$ in eqs.~(\ref{eq:Deltaq}), (\ref{eq:Deltaqkintw3}), and (\ref{eq:DeltaqF}). Since they do not contribute to the spin-dependent single-inclusive annihilation cross section, such constraints are irrelevant for this paper and for brevity we do not list them here. (They can be found in ref.~\cite{Kanazawa:2015ajw}.) We do note, however, that the EoMRs (\ref{eq:EoMQD}) and (\ref{eq:EoMQG}) are absolutely essential, as they are necessary for the restoration of gauge invariance of hard scattering cross sections at twist-3 as well as for the cancellation of infrared divergences.  We will discuss this explicitly in the following sections. Consequently, without eqs.~(\ref{eq:EoMQD}) and (\ref{eq:EoMQG}), the collinear twist-3 formalism used to describe transverse spin observables in single-inclusive processes would be flawed.  In addition, so-called Lorentz invariance relations (see the next subsection) are needed to establish the frame-independence of the cross section~\cite{Kanazawa:2015ajw}.

There are also EoMRs for gluon FFs. They can be derived from the inhomogeneous QCD EOM for gluons
\bea
D_{\mu}^{\alpha\beta}(x)\,F^{\mu \nu,\beta}(x)=-g\mu^{\varepsilon}\sum_q\,\bar{q}(x)\,\gamma^\nu\,t^\alpha\,q(x)\,,\label{eq:EoMG}
\eea
where the covariant derivative $D^{\mu;\alpha\beta}(x)=\delta^{\alpha\beta}\partial^\mu-g\,f^{\alpha\beta\gamma}A^{\mu;\gamma}(x)$ in the adjoint representation appears along with the gluonic field-strength tensor $F^{\mu\nu;\alpha}(x)=\partial^\mu A^{\nu;\alpha}(x)-\partial^{\nu}A^{\mu;\alpha}(x)+gf^{\alpha\beta\gamma}A^{\mu;\beta}(x)A^{\nu;\gamma}(x)$ and the color matrix $t^\alpha_{ab}$.

Application of eq.~(\ref{eq:EoMG}) on the matrix elements in $\Delta^g$, $\Delta^g_\partial$, $\Delta_F^{gg}$ and $\Delta_F^{q\bar{q}}$ yields the following constraints on twist-3 gluon FFs,
\bea
\frac{D_T^g(z)}{z}&=&D_{1T}^{\perp (1),g}(z)-(2-\varepsilon)\,H_1^{(1),g}(z)\nonumber\\
&&+\int_0^1\d\beta\,\frac{\Im[\hat{N}_2](z,\beta)-\Im[\hat{N}_2](z,1-\beta)\boldsymbol{-}2(1-\varepsilon)\Im[\hat{N}_1](z,\beta)}{1-\beta}\nonumber\\
& &-\frac{1} {C_F}\sum_{f=q,\bar{q}}\int_0^1\d\beta\,\Im[\hat{D}_{FT}^{f\bar{f}}](z,\beta)\,,\label{eq:EoMGD}\\[0.3cm]
\frac{G_T^g(z)}{z}&=&-G_{1T}^{\perp (1),g}(z)-\int_0^1\d\beta\,\frac{\Re[\hat{N}_2](z,\beta)-\Re[\hat{N}_2](z,1-\beta)\boldsymbol{-}2(1-\varepsilon)\Re[\hat{N}_1](z,\beta)}{1-\beta}\nonumber\\
&&+\frac{1} {C_F}\sum_{f=q,\bar{q}}\int_0^1\d\beta\,\Re[\hat{D}_{FT}^{f\bar{f}}](z,\beta)\,.\label{eq:EoMGG}
\eea
In the last lines of (\ref{eq:EoMGD}) and (\ref{eq:EoMGG}) we used the symmetry relation (\ref{eq:symDFTQQb}).

\subsection{Lorentz invariance relations}
There are also additional constraints, called Lorentz invariance relation (LIRs), derived in ref.~\cite{Kanazawa:2015ajw}, which connect the various twist-3 FFs for quarks.  The LIRs relevant for our calculation are
\begin{align}
\frac{D_T^q(z)}{z}&=-\left(1-z\frac{d}{d z}\right)\,D_{1T}^{\perp (1),q}(z)-2\int_0^1d \beta\,\frac{\Im[\hat{D}_{FT}^{qg}](z,\beta)}{(1-\beta)^2}\,,\label{eq:LIRD} \\[0.3cm]
\frac{G_T^q(z)}{z}&=\frac{G_1^q(z)}{z}+\left(1-z\frac{d}{d z}\right)\,G_{1T}^{\perp (1),q}(z)-2\int_0^1d \beta\,\frac{\Re[\hat{G}_{FT}^{qg}](z,\beta)}{(1-\beta)^2}\,.\label{eq:LIRG}
\end{align}
We emphasize that similar LIRs have not been derived so far in the literature for twist-3 FFs of gluons.

\section{Observables in single-inclusive annihilation at leading order\label{LO}}

\begin{figure}
\centering
\begin{subfigure}[b]{0.46\textwidth}
\includegraphics[width=\textwidth]{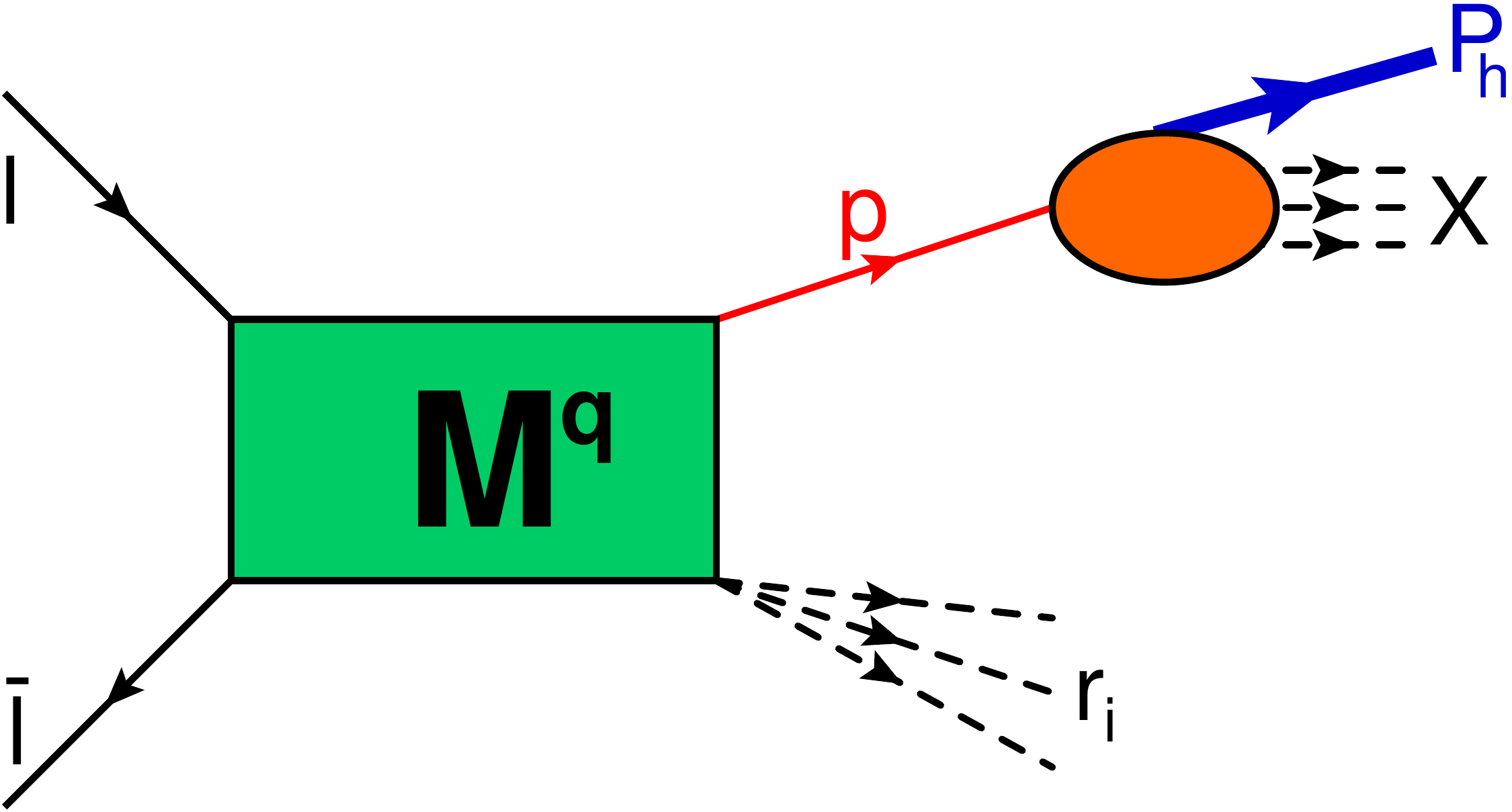}
\caption{Single quark fragmentation}
\label{fig:FragQ}
\end{subfigure}
\hskip 0.2in
\begin{subfigure}[b]{0.46\textwidth}
\includegraphics[width=\textwidth]{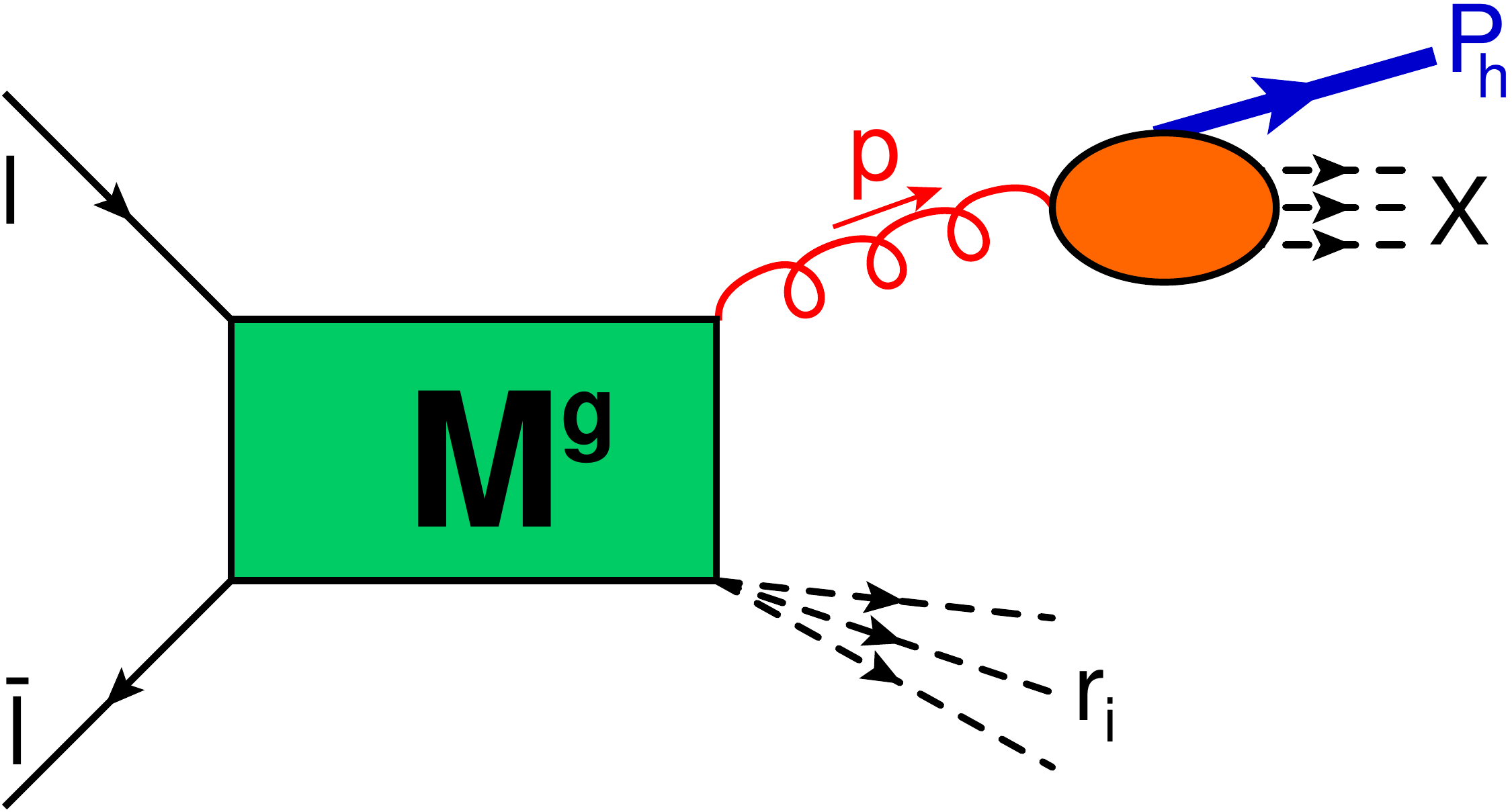}
\caption{Single gluon fragmentation}
\label{fig:FragG}
\end{subfigure}

\caption{Fragmentation mechanism in $e^+e^-$ annihilation for intrinsic and kinematical contributions.}
\label{fig:Ampq_g}
\end{figure}

After having introduced all relevant twist-3 FFs, we proceed with a discussion on transverse spin observables in the process $e^+e^-\to\Lambda\, X$. We denote the momenta of the lepton and anti-lepton by $l$ and $l^\prime$, respectively, and the momentum of the virtual photon by $q=l+l^\prime$. The typical scaling variable of this process is $z_h=2P_h\cdot q/s$. Another useful variable for the description of this reaction may be denoted as $v=P_h\cdot l^\prime /P_h\cdot q$. Throughout this paper we will work in a frame where $q^\mu$ has no {\it transverse} components, where the term ``transverse" is defined in eq.~(\ref{eq:transverseProj}). Such a choice is always possible. Consequently, the lepton and photon momentum
vectors can then be decomposed in terms of the variables $z_h$, $v$, the hard scale $s$, and the light-cone momenta $P_h$ and $n$ as follows,
\bea
q^\mu &=&\frac{1}{z_h}\,P_h^\mu+\frac{z_h}{2}s\,n^\mu\,,\nonumber\\
l^\mu &=& \frac{v}{z_h}\,P_h^\mu+\frac{z_h}{2}(1-v)\,s\,n^\mu+l_T^\mu\,,\nonumber\\
l^{\prime\mu} &=& \frac{1-v}{z_h}\,P_h^\mu+\frac{z_h}{2}v\,s\,n^\mu-l_T^\mu\,.\label{eq:ExtMomenta}
\eea
The fact that we neglect the lepton masses implies that $l_T^2=-v(1-v)\,s$.

There are several mechanisms that generate contributions to observables at twist-3, which involve the soft fragmentation matrix elements discussed in section \ref{Corr}. The amplitudes for intrinsic and kinematical twist-3 contributions to the cross section for fragmenting quarks and gluons are schematically shown in fig.~\ref{fig:Ampq_g}. In these diagrams the soft fragmentation of a quark or a gluon is already factored out from the hard scattering amplitudes $\mathcal{M}$. The fragmenting parton decays into an arbitrary hadronic final state. In addition, other $n_f$ partons carrying momenta $r_1,\,...,\,r_{n_f}$ may be emitted into the final state in the hard scattering process. Those momenta are integrated out. Note that the number of emitted partons is at least $n_f=1$. Typically, the hard scattering cross sections for fragmenting quarks and anti-quarks are the same, which is why we will not elaborate on anti-quarks separately. On the other hand, FFs for quarks and anti-quarks may very well differ.

\begin{figure}
\centering
\begin{subfigure}[b]{0.4\textwidth}
\includegraphics[width=\textwidth]{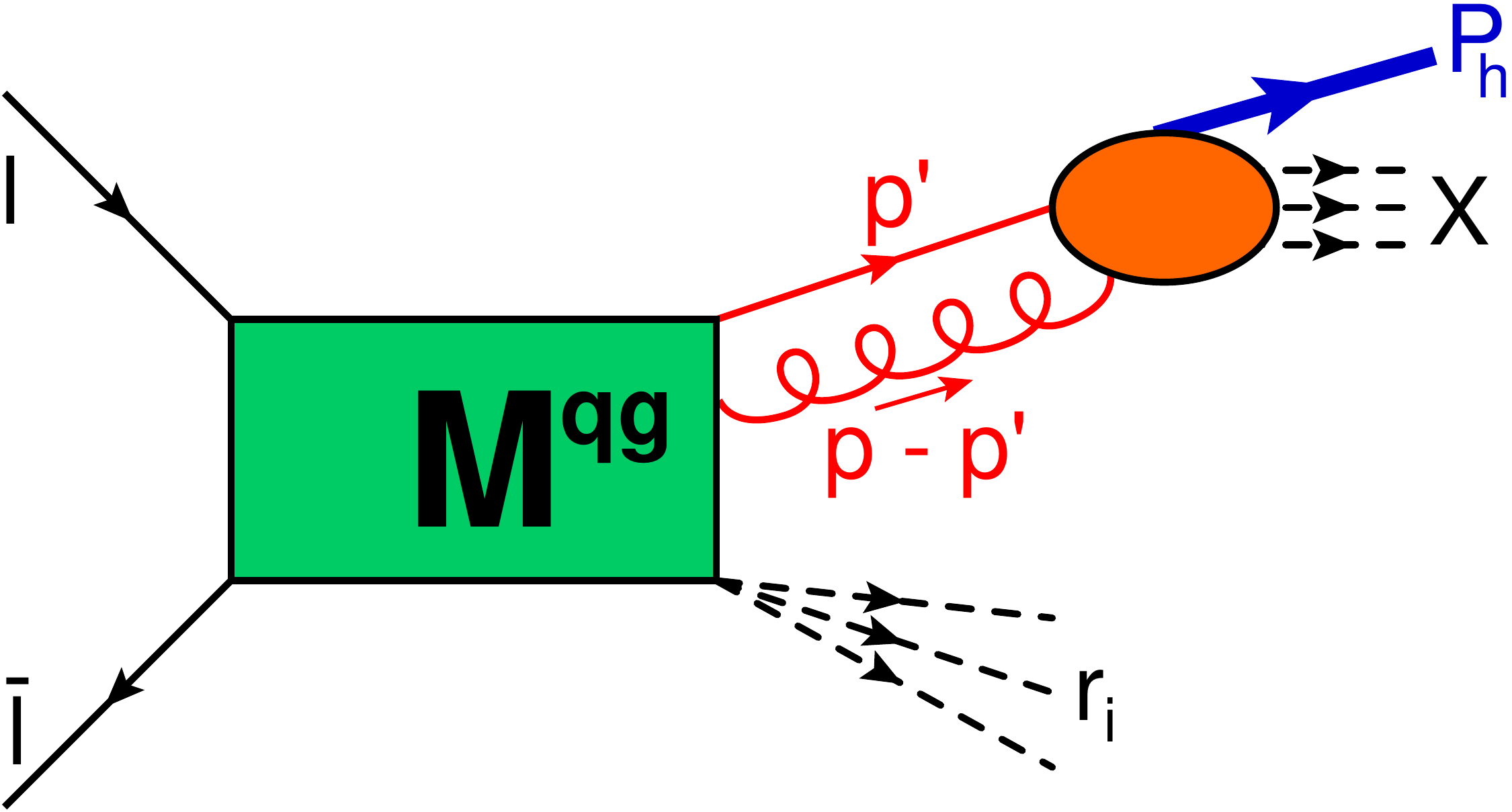}
\caption{$qg$ fragmentation}
\label{fig:FragQGQ}
\end{subfigure}
\hskip 0.2in
\begin{subfigure}[b]{0.4\textwidth}
\includegraphics[width=\textwidth]{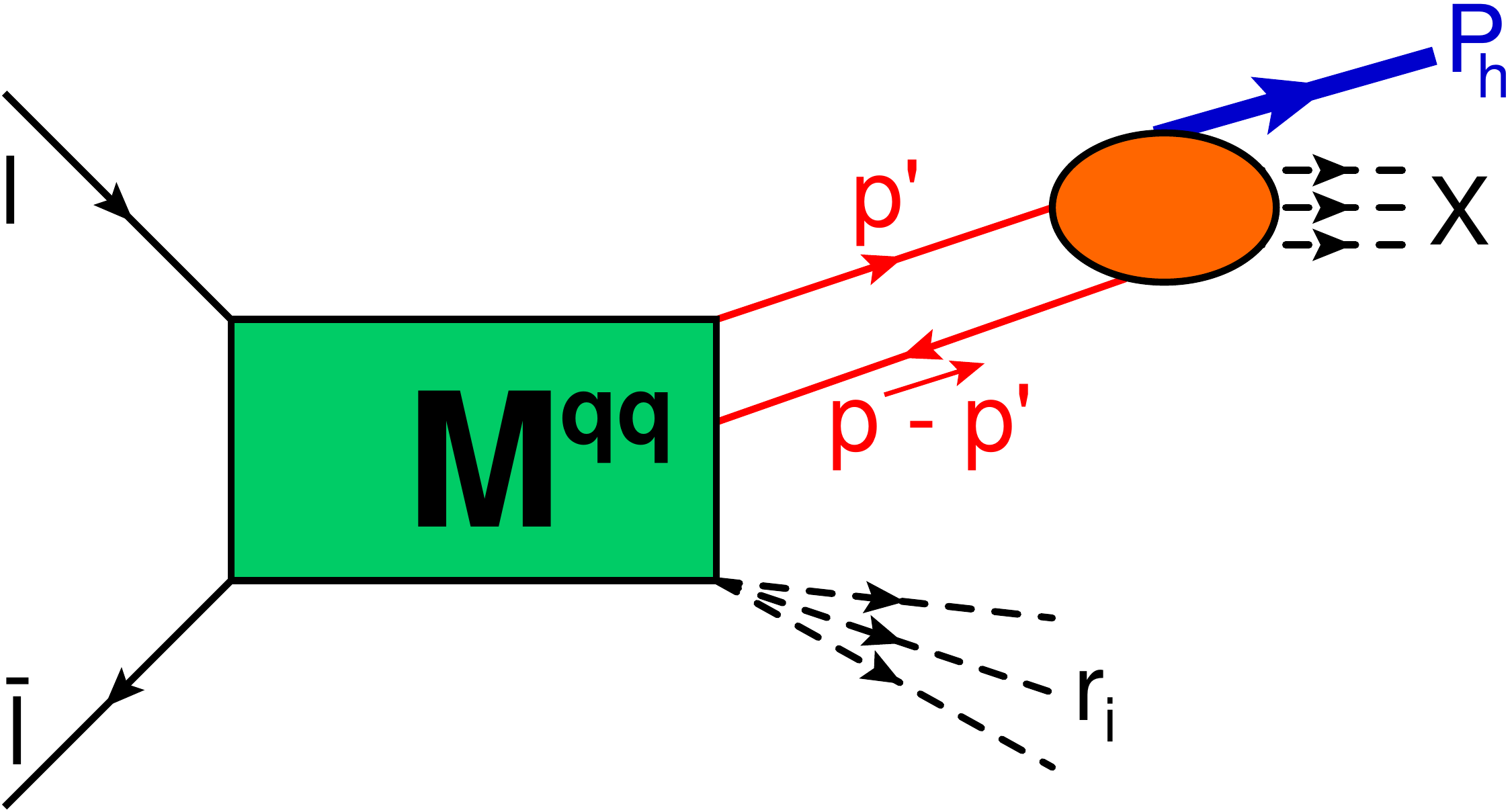}
\caption{$q\bar{q}$ fragmentation}
\label{fig:FragQQbG}
\end{subfigure}
\\[0.5cm]
\begin{subfigure}[b]{0.4\textwidth}
\includegraphics[width=\textwidth]{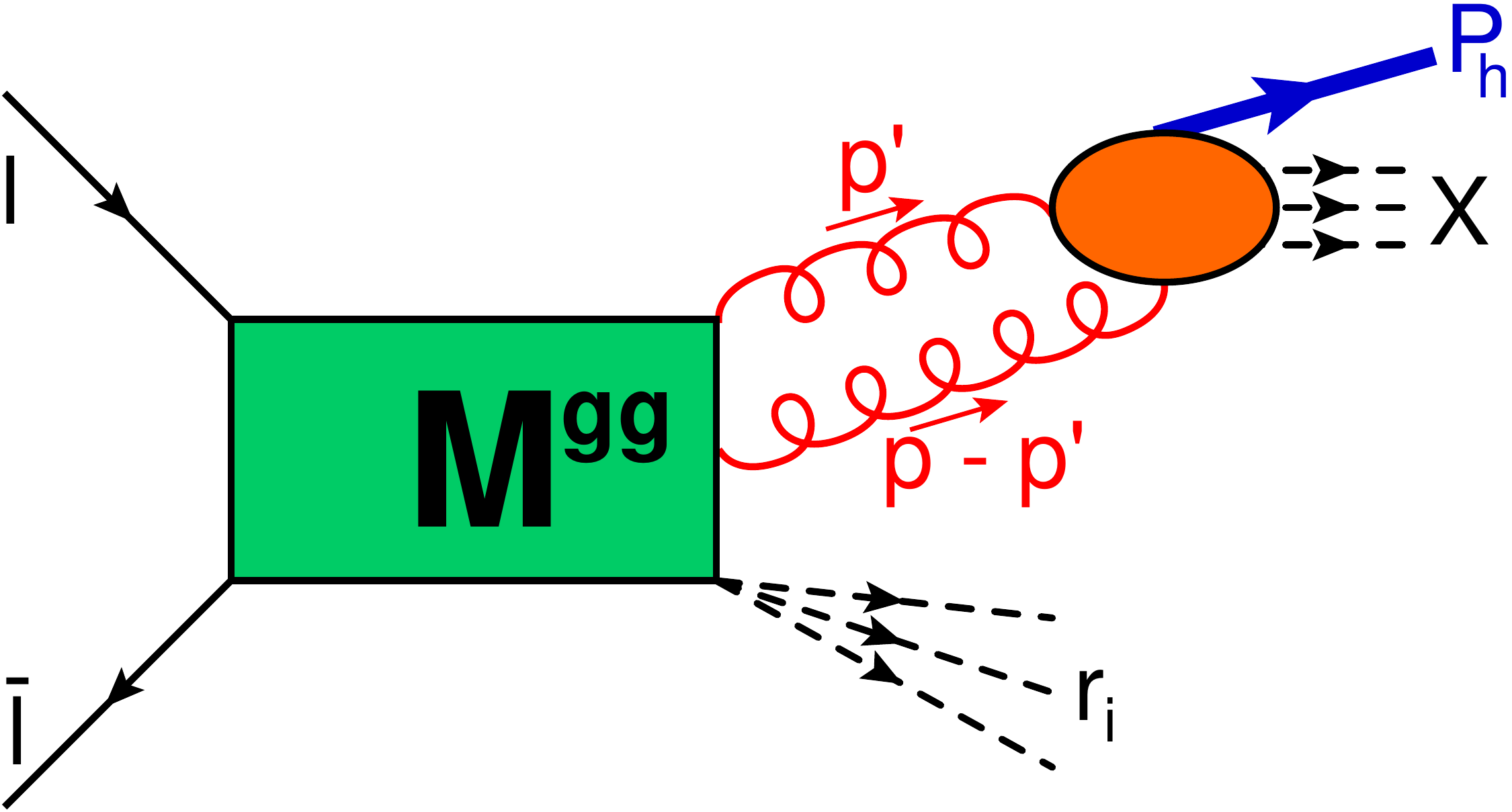}
\caption{$gg$ fragmentation}
\label{fig:FragGGG}
\end{subfigure}

\caption{Fragmentation mechanism in $e^+e^-$ annihilation for dynamical contributions.}
\label{fig:Ampqg_qq_gg}
\end{figure}

The dynamical twist-3 amplitudes are shown in fig.~\ref{fig:Ampqg_qq_gg} where two partons at the same time fragment into a hadron. Those amplitudes are meant to interfere with corresponding amplitudes in fig.~\ref{fig:Ampq_g} within a cross section formula.
The schematic diagrams in figs.~\ref{fig:Ampq_g}, \ref{fig:Ampqg_qq_gg} may be used to give a formula for the intrinsic, kinematical and dynamical twist-3 contributions to the cross section of the $e^+e^-$ single-inclusive annihilation process. Such a formula reads in $d=4-2\varepsilon$ dimensions,
\bea
E_h\frac{\d \sigma}{\d^{d-1}\vec{P}_h} & = & \int_{z_h}^1\d w\,\Bigg(\sum_{f=q,\bar{q}}\,\Tr\left[\hat{\sigma}^f(w)\,\Delta^f(\tfrac{z_h}{w})\right]+\hat{\sigma}^{g}_{\mu \nu}(w)\,\left(\tfrac{z_h}{w}\right)^2\Delta^{g;\mu \nu}(\tfrac{z_h}{w})\nonumber\\
& &+ \sum_{f=q,\bar{q}}\Tr\left[\left(\tfrac{\partial \hat{\sigma}^f}{\partial p_T^\rho}\right)\Big|_{p=\frac{w}{z_h}P_h}\,\Delta_\partial^{f;\rho}(\tfrac{z_h}{w})\right] +\left(\tfrac{\partial \hat{\sigma}^g_{\mu\nu}}{\partial p_T^\rho}\right)\Big|_{p=\frac{w}{z_h}P_h}\,\left(\tfrac{z_h}{w}\right)^2\Delta_\partial^{g;\mu \nu;\rho}(\tfrac{z_h}{w})\nonumber\\
&& +\int_0^1\d \beta\,\Bigg\{\sum_{f=q,\bar{q}}\left(\tfrac{-i}{1-\beta}\right)\,\Tr\left[\hat{\sigma}^{fg}_\rho(w,\beta)\,\Delta_F^{fg;\rho}(\tfrac{z_h}{w},\beta)\right]\nonumber\\
&& +\sum_{f=q,\bar{q}}\Tr\left[i\hat{\sigma}^{f\bar{f}}_\rho (w,\beta)\,\Delta_F^{f\bar{f};\rho}(\tfrac{z_h}{w},\beta)\right]\nonumber\\
&&+\left(\tfrac{-i z_h^2}{w^2\beta(1-\beta)}\right)\hat{\sigma}^{gg}_{\mu \nu \rho} (w,\beta)\,\Delta_F^{gg;\mu\nu\rho}(\tfrac{z_h}{w},\beta)+\mathrm{c.c.}\Bigg\}\Bigg)\,.\label{eq:SpinCSMaster}
\eea
This formula is most easily derived in light-cone gauge with asymmetric boundary conditions for the transverse gluon field components, 
\bea
n\cdot A(x)=0& \mathrm{and} & A_T(n\cdot x=+\infty)+A_T(n\cdot x=-\infty)=0.\label{eq:LCgauge}
\eea
In this gauge the fragmentation correlators $\Delta^{q,g}$, $\Delta^{qg}_\partial$ and $\Delta^{qg;q\bar{q};gg}_F$ simplify to a large extent since the Wilson lines reduce to unity and the field-strength tensors are simply $F^{n\mu}(x)={(n\cdot\partial) A_T^\mu(x)}$. The first line of (\ref{eq:SpinCSMaster}) represents the twist-2 and intrinsic twist-3 contributions of quarks and gluons to the hadron-spin dependent $e^+e^-$ cross section.  The second line gives the kinematical twist-3 contributions; the third, fourth and fifth lines give the dynamical twist-3 contributions of quark-gluon, quark-anti-quark, and gluon-gluon correlations. We note that if the process under consideration factorizes at twist-3, eq.~(\ref{eq:SpinCSMaster}) holds in any gauge.  In particular, in Feynman gauge, where $F^{n\rho}_T = (n\cdot\partial) A_T^\rho- \partial_T^\rho (n\cdot A)+...$, one is guaranteed that the term involving a matrix element with $(n\cdot\partial) A_T$ will combine with the term involving a matrix element with $\partial_T (n\cdot A)$ to give a contribution involving a gauge-invariant correlator with $F^{n\rho}_T$.  This was shown explicitly, e.g., in refs.~\cite{Eguchi:2006mc,Beppu:2010qn,Kanazawa:2013uia}.

The partonic cross sections $\hat{\sigma}$ in (\ref{eq:SpinCSMaster}) are provided by the following formulas,
\bea
\hat{\sigma}^q_{ji}(p)&=&\frac{(4\pi^2)^\varepsilon}{4(2\pi)^3\,z_h\,s} \sum_{n_f=1}^\infty \sum_{I_{n_f}}\int \d \mathrm{PS}_{n_f}\,\mathcal{M}_j^q(p)\,\bar{\mathcal{M}_i^q}(p)\,,\nonumber\\
\hat{\sigma}^{g;\mu\nu}(p)&=& \frac{(4\pi^2)^\varepsilon}{4(2\pi)^3\,z_h\,s}\sum_{n_f=2}^\infty \sum_{I_{n_f}}\int \d \mathrm{PS}_{n_f}\,\mathcal{M}^{g;\nu}(p)\,\left(\mathcal{M}^{g;\mu}\right)^\ast(p)\,,\nonumber\\
\hat{\sigma}^{qg;\rho}_{ji}(p,p^\prime)&=&\frac{(4\pi^2)^\varepsilon}{4(2\pi)^3\,z_h\,s} \sum_{n_f=1}^\infty \sum_{I_{n_f}}\int \d \mathrm{PS}_{n_f}\,\mathcal{M}_j^{qg;\rho}(p,p^\prime)\,\bar{\mathcal{M}_i^q}(p)\,,\nonumber\\
\hat{\sigma}^{q\bar{q};\rho}_{ji}(p,p^\prime)&=& \frac{(4\pi^2)^\varepsilon}{4(2\pi)^3\,z_h\,s}\sum_{n_f=2}^\infty \sum_{I_{n_f}}\int \d \mathrm{PS}_{n_f}\,\mathcal{M}_{ji}^{q\bar{q}}(p,p^\prime)\,\left(\mathcal{M}^{g;\rho}\right)^\ast(p)\,,\nonumber\\
\hat{\sigma}^{gg;\mu\nu\rho}(p,p^\prime)&=& \frac{(4\pi^2)^\varepsilon}{4(2\pi)^3\,z_h\,s}\sum_{n_f=2}^\infty \sum_{I_{n_f}}\int \d \mathrm{PS}_{n_f}\,\mathcal{M}^{gg;\nu \rho}(p,p^\prime)\,\left(\mathcal{M}^{g;\mu}\right)^\ast(p)\,.\label{eq:DefPartonicCS}
\eea
The scattering amplitudes $\mathcal{M}^{q,g}$ and $\mathcal{M}^{qg,q\bar{q},gg}$ can be calculated perturbatively by means of the usual Feynman rules with legs that connect to the soft fragmentation matrix elements being amputated. This amputation results in ``external" or ``open" Dirac- or Minkowski indices $i,j$ or $\mu,\nu,\rho$, respectively, in the scattering amplitudes in (\ref{eq:DefPartonicCS}). In addition, $ig\mu^\varepsilon$, along with a suitable color matrix, is factored out of the two-parton fragmentation scattering amplitudes $\mathcal{M}^{qg,q\bar{q},gg}$ and shifted into the definition of the three-parton fragmentation correlators $\Delta_F$. The ``barred" amplitude $\bar{\mathcal{M}}^q$ is defined as $\bar{\mathcal{M}}^q\equiv \mathcal{M}^\dagger \gamma^0$. 

The partonic factors in (\ref{eq:DefPartonicCS}) contain information on the leptonic annihilation into a virtual gauge boson (a photon in this case), and for unpolarized leptons include an average over the initial lepton helicities. One may also study the situation where one of the leptons is polarized and consider a lepton spin asymmetry. To summarize, we implicitly use the following sums or differences in the partonic cross sections in (\ref{eq:DefPartonicCS}), generically in the following form,
\bea
(\Delta)\hat{\sigma}=\frac{1}{4}\sum_{\lambda^\prime=\pm 1}\left((\mathcal{M}\mathcal{M}^\ast)(\lambda=+1,\lambda^\prime)\pm (\mathcal{M}\mathcal{M}^\ast)(\lambda=-1,\lambda^\prime)\right).\label{eq:DefLeptonHel}
\eea
The plus sign indicates the lepton spin average $\hat{\sigma}$, and the minus sign the lepton spin asymmetry $\Delta \hat{\sigma}$. On the other hand, all quantum numbers of unobserved final state partons are summed, as indicated by $\sum_{I_n}$ in (\ref{eq:DefPartonicCS}). This sum includes the $n_f$-dimensional Lorentz-invariant phase space integrals,
\bea
\int \d \mathrm{PS}_{n_f}\equiv \prod_{n=1}^{n_f}\int \frac{\d ^{d} r_n}{(2\pi)^{d-1}}\,\delta^+(r_n^2)\,(2\pi)^d\,\delta^{(d)}(q-p-R_{n_f}),\label{eq:PS}
\eea
where $\delta^+(a^2)=\theta(a\cdot n)\,\delta(a^2)$ and $R_{n_f}=\sum_nr_n$.

As already pointed out, we find that within the course of our calculations, light-cone gauge (\ref{eq:LCgauge}) is the preferable gauge for the collinear twist-3 formalism. Not only can the factorization formula (\ref{eq:SpinCSMaster}) be established in a straightforward manner, also the gauge invariance of the partonic cross sections (\ref{eq:DefPartonicCS}) can be tested. The reason for that is: wherever possible throughout our perturbative calculations we introduce a polarization sum over gluon polarization vectors of the following form,
\bea
-\sum_{\lambda=\pm1}(\epsilon_\lambda ^\mu)^\ast(p)\,\epsilon_\lambda^\nu(p)&\equiv&d^{\mu \nu}(p,n;\kappa)=g^{\mu\nu}-\kappa \frac{p^\mu n^\nu+p^\nu n^\mu}{p\cdot n+i\delta}.\label{eq:polsumLC}
\eea
This polarization sum also appears in the numerator of the gluon propagator. Switching the parameter $\kappa$ between 0 or 1 allows us to switch between covariant (Feynman) gauge and light-cone gauge. Eventually in the final result, the parameter $\kappa$ should not appear in the partonic cross sections (\ref{eq:DefPartonicCS}) if they are gauge-invariant. We consider this property as an important check for our results. While gauge-invariance is ensured in a simple way for twist-2 partonic cross sections, this is far less obvious for twist-3 partonic factors as there are many entangled contributions. We will show that gauge-invariant partonic twist-3 factors can only be obtained through application of the EoMRs (\ref{eq:EoMQD}) and (\ref{eq:EoMQG}).

One may also consider the electromagnetic (e.m.) gauge invariance of the partonic factors (\ref{eq:DefPartonicCS}) as an important check of the validity of the results. In view of this aspect one might work with a photon propagator in a general covariant gauge, i.e., a photon propagator with a numerator $g^{\mu\nu}-(1-\xi)q^{\mu}q^{\nu}/q^2$. Each partonic factor in (\ref{eq:DefPartonicCS}) can be separated into a well-known leptonic tensor $L_{\mu \nu}=\Tr[\gamma^\mu \slash{l}\gamma^{\nu}\slash{l}^\prime[\gamma_5]]$ and a hadronic tensor $W^{\mu\nu}$ such that $\hat{\sigma}\sim L_{\mu \nu}W^{\mu\nu}$. Since $q^\mu L_{\mu \nu}=q^\nu L_{\mu\nu}=0$ the leptonic part will automatically guarantee that the dependence on the gauge parameter $\xi$ drops out. However, based on {\it e.m.~current conservation} for one-photon exchange, one expects that the hadronic tensor satisfies the condition
\bea
q_\mu W^{\mu \nu}=q_\nu W^{\mu \nu}=0\,.\label{eq:emCurrentCons}
\eea
It is straightforward to see already at LO that (\ref{eq:emCurrentCons}) only holds after application of both the EoMRs (\ref{eq:EoMQD}) and (\ref{eq:EoMQG}). For this reason we consider the application of  (\ref{eq:EoMQD}) and (\ref{eq:EoMQG}) as a necessity and throughout this paper we choose to eliminate the intrinsic twist-3 FFs.

\begin{figure}
\centering
\begin{subfigure}[b]{0.37\textwidth}
\includegraphics[width=\textwidth]{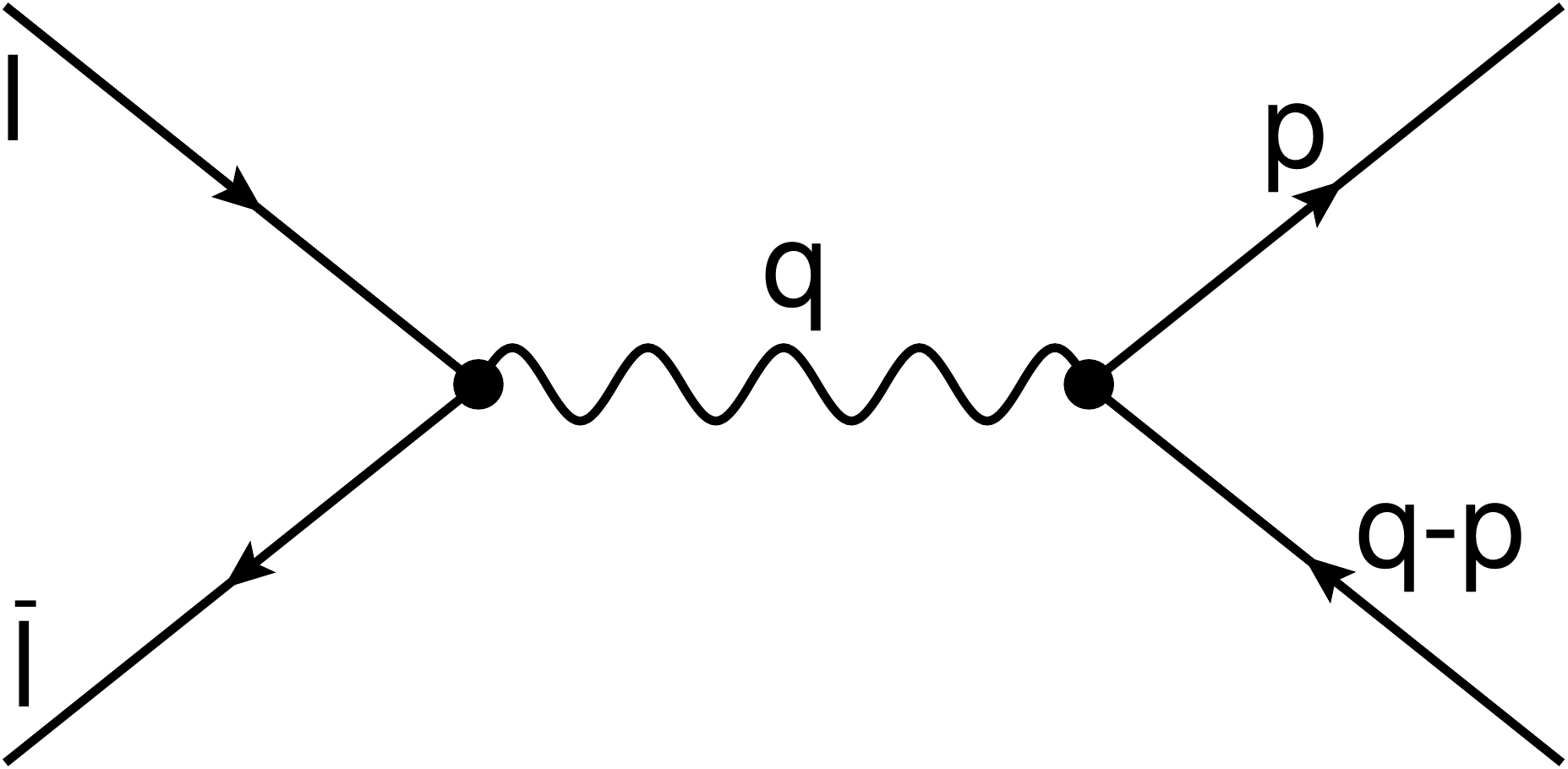}
\caption{LO quark fragmentation.}
\label{fig:LOq}
\end{subfigure}
\hskip 0.2in
\begin{subfigure}[b]{0.45\textwidth}
\includegraphics[width=\textwidth]{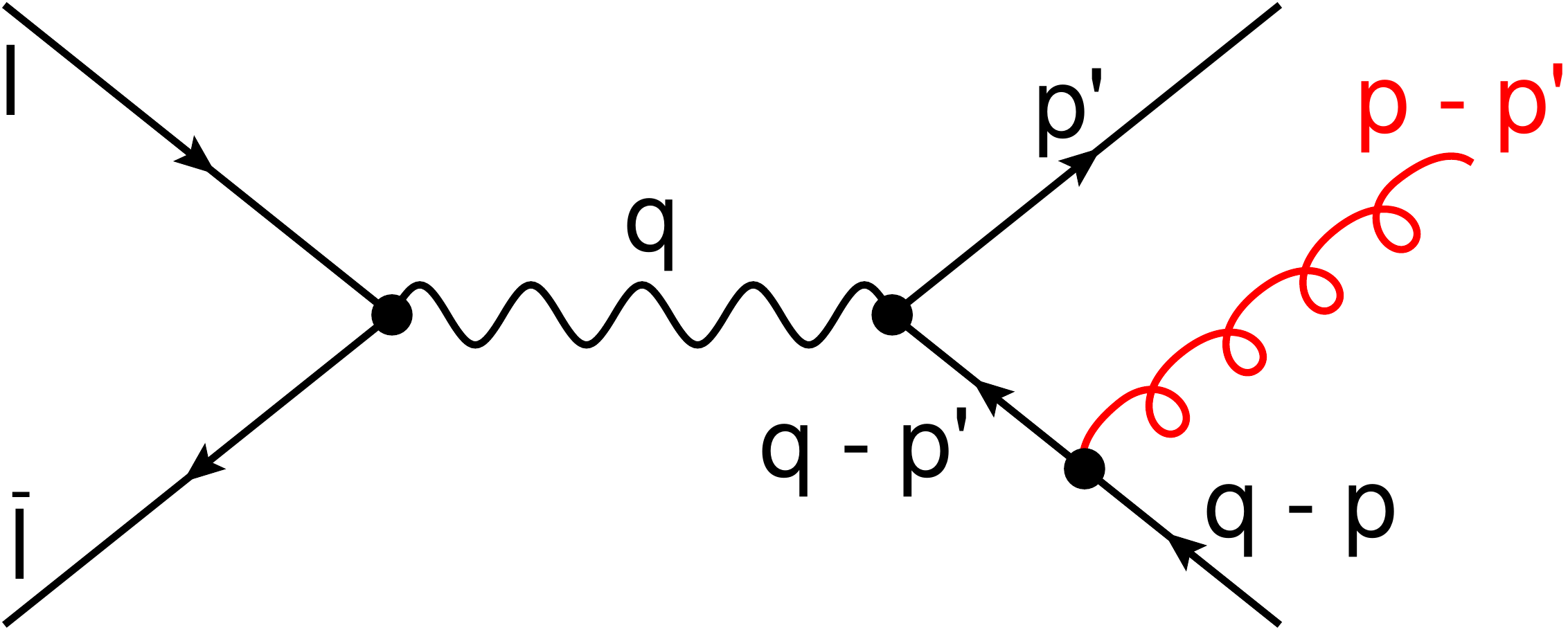}
\caption{LO quark-gluon fragmentation.}
\label{fig:LOqg}
\end{subfigure}

\caption{Leading order diagrams.}
\label{fig:LO}
\end{figure}

Below we proceed with a discussion of the leading order (LO) result without QCD corrections. The two relevant diagrams are shown in fig.~\ref{fig:LO}. We note that since a gluon polarization sum $d^{\mu\nu}(p,n;\kappa)$ does not appear in these diagrams, the gauge-invariance of LO partonic factors is automatically ensured. Also, since there is only one unobserved quark or anti-quark in the final state at LO, only the partonic cross sections $\hat{\sigma}^q$ and $\hat{\sigma}^{qg}$ in (\ref{eq:DefPartonicCS}) contribute to that order.

\subsection{Unpolarized cross section}

We first discuss the unpolarized cross section. Since this observable is leading twist, only the diagram in fig.~\ref{fig:LOq} contributes. Also, only the first term in (\ref{eq:SpinCSMaster}) is relevant. The calculation is straightforward, and we find the well-known result
\bea
\frac{E_h\,\d \sigma}{\d^{d-1}\vec{P}_h}=\sigma_0\, ((1-v)^2+v^2-\varepsilon)\sum_{f=q,\bar{q}}e_f^2\,D_1^f(z_h) + O(\alpha_s)\,,\label{eq:CSUnpolLO}
\eea
with $\sigma_0=(4\pi^2 z_h)^\varepsilon 2 N_c \alpha_{\mathrm{em}}^2/(z_h s^2)$.

\subsection{Double-longitudinally polarized cross section}
The double-longitudinal spin asymmetry for $\vec{e}^{\,+} + e^- \to \vec{\Lambda}+X$ is equally straightforward. In the notation (\ref{eq:DefLeptonHel}) we find
\bea
\Delta_{LL} \sigma=\sigma_0\,(1-2v)\,\sum_{f=q,\bar{q}}e_f^2\,G_1^f(z_h)+ O(\alpha_s)\,.\label{eq:CSLLLO}
\eea
The symbol $\Delta_{LL}$ in (\ref{eq:CSLLLO}) indicates that we have also implicitly included the asymmetry on the hadron spin, $(\sigma(S_L=1)-\sigma(S_L=-1))/2$.

\subsection{Transverse hadron-spin dependent cross section}

The transverse-spin dependent cross section will receive contributions from both diagrams in fig.~\ref{fig:LO} (see also refs.~\cite{Kanazawa:2015jxa,Koike:2017fxr}).  The calculation is straightforward, and we only present the result:
\bea
\frac{E_h\,\d \sigma(S_h)}{\d^{d-1}\vec{P}_h}&=&\sigma_0\,(1-2v)\,\frac{4M_h}{z_h\,s^2}\,\epsilon^{l l^\prime P_h S_h}\nonumber \\
&&\times\,\sum_{f=q,\bar{q}}e_f^2\,\left(\frac{D_T^f(z_h)}{z_h}-D_{1T}^{\perp (1),f}(z_h)+\int_0^1\d\beta\frac{\Im[\hat{D}_{FT}^{fg}-\hat{G}_{FT}^{fg}](z_h,\beta)}{1-\beta}\right)+ O(\alpha_s)\,\nonumber \\
&=&-\sigma_0\,(1-2v)\,\frac{8M_h}{z_h\,s^2}\,\epsilon^{l l^\prime P_h S_h}\nonumber\\
&&\times\,\sum_{f=q,\bar{q}}e_f^2\,\left(D_{1T}^{\perp (1),f}(z_h)-\int_0^1\d\beta\frac{\Im[\hat{D}_{FT}^{fg}-\hat{G}_{FT}^{fg}](z_h,\beta)}{1-\beta}\right)+ O(\alpha_s)\,\nonumber\\
&=&\sigma_0\,(1-2v)\,\frac{8M_h}{z_h\,s^2}\,\epsilon^{l l^\prime P_h S_h}\sum_{f=q,\bar{q}}e_f^2\,\frac{D_T^f(z_h)}{z_h} + O(\alpha_s)\,. \label{eq:CSUTLO}
\eea
The first equality in eq.~(\ref{eq:CSUTLO}) clearly shows the various contributions from intrinsic, kinematical, and dynamical twist-3 contributions. The EoMR (\ref{eq:EoMQD}) is used to eliminate the function $D_T$ in the second equality. On the other hand, one may choose to eliminate the kinematical and dynamical twist-3 FFs instead, as is done in the third equality. Again, we remind the reader that the LO partonic factors in (\ref{eq:DefPartonicCS}) are color gauge invariant by themselves due to the absence of a gluonic polarizations sum or propagator (\ref{eq:polsumLC}). Therefore, the use of (\ref{eq:EoMQD}) is not necessary from the point of view of color gauge invariance at LO. However, the condition (\ref{eq:emCurrentCons}) is only satisfied through the use of (\ref{eq:EoMQD}). 

The mere existence of a predicted non-zero single transverse-spin effect generated by the function $D_T^q(z)$ in the last line of eq.~\ref{eq:CSUTLO} is remarkable. In fact, the corresponding single transverse nucleon-spin asymmetry in the crossed process of inclusive DIS has been known to vanish due to time-reversal already in the 1960's in the one-photon exchange approximation \cite{Christ:1966zz}. In order to generate a non-zero effect for the single transverse nucleon-spin asymmetry in inclusive, DIS one has to deal with two-photon exchanges, cf. refs.~\cite{Metz:2006pe,Afanasev:2007ii,Metz:2012ui,Schlegel:2012ve}. The non-zero effect in the one-photon approximation in the annihilation process caused by the intrinsic twist-3 fragmentation function $D_T^q(z)$ can be attributed to the fact that fragmentation processes are not constrained by time-reversal~\cite{Boer:2003cm}. This is due to non-perturbative interactions in the {\it in} and {\it out} states in the definition of $D_T^q(x)$ in eq.~(\ref{eq:Deltaq}). On the other hand, a corresponding intrinsic twist-3 parton correlation function in the nucleon, $f_T^q(x)$, is forbidden by time-reversal~\cite{Goeke:2005hb}.

We also note that combining the LIR (\ref{eq:LIRD}) with the EoMR (\ref{eq:EoMQD}) allows us to write the spin-dependent annihilation cross section entirely in terms of dynamical twist-3 functions,
\bea
&&\frac{E_h\,\d \sigma}{\d^{d-1}\vec{P}_h}(S_h)=\sigma_0\,(1-2v)\,\frac{4M_h}{z_h\,s^2}\,\epsilon^{l l^\prime P_h S_h}\sum_{f=q,\bar{q}}e_f^2\label{eq:CSUTLOLIR}\\
&&\times\,2\int_{z_h}^1\tfrac{\d w}{w}\int_0^1\d\beta\,\left[\frac{(1+\delta(1-w))\Im[\hat{D}_{FT}^{fg}-\hat{G}_{FT}^{fg}](\tfrac{z_h}{w},\beta)}{1-\beta}+\frac{2\,\Im[\hat{D}_{FT}^{fg}](\tfrac{z_h}{w})}{(1-\beta)^2}\right]\Bigg)+\, O(\alpha_s)\,.\nonumber
\eea
This is true for any twist-3 transverse-spin observable, as first discussed in ref.~\cite{Kanazawa:2015ajw}.

\section{Observables in single-inclusive annihilation at NLO\label{NLO}}

\begin{figure}
\centering
\begin{subfigure}[b]{0.75\textwidth}
\includegraphics[width=\textwidth]{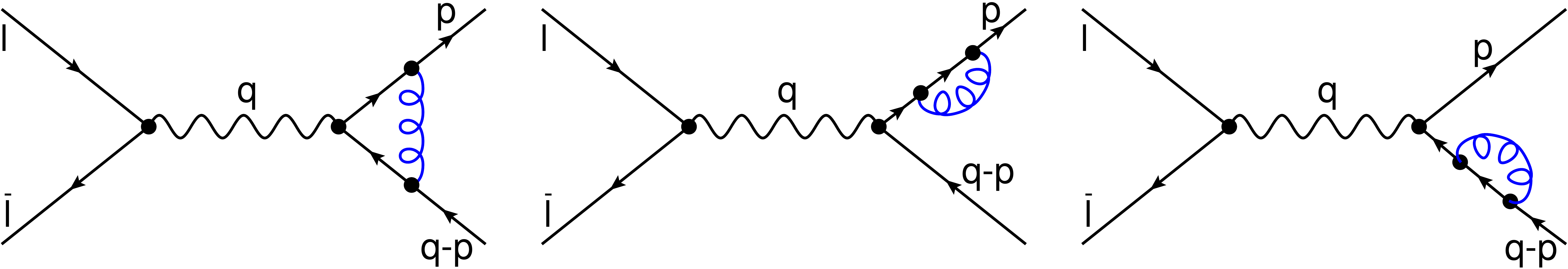}
\caption{NLO virtual diagrams for quark fragmentation}
\label{fig:NLOvirq}
\end{subfigure}
\vskip 0.3in
\begin{subfigure}[b]{0.55\textwidth}
\includegraphics[width=\textwidth]{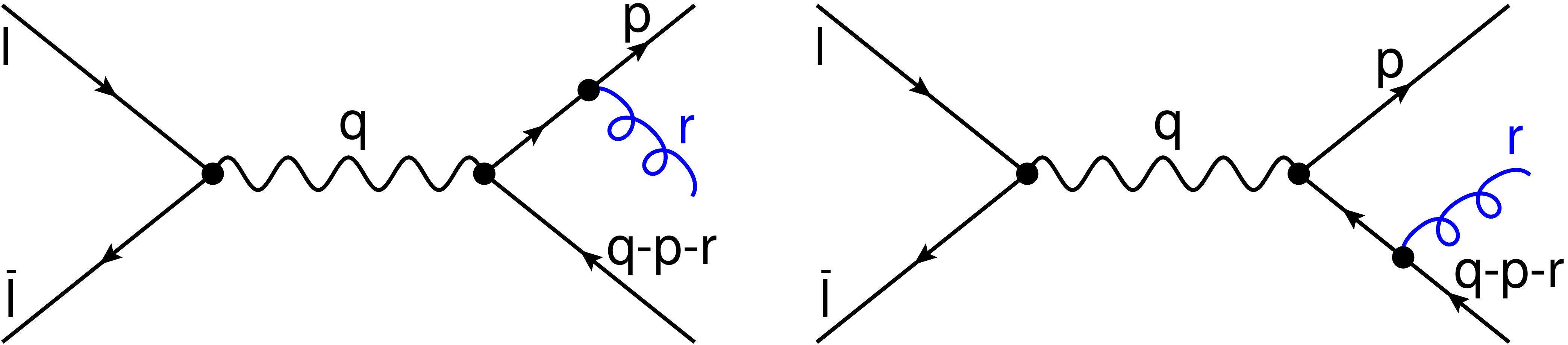}
\caption{NLO real diagrams for quark fragmentation}
\label{fig:NLOrealq}
\end{subfigure}
\vskip 0.3in
\begin{subfigure}[b]{0.55\textwidth}
\includegraphics[width=\textwidth]{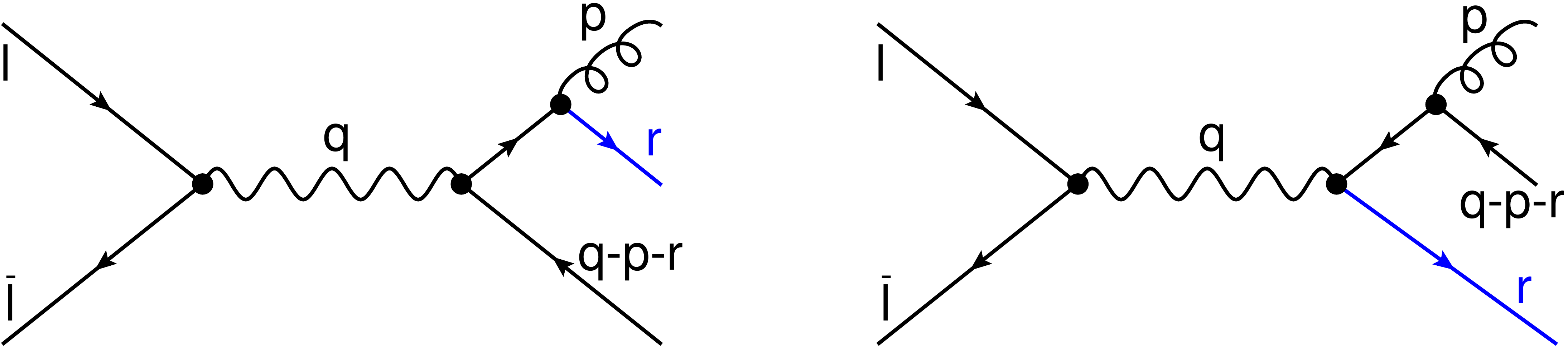}
\caption{NLO real diagrams for gluon fragmentation}
\label{fig:NLOrealg}
\end{subfigure}

\caption{Next-to-leading order diagrams relevant for 2-parton fragmentation.}
\label{fig:NLO2parton}
\end{figure}

In this section we present our results of the partonic factors in eq.~(\ref{eq:DefPartonicCS}) to NLO accuracy. We first focus on the contributions generated by the interference of single-parton fragmentation amplitudes. The relevant NLO QCD corrections are represented by the graphs in fig.~\ref{fig:NLO2parton}. The first group of diagrams in fig.~\ref{fig:NLOvirq} displays virtual corrections. Since these diagrams contain a two-particle final state just like the LO contributions, their mathematical form is similar compared to the result found in section \ref{LO}, up to corrections of order $\mathcal{O}(\alpha_s)$. However, the loop integrals in the virtual diagrams carry various sorts of divergences. The divergences are regulated throughout this paper using dimensional regularization in $d=4-2\varepsilon$ dimensions with subtractions carried out in the $\overline{{\rm MS}}$ scheme.

As mentioned in the previous section, we performed the calculations using a general gluon propagator/polarization sum as in (\ref{eq:polsumLC}) wherever possible. This allows us to perform important consistency checks on our calculations. One important check is the color gauge-invariance of the partonic factors (\ref{eq:DefPartonicCS}), as they should be independent of the ``gauge" parameter $\kappa$. We find that only after the application of the EoMRs (\ref{eq:EoMQD}) and (\ref{eq:EoMQG}), the parameter $\kappa$ drops out of the final results. This cancellation will be explained in more detail below. 

It is important to note that the form of the light-cone gauge polarization sum (\ref{eq:polsumLC}) forces us to perform integrals in a specific way.  Typically, one first has to perform the integrations over the light-cone components of a generic loop or phase space momentum $k^\mu$. Since the light-cone directions are specified by the light-cone momenta $P_h^\mu$ and $n^\mu$, we split
 the $d$-dimensional integral as
\bea
\int \d^dk = \int \d^{d-2}k_T \int \d (k\cdot n)\int \d (k\cdot P_h)\,.\label{eq:splitIntLC}
\eea
The dimension of the transverse space then regulates divergences. Working with a polarization sum (\ref{eq:polsumLC}) and $\kappa\neq 0$ induces further spurious light-cone divergences for $k\cdot n\to 0$ in (\ref{eq:splitIntLC}). Those divergences can be regulated in dimensional regularization as well by means of the well-known {\it Mandelstam-Leibbrandt} prescription \cite{Leibbrandt:1994wj}. We note that in this approach, however, it is difficult to identify the nature of the various divergences and to separate, for instance, ultraviolet (UV) from infrared (IR) divergences. It is also possible to perform the calculation in this way even if there are no explicit gluon polarizations.

The first term in the polarization sum (\ref{eq:polsumLC}) (for $\kappa=0$) refers to a calculation that is performed in Feynman gauge. One can calculate each diagram in dimensional regularization in this gauge in an alternative way, for example, by calculating the involved Feynman parameter integrals in a loop diagram directly, or by a direct evaluation of a phase space integral in an appropriate frame. We performed the calculation for $\kappa=0$ for each diagram in both aforementioned ways, and obtained the same analytical results for $\kappa=0$. This gives us confidence that our final results are correct.

In Feynman gauge, one can explicitly show that the UV-divergences between the vertex graph and the self-energy graphs in fig.~\ref{fig:NLOvirq} cancel; hence, no direct UV-counterterms are needed here. One can also show in general that the gauge parameter $\kappa$ drops out in the virtual diagrams in fig.~\ref{fig:NLOvirq}.

The real graphs in fig.~\ref{fig:NLOrealq} do contain IR- and collinear divergences that can be regulated by analytically continuing to negative values of $\epsilon$. Typically, one separates a collinear divergence through a {\it plus-prescription},
\bea
(1-w)^{-1-\varepsilon}=-\frac{1}{\varepsilon}\,\delta(1-w)+\frac{1}{(1-w)_+}-\varepsilon \,\left(\frac{\ln(1-w)}{1-w}\right)_+ +\mathcal{O}(\varepsilon^2)\,,\label{eq:CollDivSep}
\eea
along with the usual definition,
\bea
\int_0^1\d w\,f(w)\,[g(w)]_+=\int_0^1\d w\,[f(w)-f(1)]\,g(w)\,.\label{eq:DefPlus}
\eea
Unlike IR singularities that show up as $1/\varepsilon^2$-poles and cancel between real and virtual diagrams, $1/\varepsilon$ collinear divergences in a first step remain in NLO cross sections. For both the twist-2 and twist-3 observables analyzed in the following sections, we will follow the subtraction procedure of Collins~\cite{Collins:2011zzd} as our method to remove these collinear singularities and obtain finite results.

\subsection{Unpolarized cross section at NLO}
As a test case for our calculation of the interference effects of single-parton fragmentation amplitudes shown in fig.~\ref{fig:NLO2parton}, we use the twist-2 unpolarized cross section. We find that all types of partonic cross sections that contribute at NLO in figs.~\ref{fig:NLOvirq}, \ref{fig:NLOrealq}, \ref{fig:NLOrealg} are individually gauge invariant, as expected.  The full result takes the following form~\cite{Collins:2011zzd},
\bea
\frac{E_h\,\d \sigma}{\d^{d-1}\vec{P}_h}&=&\sigma_0\, ((1-v)^2+v^2)\nonumber\\
&&\times\sum_{f=q,\bar{q}}e_f^2\,\int_{z_h}^1\tfrac{\d w}{w}\,w^{-2\varepsilon}\left[\hat{\sigma}_{D_1}^{1;f}(w)\,D_{1}^{f[0]}(\tfrac{z_h}{w})+\hat{\sigma}_{D_1}^{1;g}(w)\,D_{1}^{g[0]}(\tfrac{z_h}{w})\right]\,\nonumber\\[0.3cm]
&&+\,\sigma_0\,4\,v\,(1-v)\nonumber\\
&&\times\sum_{f=q,\bar{q}}e_f^2\,\int_{z_h}^1\tfrac{\d w}{w}\,w^{-2\varepsilon}\left[\hat{\sigma}_{D_1}^{2;f}(w)\,D_{1}^{f[0]}(\tfrac{z_h}{w})+\hat{\sigma}_{D_1}^{2;g}(w)\,D_{1}^{g[0]}(\tfrac{z_h}{w})\right]\nonumber\\[0.3cm]
&&-\,\sigma_0\, ((1-v)^2+v^2-\varepsilon)\sum_{f=q,\bar{q}}e_f^2\,D_1^{f[1]}(z_h) + O(\alpha_s^2)\,,\label{eq:CSUnpolNLO}
\eea
where
\bea
\hat{\sigma}_{D_1}^{1;f}(w) &=& (1-\varepsilon)\delta(1-w)+\frac{C_F\,\alpha_s}{2\pi}S_\varepsilon \left(\frac{s}{\mu^2}\right)^{-\varepsilon}\Bigg[-\frac{1}{\varepsilon}\left(\frac{1+w^2}{(1-w)_+}+\frac{3}{2}\,\delta(1-w)\right)\nonumber\\
&&+\,\delta(1-w)\, \left(-3+\frac{2\pi^2}{3} \right)+\left(\frac{\ln(1-w)}{1-w}\right)_+(1+w^2)+\frac{1-3w+\frac{5}{2}w^2}{(1-w)_+}\Bigg],\label{eq:partSigD11q}\\[0.3cm]
\hat{\sigma}_{D_1}^{2;f}(w) &=&-\frac{\varepsilon} {2}\delta(1-w)+\frac{C_F\alpha_s}{2\pi}S_\varepsilon \left(\frac{s}{\mu^2}\right)^{-\varepsilon} \left[\frac{3}{4}\,\delta(1-w)+\frac{3-2w+w^2}{2(1-w)_+}\right],\label{eq:partSigD12q}\\[0.3cm]
\hat{\sigma}_{D_1}^{1;g}(w) &=&\frac{C_F\alpha_s}{2\pi}S_\varepsilon \left(\frac{s}{\mu^2}\right)^{ -\varepsilon} \left[-\frac{1}{\varepsilon}\frac{1+(1-w)^2}{w}+\frac{1+(1-w)^2}{w}\ln(1-w)+w\right],\label{eq:partSigD11g}\\[0.3cm]
\hat{\sigma}_{D_1}^{2;g}(w) &=&\frac{C_F\alpha_s}{2\pi}S_\varepsilon \left(\frac{s}{\mu^2}\right)^{-\varepsilon} \left[\frac{6-6w+w^2}{2w}\right].\label{eq:partSigD12g}
\eea
The functions $D_{1}^{(f,g)[n]}(z)$ in (\ref{eq:CSUnpolNLO}) are $n$-loop renormalized FFs. The color factor $C_F$ is the usual $C_F=(N_c^2-1)/(2N_c)$, with $N_c=3$ the number of colors. The renormalization scale $\mu$ also appears through the dimensional regularization approach, along with the $\overline{\mathrm{MS}}$ factor $S_\varepsilon=(4\pi)^\varepsilon/\Gamma(1-\varepsilon)$ from ref.~\cite{Collins:2011zzd}. Note that there are two structures in (\ref{eq:CSUnpolNLO}), one proportional to $(1-v)^2+v^2$, and the other to $4v(1-v)$. Those structures can be attributed to different structure functions of the unpolarized cross section.

The $1/\varepsilon$-terms in (\ref{eq:partSigD11q}) and (\ref{eq:partSigD11g}) are the well-known collinear singularities that one encounters in NLO calculations, and they arise in the first and second terms of eq.~(\ref{eq:CSUnpolNLO}). The last term in eq.~(\ref{eq:CSUnpolNLO}) is the ``subtraction term'' designed to remove these $1/\varepsilon$-poles~\cite{Collins:2011zzd}, {\it if a given process factorizes.}  The function $D_1^{f[1]}(z)$ in that term takes the form~\cite{Collins:2011zzd},
\bea
D_{1}^{f[1]}(z)=D_{1(0)}^{f[1]}(z)+\sum_{f'=f,g}\int_z^1\tfrac{\d w}{w}\,Z_{f\to f'}^{[1]}(w)\,D_{1}^{f'[0]}(\tfrac{z}{w})\,,\label{eq:renD1}
\eea
where $D_{1(0)}^{f[1]}(z)$ is the one-loop bare FF, and
\bea
Z_{f\to f}^{[1]}(w)&=&-\frac{C_F \alpha_s}{2\pi}\frac{S_\varepsilon}{\varepsilon}\left(\frac{1+w^2}{(1-w)_+}+\frac{3}{2}\,\delta(1-w)\right)\,,\label{eq:SplitFuncD1}\\[0.3cm]
Z_{f\to g}^{[1]}(w)&=&-\frac{C_F \alpha_s}{2\pi}\frac{S_\varepsilon}{\varepsilon}\left(\frac{1+(1-w)^2}{w}\right)\,.
\eea
We note that in massless QCD, $D_{1(0)}^{q[1]}(z)$ involves scaleless $k_T$-integrals and therefore vanishes in dimensional regularization. 

The last term in eq.~(\ref{eq:CSUnpolNLO}), after one inserts (\ref{eq:renD1}), cancels the $1/\varepsilon$-poles in (\ref{eq:partSigD11q}) and (\ref{eq:partSigD11g}).
We note that the cancellation of the collinear poles of the NLO cross section through this procedure is a necessary condition for factorization of the observable.  That is, the collinear singularities of the NLO cross section (without a subtraction term) must match those from a direct NLO calculation of the correlator.  If only one of the collinear singularities in the NLO (unsubtracted) partonic cross sections does not have a corresponding divergence in the correlator, then this mismatch directly proves the observable does not factorize.

The final result for the unpolarized cross section in the limit $\varepsilon\to 0$ is
\bea
\frac{E_h\,\d \sigma}{\d^{3}\vec{P}_h}&=&\frac{2 N_c \alpha_{\mathrm{em}}^2} {z_h s^2} \Bigg\{ ((1-v)^2+v^2)\sum_{f=q,\bar{q}}e_f^2\,D_1^{f}(z_h,\mu)\nonumber\\[0.3cm]
&&+\,((1-v)^2+v^2)\nonumber\\
&&\times\sum_{f=q,\bar{q}}e_f^2\,\int_{z_h}^1\tfrac{\d w}{w}\left[\hat{c}_{D_1}^{1;f}(w)\,D_1^{f}(\tfrac{z_h}{w};\mu)+\hat{c}_{D_1}^{1;g}(w)\,D_1^{g}(\tfrac{z_h}{w};\mu)\right]\,\nonumber\\[0.3cm]
&&+\,4\,v\,(1-v)\nonumber\\
&&\times\sum_{f=q,\bar{q}}e_f^2\,\int_{z_h}^1\tfrac{\d w}{w}\left[\hat{c}_{D_1}^{2;f}(w)\,D_1^{f}(\tfrac{z_h}{w};\mu)+\hat{c}_{D_1}^{2;g}(w)\,D_1^{g}(\tfrac{z_h}{w};\mu)\right]\!\Bigg\}\,,
\eea
where the finite partonic cross sections at order $\mathcal{O}(\alpha_s)$ read,
\bea
\hat{c}_{D_1}^{1;f}(w) &=&\frac{C_F\,\alpha_s}{2\pi}\Big[\delta(1-w)\, \left(\tfrac{3}{2}\ln(\tfrac{s}{\mu^2})-\tfrac{9}{2}+\tfrac{2\pi^2}{3} \right)\nonumber\\
&&+\left(\frac{\ln(1-w)}{1-w}\right)_+(1+w^2)+\frac{1+w^2}{(1-w)_+}\ln(w^2\tfrac{s}{\mu^2})-\frac{3w(2-w)}{2(1-w)_+}\Big]\,,\label{eq:partSigD11qren}\\[0.3cm]
\hat{c}_{D_1}^{2;f}(w) &=&\frac{C_F\alpha_s}{2\pi}\,,\label{eq:partSigD12qren}\\[0.3cm]
\hat{c}_{D_1}^{1;g}(w) &=&\frac{C_F\alpha_s}{2\pi} \left[\frac{1+(1-w)^2}{w}\ln(w^2(1-w)\tfrac{s}{\mu^2})-\frac{2(1-w)}{w}\right]\,,\label{eq:partSigD11gren}\\[0.3cm]
\hat{c}_{D_1}^{2;g}(w) &=&\frac{C_F\alpha_s}{2\pi} \left[\frac{2(1-w)}{w}\right]\,.\label{eq:partSigD12gren}
\eea
These results are in agreement with earlier works (see refs.~\cite{deFlorian:1997zj,deFlorian:1998ba} and references therein).  Throughout the paper we will denote partonic cross sections before collinear divergences are subtracted by $\hat{\sigma}$'s and finite partonic cross sections after subtraction by $\hat{c}$'s.

Note that the partonic cross sections (\ref{eq:partSigD11qren}), (\ref{eq:partSigD11gren}) depend on the arbitrary renormalization scale $\mu$. The fact that the physical, measurable cross section $E_h \d\sigma/\d^3\vec{P}_h$ does not depend on $\mu$ allows one to deduce an evolution equation for the unpolarized FF,
\bea
&&\frac{\partial}{\partial \ln\mu^2}\left(E_h\frac{\d\sigma}{\d^3\vec{P}_h}\right)=0 \\[0.3cm]
&&\Longrightarrow \frac{\partial D_1^f(z;\mu)}{\partial\ln\mu^2}=\sum_{f'=f,g}\int_z^1\tfrac{\d w}{w}\,P_{f\to f'}^{[1]}(w)\,D_1^{f'}(\tfrac{z}{w};\mu)\,,\label{eq:EvolD1}
\eea
where the well-known LO splitting functions $P_{f\to f}^{[1]}(w)$ and $P_{f\to g}^{[1]}(w)$ are given by
\bea
P_{f\to f}^{[1]}(w)&=&\frac{C_F \alpha_s}{2\pi}\left(\frac{1+w^2}{(1-w)_+}+\frac{3}{2}\,\delta(1-w)\right)\,,\\[0.3cm]
P_{f\to g}^{[1]}(w)&=&\frac{C_F \alpha_s}{2\pi}\left(\frac{1+(1-w)^2}{w}\right)\,.\nonumber
\eea
The expression in eq.~(\ref{eq:EvolD1}) is the standard LO DGLAP-evolution equation. The reason for the detailed discussion of the well-known twist-2 unpolarized cross section is that we will use similar strategies for the more complicated and not as well-known twist-3 observables.

\subsection{The double-longitudinally polarized cross section at NLO}
 For completeness we also include a discussion of the double-longitudinally polarized cross section (\ref{eq:CSLLLO}), extended to NLO. The calculation for this twist-2 observable is similar to the one discussed in the previous section, but with two distinctions: firstly, we have to deal with the Dirac-matrix $\gamma_5$ in $d$ dimensions. This requires a special procedure, and throughout this paper we apply the so-called {\it HVBM-scheme} \cite{Hooft:1972fi,Breitenlohner:1977hr}. Secondly, it is customary to include a term $+4\varepsilon(1-w)$ in the polarized renormalization factor $\Delta Z^{qq}=Z^{qq}+4C_F\varepsilon (1-w)$ in order to preserve helicity conservation at the quark-gluon vertex in $d$ dimensions \cite{deFlorian:1997zj,deFlorian:1998ba}. Otherwise, all comments made in the previous section on the NLO unpolarized cross section also apply here. 

We find for the double-longitudinal spin asymmetry in $d=4$ dimensions after inclusion of the subtraction graphs,
\bea
\Delta_{LL} \sigma &=& \sum_{f=q,\bar{q}}e_f^2\,\int_{z_h}^1\tfrac{\d w}{w}\,\left[\Delta\hat{c}_{G_1}^{f}(w)\,G_{1}^{f}(\tfrac{z_h}{w};\mu)+\Delta\hat{c}_{G_1}^{g}(w)\,G_1^{g}(\tfrac{z_h}{w};\mu)\right]\Bigg\}+O(\alpha_s^2)\,,\nonumber
\eea
with the finite polarized partonic cross sections,
\bea
\Delta\hat{c}_{G_1}^{f}(w) &=&\delta(1-w)+\frac{C_F\,\alpha_s}{2\pi}\Bigg[\delta(1-w)\, \left(\tfrac{3}{2}\ln(\tfrac{s}{\mu^2})-\tfrac{9}{2}+\tfrac{2\pi^2}{3} \right)\nonumber\\
&&+\left(\frac{\ln(1-w)}{1-w}\right)_+(1+w^2)+\frac{1+w^2}{(1-w)_+}\ln(w^2\tfrac{s}{\mu^2})-\frac{2+2w-w^2}{2(1-w)_+}\Bigg],\label{eq:partDSigG11qren}\\[0.3cm]
\Delta\hat{c}_{G_1}^{g}(w) &=&\frac{C_F\alpha_s}{2\pi} \Bigg[(2-w)\ln(w^2(1-w)\tfrac{s}{\mu^2})-4+3w\Bigg]\,.\label{eq:partGSigG11gren}
\eea
Again, our calculation agrees with refs.~\cite{deFlorian:1997zj,deFlorian:1998ba}.

\subsection{Transverse hadron-spin dependent cross section at NLO}
We are now in a position to analyze the twist-3 observables. In this section we discuss the spin-dependent cross section for unpolarized leptons and a transversely polarized hadron. Note that we omit the subtraction graphs in the calculations that follow, and instead postpone a discussion of this term until section~\ref{Evol}.  

\subsubsection{Quark-Quark \& Quark-Gluon-Quark fragmentation}
\paragraph{Intrinsic \& Kinematical Twist-3}
We first focus on the intrinsic and kinematical twist-3 contributions from quarks. 
Like the twist-2 observables, the intrinsic and kinematical twist-3 partonic cross sections at NLO can be calculated in a straightforward fashion from the same diagrams in fig.~(\ref{fig:NLO2parton}). Since we have already noticed the importance of the EoMR (\ref{eq:EoMQD}) for the LO results, we also apply this relation immediately at NLO and replace the intrinsic twist-3 functions with kinematical and dynamical twist-3 functions. We find the following results for the spin-dependent cross section,
\bea
\frac{E_h\,\d \sigma^{\mathrm{intr\& kin}}(S_h)}{\d^{d-1}\vec{P}_h}&=&\sigma_0\,(1-2v)\,\frac{4M_h}{z_h\,s^2}\,\epsilon^{l l^\prime P_h S_h}\sum_{f=q,\bar{q}}e_f^2\,\int_{z_h}^1\tfrac{\d w}{w^2}w^{-2\varepsilon}\nonumber\\
&&\times\Bigg(\hat{\sigma}_{D_{1T}^{\perp (1)}}^{f}(w)\,D_{1T}^{\perp (1),f[0]}(\tfrac{z_h}{w})+\hat{\sigma}^f_{D_T}(w)\,\int_0^1\d \beta\,\frac{\Im[\hat{D}_{FT}^{fg[0]}-\hat{G}_{FT}^{fg[0]}](\tfrac{z_h}{w},\beta)}{1-\beta}\nonumber\\
&&+\,\hat{\sigma}_{G_T}^{f}(w)\,\int_0^1\d \beta\,\frac{\Re[\hat{D}_{FT}^{fg[0]}-\hat{G}_{FT}^{fg[0]}](\tfrac{z_h}{w},\beta)}{1-\beta}\,\Bigg)+O(\alpha_s^2),\label{eq:CSUTNLOintkin}
\eea
where we refrain from giving the explicit form of the relevant partonic cross sections at this point but rather wait until the dynamical graphs are included. We mention again that the superscript ``$[0]$'' indicates LO renormalized functions.

Note that the loop diagrams in fig.~\ref{fig:NLOvirq} generate an imaginary part. As a consequence, there are also contributions to intrinsic- and kinematical twist-3 parts that are generated by the FFs $G_T$ and $G_{1T}^{\perp (1)}$. By virtue of (\ref{eq:EoMQG}) we can eliminate these functions in favor of the real parts of the quark-gluon-quark FFs ($\hat{D}_{FT}-\hat{G}_{FT}$) in (\ref{eq:CSUTNLOintkin}). We find that the vertex diagram in fig.~\ref{fig:NLOvirq} is gauge invariant, i.e., the dependence on the parameter $\kappa$, introduced in eq.~(\ref{eq:polsumLC}), drops out. This is different for the real diagrams in fig.~\ref{fig:NLOrealq} where we find an explicit gauge dependence in both partonic factors for the intrinsic and kinematical twist-3 functions $D_T^q$ and $D_{1T}^{\perp (1),q}$ {\it before} application of (\ref{eq:EoMQD}). {\it After} application of (\ref{eq:EoMQD}), the partonic cross sections can be combined, and we find that the gauge-dependence for the kinematical twist-3 function $D_{1T}^{\perp (1),q}$ drops out in eq.~(\ref{eq:partSigqD1Tp}) of appendix A. However, there is a remaining gauge dependence in $\hat{\sigma}_{D_T}$ of the following form,
\bea
\hat{\sigma}_{D_T}^f(w)&=&\dots+\kappa \Big\{-\delta(1-w)\,\left(\frac{1}{\varepsilon^2}-\frac{\pi^2}{6}\right)+\frac{1}{\varepsilon}\frac{w(2-w)}{(1-w)_+}-\left(\frac{\ln(1-w)}{1-w}\right)_+w(2-w)\nonumber\\
&&+w(6-w)\Big\}\Big]\,,\label{eq:GaugeDepDT}
\eea
where the dots indicate the gauge-independent part of $\hat{\sigma}_{D_T}$. The explicit gauge dependence must cancel a corresponding gauge dependence in the $qgq$ fragmentation.
We also mention that the $1/\varepsilon^2$-poles cancel individually (except for the gauge-dependent ones in (\ref{eq:GaugeDepDT})) for each partonic cross section even before application of (\ref{eq:EoMQD}). 

\paragraph{Dynamical Twist-3}
\begin{figure}
\centering
\begin{subfigure}[b]{0.95\textwidth}
\includegraphics[width=\textwidth]{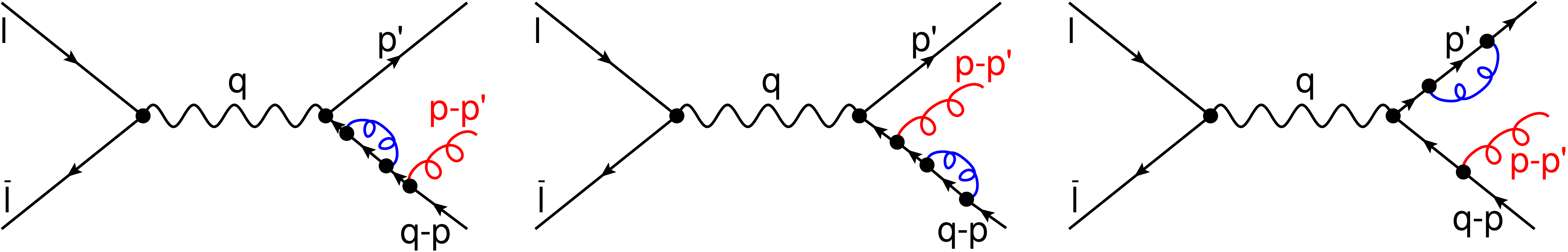}
\caption{Self-energy corrections to the quark lines}
\label{fig:NLOSE1}
\end{subfigure}
\vskip 0.3in
\begin{subfigure}[b]{0.75\textwidth}
\includegraphics[width=\textwidth]{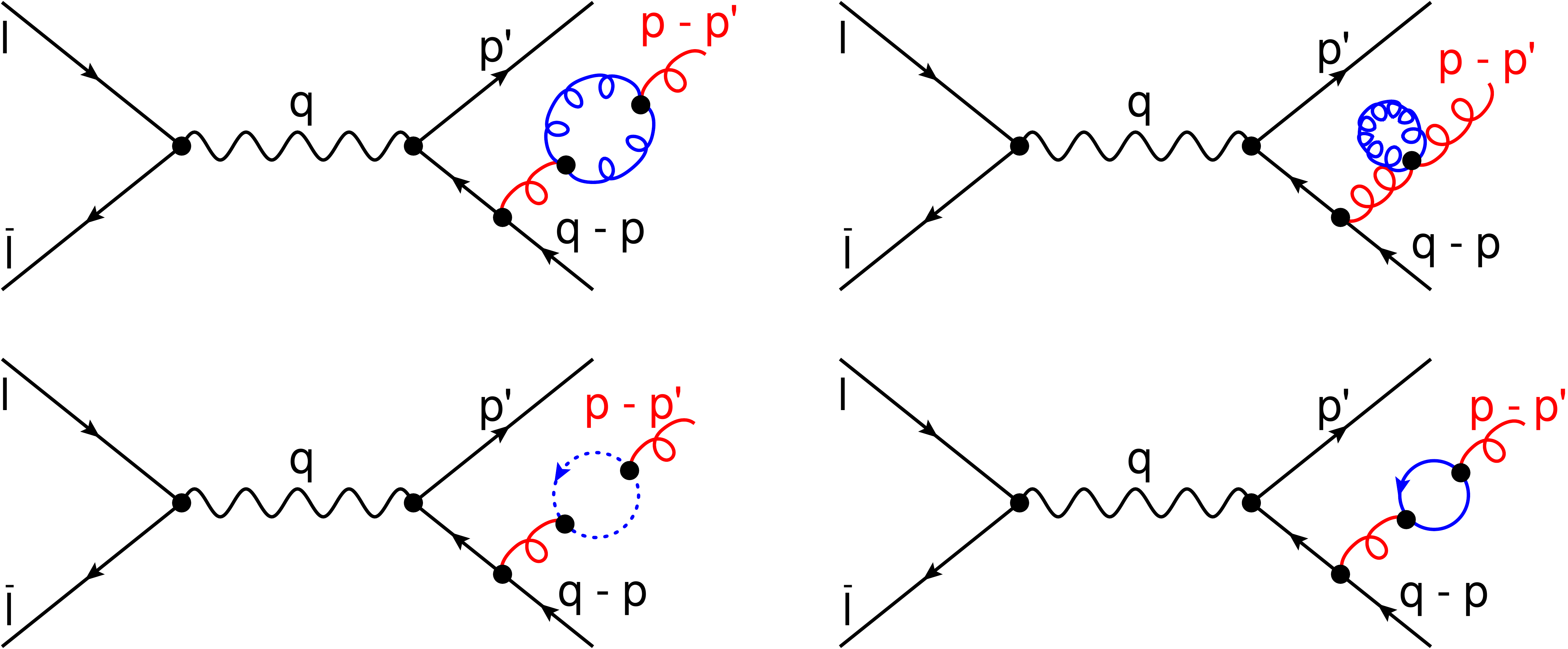}
\caption{Self-energy corrections on the gluon line}
\label{fig:NLOSE2}
\end{subfigure}
\vskip 0.3in
\begin{subfigure}[b]{0.95\textwidth}
\includegraphics[width=\textwidth]{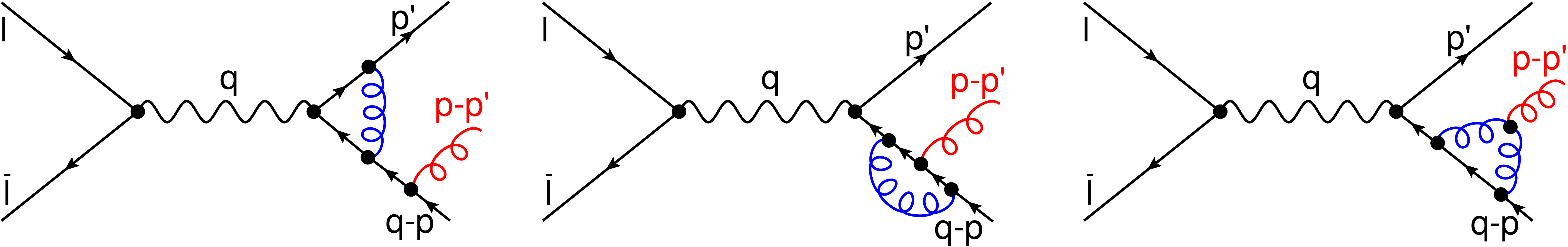}
\caption{Vertex corrections}
\label{fig:NLOV}
\end{subfigure}
\vskip 0.3in
\begin{subfigure}[b]{0.95\textwidth}
\includegraphics[width=\textwidth]{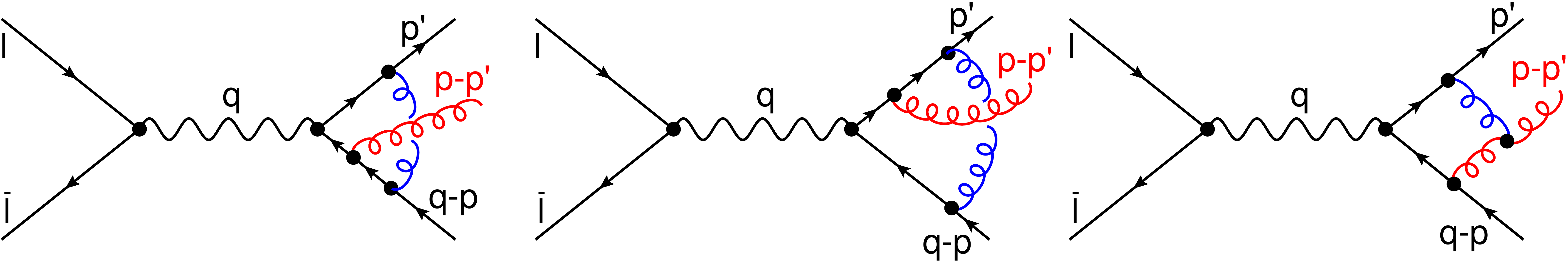}
\caption{Box diagram}
\label{fig:NLOB}
\end{subfigure}
\caption{Virtual one-loop diagrams.}
\label{fig:NLOvir}
\end{figure}

Of course we expect also contributions from dynamical $qgq$ twist-3 fragmentation at NLO. We first analyze the virtual loop corrections shown in fig.~\ref{fig:NLOvir}. Again, we emphasize that we calculated the loops both in Feynman gauge ($\kappa=0$) and light-cone gauge ($\kappa=1$), see eq.~(\ref{eq:polsumLC}). As before, we use dimensional regularization to deal with UV, IR, collinear divergences, as well as light-cone divergences in light-cone gauge upon application of the Mandelstam-Leibbrandt prescription \cite{Leibbrandt:1994wj}. We find that the sum of all loop diagrams in fig.~\ref{fig:NLOvir} is gauge-invariant $\--$ while individual diagrams yield very different expressions in both gauges. This is a very important feature as it concerns the subtraction of UV-divergences by means of renormalization counterterms. If the sum of all diagrams in fig.~\ref{fig:NLOvir} is gauge-invariant, so is its UV behavior. Consequently, also the sum of all counterterms derived from a renormalized QCD Lagrangian must be gauge-invariant. Hence, we may use the well-known counterterms in covariant gauge, see, e.g., \cite{Collins:2011zzd}. We apply counterterms for the self-energy diagrams in fig.~\ref{fig:NLOSE1} and \ref{fig:NLOSE2}, and the vertex graphs in fig.~\ref{fig:NLOV}. Self-energy corrections for external lines (the last two diagrams in fig.~\ref{fig:NLOSE1} and \ref{fig:NLOSE2}) vanish in dimensional regularization as they are given by massless integrals. However, counterterms for those diagrams must be included, and they come with a well-known factor $1/2$ \cite{Collins:1988wj}. Similar to the twist-2 observables, the counterterms for the last two diagrams in fig.~\ref{fig:NLOSE1} cancel with that for the first vertex correction in fig.~\ref{fig:NLOV}. The remaining $\overline{\mathrm{MS}}$ UV counterterm has the following form,
\bea
-\frac{\alpha_s}{2\pi}\frac{S_\varepsilon}{\varepsilon}\left[\frac{11}{12}N_c-\frac{1}{6}n_f\right]\equiv-\frac{\alpha_s}{2\pi}\frac{S_\varepsilon}{\varepsilon}\delta_{\mathrm{UV}}.\label{eq:UVCT}
\eea
We explicitly checked that it is indeed gauge-invariant, i.e., there is no dependence on a gauge-parameter $\xi$ in covariant gauge. Also, $n_f$ is the number of active flavors, inherited from the counterterm for the last self-energy graph in fig.~\ref{fig:NLOSE2}.
The loops in fig.~\ref{fig:NLOvir} generate imaginary parts as well. As in the case of quark-quark fragmentation, this induces an additional sensitivity on the real parts of the dynamical twist-3 FFs.

\begin{figure}
\centering
\includegraphics[width=0.9\textwidth]{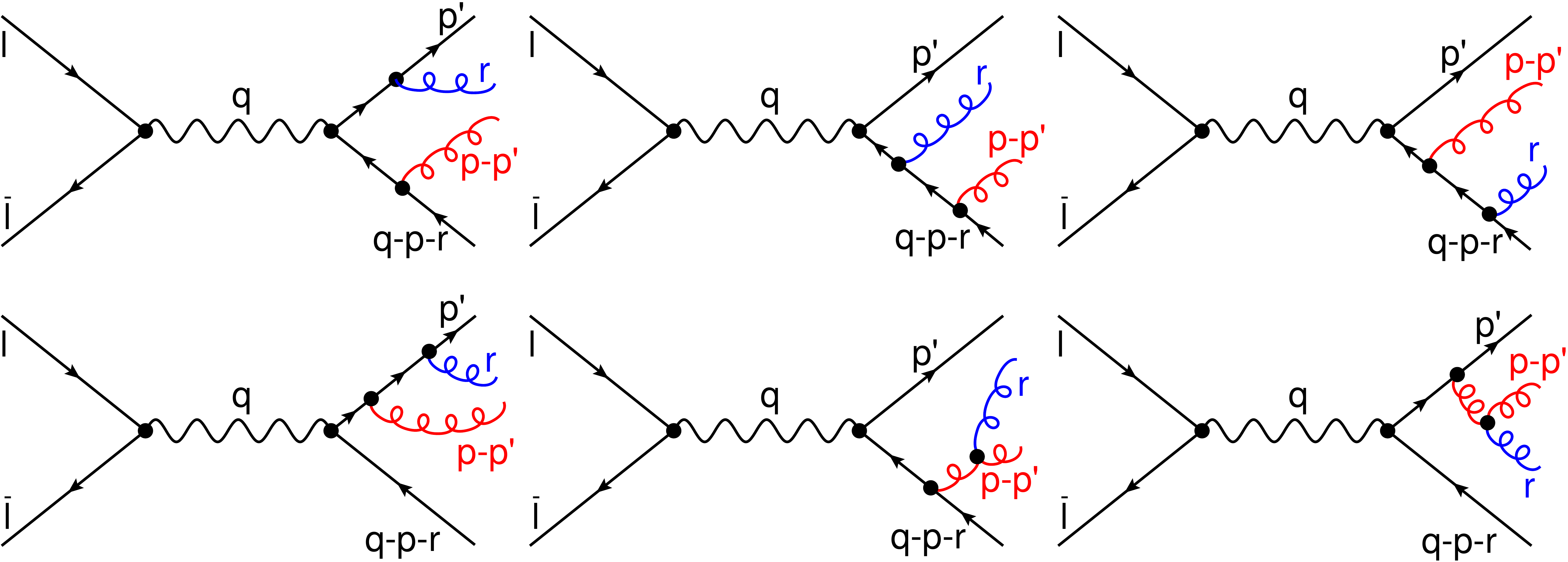}
\caption{NLO corrections from real gluon radiation to quark-gluon fragmentation.}
\label{fig:NLOrealqg}
\end{figure}

We also need to include the radiative corrections from real diagrams shown in fig.~\ref{fig:NLOrealqg}. It turns out that this class of corrections is not gauge-invariant, i.e., we find different results in Feynman gauge and light-cone gauge. Collinear and IR-divergences are handled by dimensional regularization, and we use eq.~(\ref{eq:CollDivSep}) to introduce the plus-prescription (\ref{eq:DefPlus}). We again find that $1/\varepsilon^2$-poles cancels when adding virtual and real diagrams.

The $qgq$ fragmentation channel assumes the following form at NLO,
\bea
\frac{E_h\,\d \sigma^{\mathrm{dyn}}(S_h)} {\d^{d-1}\vec{P}_h}&=&\sigma_0\,(1-2v)\,\frac{4M_h}{z_h\,s^2}\,\epsilon^{l l^\prime P_h S_h}\sum_{f=q,\bar{q}}e_f^2\,\int_{z_h}^1\tfrac{\d w}{w^2}w^{-2\varepsilon}\nonumber\\
&&\times\,\int_0^1 \d \beta \Bigg(\hat{\sigma}_{1}^{qgg}(w,\beta)\,\frac{\Im[\hat{D}_{FT}^{fg[0]}-\hat{G}_{FT}^{fg[0]}](\tfrac{z_h}{w},\beta)}{1-\beta}\nonumber\\
&&+\hat{\sigma}_{2}^{qgg}(w,\beta)\,\frac{2\,\Im[\hat{D}_{FT}^{fg[0]}](\tfrac{z_h}{w},\beta)}{\beta\,(1-\beta)^2}\nonumber\\
&& +\hat{\sigma}_{3}^{qgg}(w,\beta)\,\frac{\Re[\hat{D}_{FT}^{fg[0]}-\hat{G}_{FT}^{fg[0]}](\tfrac{z_h}{w},\beta)}{1-\beta}\Bigg)+O(\alpha_s^2)\,,\label{eq:CSUTNLOdyn}
\eea
where again we refrain from stating the explicit form of the partonic cross sections until eq.~(\ref{eq:CSUTNLOdyn}) is added to eq.~(\ref{eq:CSUTNLOintkin}).
We note that the same gauge dependence as in eq.~(\ref{eq:GaugeDepDT}) $\--$ but with a different sign $\--$ appears in $\hat{\sigma}_1^{qgq}$. Hence, the gauge dependence cancels when adding all twist-3 contributions. We also note that the $1/\varepsilon$-pole in the partonic cross sections generated by imaginary parts of loop integrals, $\hat{\sigma}_{G_T}^{f}$ of (\ref{eq:CSUTNLOintkin}) and $\hat{\sigma}_3^{qgq}$ of (\ref{eq:CSUTNLOdyn}), cancel when added together.

\paragraph{Result for Quark-Quark \& Quark-Gluon-Quark Fragmentation at NLO}

The full result for quark-quark and quark-gluon-quark fragmentation is given by the sum of eq.~(\ref{eq:CSUTNLOintkin}) and eq.~(\ref{eq:CSUTNLOdyn}),
\bea
&&\frac{E_h\,\d \sigma^{\mathrm{qq\& qgq}}(S_h)}{\d^{d-1}\vec{P}_h}=\sigma_0\,(1-2v)\,\frac{4M_h}{z_h\,s^2}\,\epsilon^{l l^\prime P_h S_h}\sum_{f=q,\bar{q}}e_f^2\,\int_{z_h}^1\tfrac{\d w}{w^2}w^{-2\varepsilon}\int_0^1 \d\beta\label{eq:CSUTNLOqqqgq}\\
&& \times\Bigg(\hat{\sigma}_{D_{1T}^{\perp (1)}}^f(w)\,D_{1T}^{\perp (1),f[0]}(\tfrac{z_h}{w})+\hat{\sigma}_{1}^{fg}(w,\beta)\,\frac{\Im[\hat{D}_{FT}^{fg[0]}-\hat{G}_{FT}^{fg[0]}](\tfrac{z_h}{w},\beta)}{1-\beta}\nonumber\\
&&+\,\hat{\sigma}_{2}^{fg}(w,\beta)\,\frac{2\,\Im[\hat{D}_{FT}^{fg[0]}](\tfrac{z_h}{w},\beta)}{(1-\beta)^2}+\hat{\sigma}_{3}^{fg}(w,\beta)\,\frac{\Re[\hat{D}_{FT}^{fg[0]}-\hat{G}_{FT}^{fg[0]}](\tfrac{z_h}{w},\beta)}{1-\beta}\Bigg)+O(\alpha_s^2)\,,\nonumber
\eea
where the gauge-invariant partonic cross sections are given in Appendix A, eqs.~(\ref{eq:partSigqD1Tp})--(\ref{eq:partSigqgdyn3}).
Note that there are no divergences as $\beta\to 1$ because of the support properties of the fragmentation correlators discussed before \ref{qgqcor}.

As with the LO result (cf.~eq.~(\ref{eq:CSUTLOLIR})), we can replace the kinematical twist-3 FF $D_{1T}^{\perp (1),q}$ in (\ref{eq:CSUTNLOqqqgq}) by combining both the EoMR (\ref{eq:EoMQD}) and the LIR (\ref{eq:LIRD}). This leads to the following equations~\cite{Kanazawa:2015ajw},
\bea
\frac{\d}{\d z}D_{1T}^{\perp (1),q}(z)&=&\frac{1}{z}\int_0^1\d\beta\,\left[\frac{\Im[\hat{D}_{FT}^{qg}-\hat{G}_{FT}^{qg}](z,\beta)}{1-\beta}+2\frac{\Im[\hat{D}_{FT}^{qg}](z,\beta)}{(1-\beta)^2}\right]\,,\label{eq:DerivDperp}\\[0.3cm]
D_{1T}^{\perp (1),q}(z)&=&-\int_z^1\tfrac{\d w}{w}\int_0^1\d\beta\,\left[\frac{\Im[\hat{D}_{FT}^{qg}-\hat{G}_{FT}^{qg}](\tfrac{z}{w},\beta)}{1-\beta}+2\frac{\Im[\hat{D}_{FT}^{qg}](\frac{z}{w},\beta)}{(1-\beta)^2}\right]\,.\label{eq:IntDperp}
\eea
In the second line, the usual boundary condition $D_{1T}^{\perp (1),q}(z=1)=0$ was applied. 

Since the function $D_{1T}^{\perp (1),q}$ appears convoluted under an integral in (\ref{eq:CSUTNLOqqqgq}), the replacement of it is a bit more subtle than at LO (\ref{eq:CSUTLOLIR}). We first realize that we need to split the partonic cross section $w^{-2-2\varepsilon}\hat{\sigma}_{D_{1T}^{\perp (1)}}(w)$ in eq.~(\ref{eq:partSigqD1Tp}) into two parts, one that is proportional to the delta function $\delta(1-w)$, and one that is proportional to plus distributions,
\bea
w^{-2-2\varepsilon}\hat{\sigma}_{D_{1T}^{\perp (1)}}(w) = \hat{\sigma}_{\delta}\,\delta(1-w) + w^{-2-2\varepsilon}\hat{\sigma}_{+}(w)\,.\label{eq:LIRsplit}
\eea
The part proportional to $\delta(1-w)$ requires the replacement (\ref{eq:IntDperp}), while for the other part we need to integrate by parts and apply both identities (\ref{eq:IntDperp}) and (\ref{eq:DerivDperp}). We find after a straightforward calculation,
\bea
\int_{z_h}^1 \tfrac{\d w}{w^2}\,w^{-2\varepsilon}\,\hat{\sigma}_{D_{1T}^{\perp (1)}}^f(w)\,D_{1T}^{\perp (1),f}(\tfrac{z_h}{w})&=&\int_{z_h}^1\tfrac{\d w}{w}\int_0^1\d \beta\,\hat{\Sigma}_+(w)\times\label{eq:LIRNLO}\\
&&\,\left[\frac{\Im[\hat{D}_{FT}^{fg}-\hat{G}_{FT}^{fg}](\tfrac{z_h}{w},\beta)}{1-\beta}+2\frac{\Im[\hat{D}_{FT}^{fg}](\tfrac{z_h}{w},\beta)}{(1-\beta)^2}\right]\,,\nonumber
\eea
where $\hat{\Sigma}_+(w)$ is related to the principal function of $w^{-2-2\varepsilon}\hat{\sigma}_+(w)$ $\--$ with the plus prescription removed in (\ref{eq:partSigqD1Tp}). Eventually, we find,
\bea
\hat{\Sigma}_+(w)&=&2+2\,\frac{C_F\,\alpha_s}{2\pi}S_\varepsilon \left(\frac{s}{\mu^2}\right)^{-\varepsilon}\Big\{-\frac{1}{\varepsilon }\left[\tfrac{1}{w}+\tfrac{1}{2}+2 \ln (1-w)-\ln
   (w)\right]+5\, \text{Li}_2(w)\nonumber\\
&&\hspace{-2cm}+ \ln ^2(1-w)+(-\tfrac{5}{2}+4 \ln (w)+\tfrac{1}{w})\, \ln (1-w)+ (4-\ln (w)+\tfrac{2}{w})\,\ln (w)-\tfrac{9}{2}-\tfrac{\pi^2}{6}\Big\}.\label{eq:partInt}
\eea
One may readily replace the kinematical twist-3 contributions in (\ref{eq:CSUTNLOqqqgq}) with dynamical functions by means of (\ref{eq:LIRNLO}), and, as a result, add the function $w^{1+2\varepsilon}\hat{\Sigma}_+(w)$ in (\ref{eq:partInt}) to the partonic cross sections $\hat{\sigma}_1^{fg}(w,\beta)$ and $\hat{\sigma}_2^{fg}(w,\beta)$ in eqs.~(\ref{eq:partSigqgdyn1}),~(\ref{eq:partSigqgdyn2}).  Thus, one can obtain a result solely in terms of quark-gluon-quark correlators.  In fact, these dynamical functions are what one probes in a measurement of this observable, rather than the often discussed polarizing FF $D_{1T}^\perp$~\cite{Anselmino:2000vs,Anselmino:2001js,Boer:2010ya}.

\subsubsection{Quark Anti-quark Gluon Fragmentation}

\begin{figure}
\centering
\includegraphics[width=0.9\textwidth]{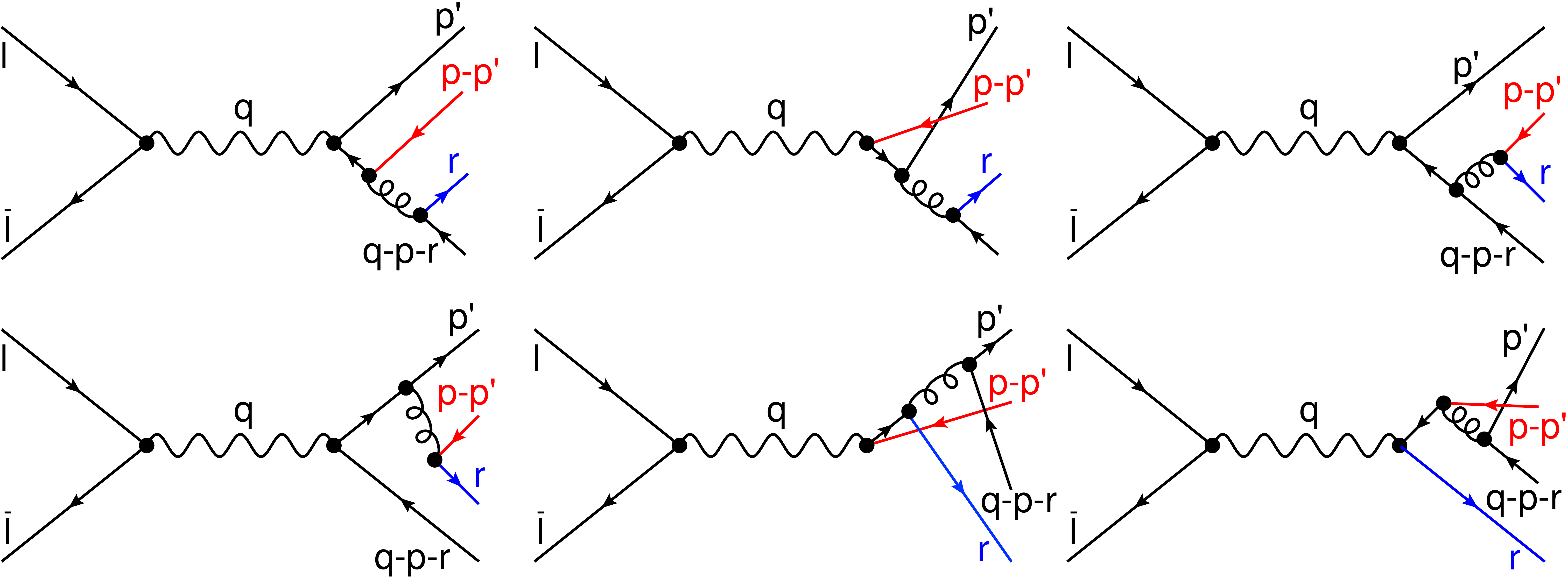}
\caption{NLO corrections from quark radiation to quark/anti-quark fragmentation.}
\label{fig:NLOrealqqb}
\end{figure}
In the same way we can study another reaction channel induced by quark/anti-quark-gluon fragmentation. The relevant diagrams are shown in fig.~\ref{fig:NLOrealqqb} and they interfere with the gluon fragmentation diagrams in fig.~\ref{fig:NLOrealg}. We find that the contributions coming from the first and second diagram in fig.~\ref{fig:NLOrealqqb} cancel when summed. The other diagrams contribute and the resulting cross section acquires the following form,
\bea
\frac{E_h\,\d \sigma^{\mathrm{q\bar{q}g}}(S_h)}{\d^{d-1}\vec{P}_h}&=&\sigma_0\,(1-2v)\,\frac{4M_h}{z_h\,s^2}\,\epsilon^{l l^\prime P_h S_h}\sum_{f=q,\bar{q}}e_f^2\,\int_{z_h}^1\tfrac{\d w}{w^2}w^{-2\varepsilon}\int_0^1 \d\beta\,\label{eq:CSUTNLOqqbg}\\
&&\times \Bigg(\hat{\sigma}_{4}^{f\bar{f}}(w,\beta)\,\Im[\hat{D}_{FT}^{f\bar{f}[0]}](\tfrac{z_h}{w},\beta)+\hat{\sigma}_{5}^{f\bar{f}}(w,\beta)\,\Im[\hat{G}_{FT}^{f\bar{f}[0]}](\tfrac{z_h}{w},\beta)\Bigg)+O(\alpha_s^2)\,,\nonumber
\eea
where the partonic cross sections are given in Appendix A, eqs.~(\ref{eq:partSigqqbdyn4}), (\ref{eq:partSigqqbdyn5}).
We again find that the $\kappa$ dependence completely drops out in the partonic cross sections $\hat{\sigma}_{4}$ and $\hat{\sigma}_{5}$. Also, we note that $\hat{\sigma}_{4}$ is symmetric and $\hat{\sigma}_{5}$ antisymmetric under a transformation $\beta\to 1-\beta$. This means that quark/anti-quark/gluon and anti-quark/quark/gluon fragmentation contribute equally.

\subsubsection{Gluon-Gluon \& Tri-Gluon Fragmentation}

\begin{figure}
\centering
\includegraphics[width=0.9\textwidth]{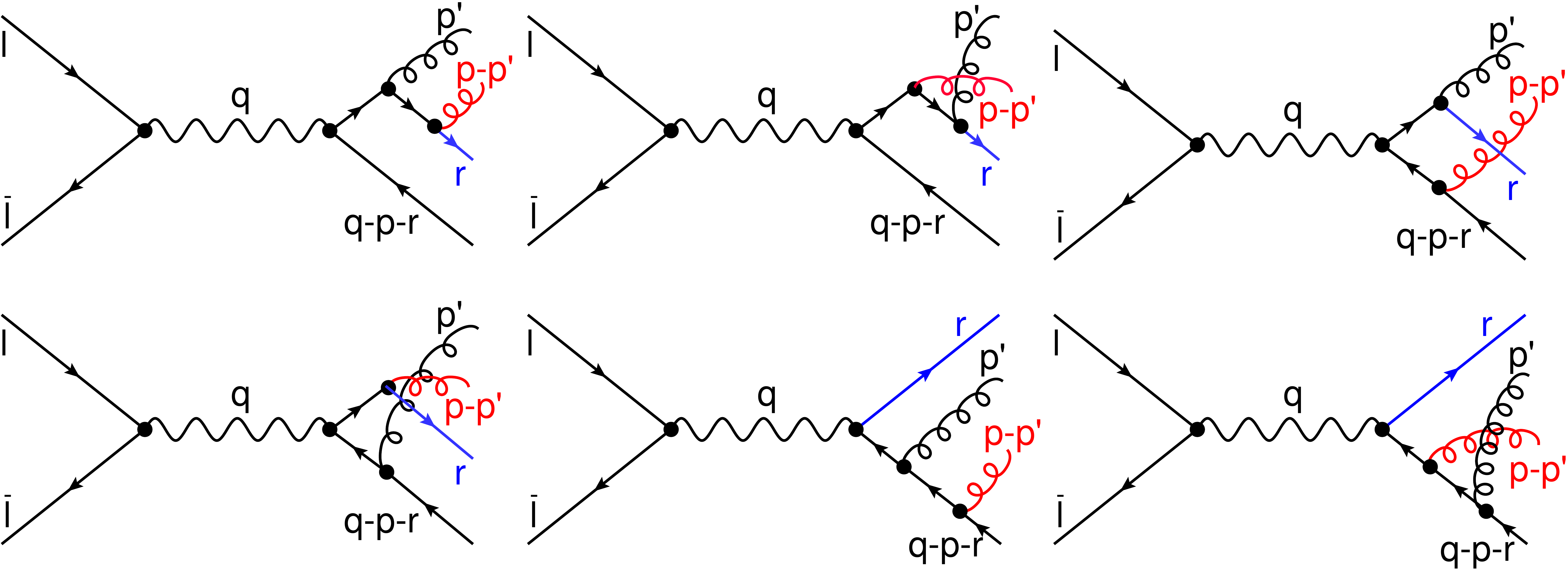}
\caption{NLO corrections from quark radiation to gluon-gluon fragmentation.}
\label{fig:NLOrealgg}
\end{figure}
At last we calculate the remaining contributions from two-gluon and tri-gluon fragmentation. The first contributions originate from the squared sum of the diagrams in fig.~\ref{fig:NLOrealg} while the latter is generated from an interference of the diagrams in fig.~\ref{fig:NLOrealgg} with those in fig.~\ref{fig:NLOrealg}. We note that none of these diagrams involves a gluon propagator or polarization sum. Hence, we cannot explicitly check that the $\kappa$-dependence vanishes, as we did for the other channels. Nonetheless we calculated the partonic cross sections using two different methods, as discussed below eq.~(\ref{eq:splitIntLC}), and find exact agreement for each of the perturbative cross sections.

\paragraph{Intrinsic \& Kinematical Twist-3 for Gluons}
The contributions from gluon intrinsic- and kinematical twist-3 functions read 
\bea
\frac{E_h\,\d \sigma^{\mathrm{intr\& kin}}(S_h)}{\d^{d-1}\vec{P}_h}&=&\sigma_0\,(1-2v)\,\frac{4M_h}{z_h\,s^2}\,\epsilon^{l l^\prime P_h S_h}\,\mathcal{Q}\,\int_{z_h}^1\tfrac{\d w}{w^2}w^{-2\varepsilon}\,\label{eq:CSUTgNLOintkin}\\
&&\hspace{-1.5cm}\times\,\Bigg(\hat{\sigma}_{D_{1T}^{\perp (1)}}^g(w)\,D_{1T}^{\perp (1),g[0]}(\tfrac{z_h}{w})+\hat{\sigma}_{H_{1}}^g(w)\,H_{1}^{(1),g[0]}(\tfrac{z_h}{w})\nonumber\\
&&+\hat{\sigma}^g_{D_T}(w)\,\int_0^1\d \beta \,\frac{2\,\Im[\hat{N}_2^{a[0]}](\tfrac{z_h}{w},\beta)-2(1-\varepsilon)\,\Im[\hat{N}_1^{[0]}](\tfrac{z_h}{w},\beta)}{1-\beta}\nonumber\\
&&-\frac{1}{C_F}\,\hat{\sigma}^g_{D_T}(w)\,\sum_{f=q,\bar{q}}\int_0^1\d\beta\,\Im[\hat{D}_{FT}^{f\bar{f}[0]}](\tfrac{z_h}{w},\beta)\Bigg)+O(\alpha_s^2)\,,\nonumber
\eea
where again we have already applied the EoMR (\ref{eq:EoMGD}) for gluons in order to eliminate the intrinsic twist-3 gluon function $D_T^g(z)$. In addition, we introduced a notation for the sum of fractional active quark charges, $\mathcal{Q}\equiv 2 \sum_{q}e_q^2$, and $\hat{N}_2^{a}$ is the antisymmetric combination $\hat{N}_2^{a}(z,\beta)\equiv (\hat{N}_2(z,\beta)- \hat{N}_2(z,1-\beta))/2$.  As before, we refrain from explicitly stating the relevant partonic cross sections at this point until the dynamical graphs are included.

\paragraph{Dynamical Twist-3 for Gluons}
The contribution from tri-gluon fragmentation takes the following form,
\bea
\frac{E_h\,\d \sigma^{\mathrm{dyn}}(S_h)}{\d^{d-1}\vec{P}_h}&=&\sigma_0\,(1-2v)\,\frac{4M_h}{z_h\,s^2}\,\epsilon^{l l^\prime P_h S_h}\,\mathcal{Q}\,\int_{z_h}^1\tfrac{\d w}{w^2}w^{-2\varepsilon}\int_0^1\d \beta\,\Bigg(\hat{\sigma}^g_{N_2^s}(w,\beta)\,\frac{\Im[\hat{N}^{s[0]}_2](\tfrac{z_h}{w},\beta)}{\beta^2(1-\beta)^2}\nonumber\\
&&+\,\hat{\sigma}^g_{N_2^a}(w,\beta)\,\frac{\Im[\hat{N}_2^{a[0]}](\tfrac{z_h}{w},\beta)}{\beta^2(1-\beta)^2}+\hat{\sigma}_{N_1}^g(w\beta)\frac{\Im[\hat{N}_1^{[0]}](\tfrac{z_h}{w},\beta)}{\beta^2(1-\beta)^2}\Bigg)+O(\alpha_s^2)\,\label{eq:CSUTNLOgdyn}
\eea
with $\hat{N}_2^{s,a}(z,\beta)\equiv (\hat{N}_2(z,\beta)\pm \hat{N}_2(z,1-\beta))/2$ the symmetric and antisymmetric part of $\hat{N}_2(z,\beta)$.
Note again that there are no divergences as $\beta\to 0\;{\rm or}\;1$ because of the support properties of the fragmentation correlators discussed before \ref{qgqcor}.

\paragraph{Result for the combined $gg\,\&\,ggg$ channel}
The full result for gluon-gluon and tri-gluon fragmentation is given by the sum of eq.~(\ref{eq:CSUTgNLOintkin}) and eq.~(\ref{eq:CSUTNLOgdyn}),
\bea
\frac{E_h\,\d \sigma^{\mathrm{gg\& ggg}}(S_h)}{\d^{d-1}\vec{P}_h}&=&\sigma_0\,(1-2v)\,\frac{4M_h}{z_h\,s^2}\,\epsilon^{l l^\prime P_h S_h}\,\mathcal{Q}\,\int_{z_h}^1\tfrac{\d w}{w^2}w^{-2\varepsilon}\int_0^1\d\beta\label{eq:CSUTgNLO5g}\\
&&\hspace{-2.5cm}\times\Bigg(\hat{\sigma}_{D_{1T}^{\perp (1)}}^g(w)\,D_{1T}^{\perp (1),g[0]}(\tfrac{z_h}{w})+\hat{\sigma}_{H_{1}}^g(w)\,H_{1}^{(1),g[0]}(\tfrac{z_h}{w})\nonumber\\
&&\hspace{-2.5cm}+\hat{\sigma}^g_{1}(w,\beta)\,\frac{\Im[\hat{N}^{s[0]}_2](\tfrac{z_h}{w},\beta)}{\beta^2(1-\beta)^2}+\hat{\sigma}^g_{2}(w,\beta)\,\frac{\Im[\hat{N}_2^{a[0]}](\tfrac{z_h}{w},\beta)}{\beta^2(1-\beta)^2}+\hat{\sigma}^g_{3}(w,\beta)\,\frac{\Im[\hat{N}^{[0]}_1](\tfrac{z_h}{w},\beta)}{\beta^2(1-\beta)^2}\nonumber\\
&&-\frac{1}{C_F}\,\hat{\sigma}^g_{D_T}(w)\,\sum_{f=q,\bar{q}}\Im[\hat{D}_{FT}^{f\bar{f}[0]}](\tfrac{z_h}{w},\beta)\Bigg),\nonumber
\eea
where the explicit form of the partonic cross sections is given in Appendix A, eqs.~\eqref{eq:DTg}--\eqref{eq:partSiggg3}.

Note that the collinear divergence for $\hat{\sigma}_{H_{1}}^g(w)$ cancels. Hence, it does not appear in the evolution of the twist-3 fragmentation functions at LO. Also, we mention that the collinear pole for the kinematical gluon twist-3 function $D_{1T}^{\perp (1), g}$ is just the usual twist-2 $qg$-splitting function. We conjecture that the term in (\ref{eq:CSUTgNLO5g}) generated by $D_{1T}^{\perp (1), g}$ can be converted to $ggg$- and $q\bar{q}g$ dynamical twist-3 functions by means of gluon LIRs, just like we did in (\ref{eq:LIRNLO}). Unfortunately, LIRs for gluons have not been derived in the literature, to the best of our knowledge. We leave this subject as future work.
In addition, note that $\hat{\sigma}^g_{1}$ is symmetric under $\beta \to 1-\beta$, while $\hat{\sigma}^g_{2,3}$ are antisymmetric.
The last term in (\ref{eq:CSUTgNLO5g}) is proportional to $\Im[\hat{D}_{FT}^{q\bar{q}}]$, which is generated non-perturbatively by the gluon QCD EoMR (\ref{eq:EoMGD}). Note the difference with the term proportional to $\Im[\hat{D}_{FT}^{q\bar{q}}]$ in (\ref{eq:CSUTNLOqqbg}). The partonic cross sections $\hat{\sigma}_{D_{T}}^g$ in  (\ref{eq:CSUTgNLO5g}) and $\hat{\sigma}_4^{f\bar{f}}$ in (\ref{eq:CSUTNLOqqbg}) carry different color factors and also different charge factors. Nevertheless, in principle, they may be combined when adding all twist-3 contributions (\ref{eq:CSUTNLOqqqgq}), (\ref{eq:CSUTNLOqqbg}) and (\ref{eq:CSUTgNLO5g}).

\section{Evolution equation for $\boldsymbol{D_T(z)}$ \label{Evol}}

In section \ref{NLO} we calculated terms relevant for the transverse-spin dependent $e^+e^-\to \Lambda^\uparrow X$ cross section at NLO accuracy, where we have shown how to obtain gauge-invariant partonic cross sections free of $1/\varepsilon^2$-poles.  We can collect these results and write down the total cross section as
\begin{align}
\frac{E_h\,\d \sigma(S_h)}{\d^{d-1}\vec{P}_h}&=(\ref{eq:CSUTNLOqqqgq})+(\ref{eq:CSUTNLOqqbg})+(\ref{eq:CSUTgNLO5g})-\sigma_0\,(1-2v)\,\frac{4M_h}{z_h\,s^2}\epsilon^{l l^\prime P_h S_h}\!\!\!\sum_{f=q,\bar{q}}\!e_f^2\,\frac{2D_T^{f[1]}(z_h)}{z_h}+ O(\alpha_s^2)\,, \nonumber\\ \label{eq:CSUTNLO}
\end{align} 
where  $\sigma_0=(4\pi^2 z_h)^\varepsilon 2 N_c \alpha_{\mathrm{em}}^2/(z_h s^2)$.  The last term in eq.~(\ref{eq:CSUTNLO}), where the function $D_T^{f[1]}(z)$ is the one-loop renormalized intrinsic FF, is the subtraction term that should remove the collinear divergences in the partonic cross sections (\ref{eq:partSigqD1Tp})--(\ref{eq:partSiggg3}).  This is in full analogy to the unpolarized case (cf.~eq.~(\ref{eq:CSUnpolNLO})).  Note again, due to the EoMR (\ref{eq:EoMQD}) and LIR (\ref{eq:LIRD}), the function $D_T^f(z)$ can be written as 
\begin{align}
D_T^f(z)
&=-z\,\left(D_{1T}^{\perp (1),f}(z)-\int_0^1\d\beta \frac{\Im[\hat{D}_{FT}^{fg}-\hat{G}_{FT}^{fg}](z,\beta)}{1-\beta}\right)\\[0.1cm]
&=z\int_{z}^1\tfrac{\d w}{w}\int_0^1\d\beta\,\left[\frac{(1+\delta(1-w))\Im[\hat{D}_{FT}^{fg}-\hat{G}_{FT}^{fg}](\tfrac{z}{w},\beta)}{1-\beta}+\frac{2\,\Im[\hat{D}_{FT}^{fg}](\tfrac{z}{w})}{(1-\beta)^2}\right].\label{eq:DefDren}
\end{align}
Unfortunately, the result for $D_T^{f[1]}(z)$ has not been derived in the literature so far, to the best of our knowledge. There have only been NLO calculations of chiral-odd collinear twist-3 FFs~\cite{Belitsky:1996hg,Kang:2010xv,Ma:2017upj} that are relevant for unpolarized hadrons as well as $D_{1T}^{\perp(1)}(z)$ for transverse polarization~\cite{Kang:2010xv}. 
The derivation of $D_T^{f[1]}(z)$ (along with the renormalization of the dynamical twist-3 FFs in eq.~(\ref{eq:DefDren})) needed for the subtraction term in eq.~(\ref{eq:CSUTNLO}) is beyond the scope of this paper and will be the subject of future work.

However, {\it if collinear twist-3 factorization holds for $e^+e^-\to \Lambda^\uparrow X$}, we can read off the renormalization counterterms from the unsubtracted partonic cross sections (\ref{eq:partSigqD1Tp})--(\ref{eq:partSigqgdyn2}), (\ref{eq:partSigqqbdyn4})--(\ref{eq:partSigqqbdyn5}), and (\ref{eq:DTg})--(\ref{eq:partSiggg3}).  The one-loop renormalized intrinsic FF then takes the form,
\begin{align}
D_{T}^{f[1]}(z)&=D_{T(0)}^{f[1]}(z)\nonumber\\[0.3cm]
&+\frac{z}{2}\,\int_z^1\tfrac{\d w}{w^2}\int_0^1 \d\beta\,\Bigg[\,Z_{1,f\to f}^{[1]}(w)\,D_{1T}^{\perp (1),f[0]}(\tfrac{z}{w})+Z_{1,f\to g}^{[1]}(w)\,D_{1T}^{\perp (1),g[0]}(\tfrac{z}{w})\,\nonumber\\[0.3cm]
&+Z_{2,f\to fg}^{[1]}(w,\beta)\,\frac{\Im[\hat{D}_{FT}^{fg[0]}-\hat{G}_{FT}^{fg[0]}](\tfrac{z}{w},\beta)}{1-\beta}+Z_{3,f\to fg}^{[1]}(w,\beta)\,\frac{2\,\Im[\hat{D}_{FT}^{fg[0]}](\tfrac{z}{w},\beta)}{(1-\beta)^2}\nonumber\\[0.3cm]
&+\sum_{f'=q^\prime,\bar{q}^\prime}Z_{4,f\to f'\bar{f'}}^{[1]}(w,\beta)\,\Im[\hat{D}_{FT}^{f'\bar{f'}[0]}(\tfrac{z}{w},\beta)]+\sum_{f'=q^\prime,\bar{q}^\prime}Z_{5,f\to f'\bar{f'}}^{[1]}(w,\beta)\,\Im[\hat{G}_{FT}^{f'\bar{f'}[0]}(\tfrac{z}{w},\beta)]\nonumber\\[0.3cm]
&+Z_{6,f\to gg}^{[1]}(w,\beta)\,\frac{\Im[\hat{N}_{2}^{s[0]}(\tfrac{z}{w},\beta)]}{\beta^2(1-\beta)^2}+Z_{7,f\to gg}^{[1]}(w,\beta)\,\frac{\Im[\hat{N}_{2}^{a[0]}(\tfrac{z}{w},\beta)]}{\beta^2(1-\beta)^2}\nonumber\\[0.3cm]
&+Z_{8,f\to gg}^{[1]}(w,\beta)\,\frac{\Im[\hat{N}_{1}^{[0]}(\tfrac{z}{w},\beta)]}{\beta^2(1-\beta)^2}\,\Bigg]\,,\label{eq:renDT}
\end{align}
where the UV counterterms $Z$ can be found in appendix B, eqs.~(\ref{eq:ZDqq})--(\ref{eq:ZDT8gg}).
We again emphasize that we have simply postulated the form of $D_{T}^{f[1]}(z)$, and this is {\it not} a proof of twist-3 factorization at one loop for this process.  Rather, one would have to directly calculate $D_{T}^{f[1]}(z)$ and confirm eq.~(\ref{eq:renDT}) and the UV counterterms (\ref{eq:SplitFuncD})--(\ref{eq:ZDT8gg}).
Nevertheless, we proceed with the evaluation of eq.~(\ref{eq:CSUTNLO}) to determine the cross section for $e^+e^-\to \Lambda^\uparrow\, X$ at NLO in $d=4$ dimensions,
\begin{align}
\frac{E_h\,\d \sigma(S_h)}{\d^{3}\vec{P}_h}&=\frac{8M_h N_c \alpha_{\mathrm{em}}^2} {(z_h s^2)^2}(1-2v)\,\epsilon^{l l^\prime P_h S_h}\sum_{f=q,\bar{q}}e_f^2\,\int_{z_h}^1\tfrac{\d w}{w^2}\int_0^1 \d\beta \nonumber\\[0.2cm]
&\hspace{-1cm}\times\Bigg\{\Bigg(\hat{c}_{D_{1T}^{\perp (1)}}^f(w)\,D_{1T}^{\perp (1),f}(\tfrac{z_h}{w};\mu)+\hat{c}_{1}^{fg}(w,\beta)\,\frac{\Im[\hat{D}_{FT}^{fg}-\hat{G}_{FT}^{fg}](\tfrac{z_h}{w},\beta;\mu)}{1-\beta}\nonumber\\[0.2cm]
&+\,\hat{c}_{2}^{fg}(w,\beta)\,\frac{2\,\Im[\hat{D}_{FT}^{fg}](\tfrac{z_h}{w},\beta;\mu)}{(1-\beta)^2}+\hat{c}_{3}^{fg}(w,\beta)\,\frac{\Re[\hat{D}_{FT}^{fg}-\hat{G}_{FT}^{fg}](\tfrac{z_h}{w},\beta;\mu)}{1-\beta}\Bigg)\nonumber\\[0.2cm]
&+\,\Bigg(\hat{c}_{4}^{f\bar{f}}(w,\beta)\,\Im[\hat{D}_{FT}^{f\bar{f}}](\tfrac{z_h}{w},\beta;\mu)+\hat{c}_{5}^{f\bar{f}}(w,\beta)\,\Im[\hat{G}_{FT}^{f\bar{f}}](\tfrac{z_h}{w},\beta;\mu)\Bigg)\nonumber\\[0.2cm]
&+\,\Bigg(\hat{c}_{D_{1T}^{\perp (1)}}^g(w)\,D_{1T}^{\perp (1),g}(\tfrac{z_h}{w};\mu)+\hat{c}_{H_{1}}^g(w)\,H_{1}^{(1),g}(\tfrac{z_h}{w};\mu)\nonumber\\[0.2cm]
&+\hat{c}^g_{1}(w,\beta)\,\frac{\Im[\hat{N}^{s}_2](\tfrac{z_h}{w},\beta;\mu)}{\beta^2(1-\beta)^2}+\hat{c}^g_{2}(w,\beta)\,\frac{\Im[\hat{N}_2^{a}](\tfrac{z_h}{w},\beta;\mu)}{\beta^2(1-\beta)^2}+\hat{c}^g_{3}(w,\beta)\,\frac{\Im[\hat{N}_1](\tfrac{z_h}{w},\beta;\mu)}{\beta^2(1-\beta)^2}\nonumber\\[0.2cm]
&-\frac{1}{C_F}\,\hat{c}^g_{D_T}(w)\,\sum_{f'=q',\bar{q}'}\Im[\hat{D}_{FT}^{f'\bar{f'}}](\tfrac{z_h}{w},\beta;\mu)\Bigg)\Bigg\}+ O(\alpha_s^2)\,,  \label{eq:CSUTNLOfinite}
\end{align} 
where the finite partonic cross sections in the ${\rm \overline{MS}}$ scheme are given by
\bea
\hat{c}_{D_{1T}^{\perp (1)}}^f(w) &=&-2\delta(1-w)+2\frac{C_F\alpha_s}{2\pi}\Bigg[\frac{1+w^2}{(1-w)_+}\ln\!\left(\frac{\mu^2}{sw^2}\right)\nonumber\\
&&\hspace{-1.5cm}+\,\delta(1-w)\left(\frac{3}{2}\ln\!\left(\frac{\mu^2}{s}\right)+\frac{9}{2}-\frac{2\pi^2}{3}\right)-\left(\frac{\ln(1-w)}{1-w}\right)_+(1+w^2)+\frac{4+2w-3w^2}{2\,(1-w)_+}\Bigg],\label{eq:partSigqD1Tpfinite} \\[0.3cm]
\hat{c}_{1}^{fg}(w,\beta) &=&2\,\delta(1-w)\nonumber\\
&&\hspace{-1.5cm}+\,2\frac{\alpha_s}{2\pi}\Bigg[\delta(1-w)\,\Big[(C_F-\tfrac{N_c}{2})\,\frac{\ln(\beta)}{1-\beta}\ln\!\left(\frac{\mu^2}{sw^2}\right)+\delta_{\mathrm{UV}}\ln w\Big]+C_F\frac{1-2w-w^2}{(1-w)_+}\ln\!\left(\frac{\mu^2}{sw^2}\right)\nonumber\\
&&+\,C_F\,\Bigg\{\delta(1-w)\left[\frac{2\ln(\beta)-\frac{1}{2}\ln^2(\beta)}{1-\beta}-\tfrac{1}{2}\ln(1-\beta)-\tfrac{3}{2}+\frac{2\pi^2}{3}\right]\nonumber\\
&&\hspace{1.5cm}-\,(1-2w-w^2)\left(\frac{\ln(1-w)}{1-w}\right)_++\frac{2-\frac{11}{2}w+2w^2}{(1-w)_+}-\frac{w}{\beta}\Bigg\}\nonumber\\
&&-\,\frac{N_c}{2}\Bigg\{\delta(1-w)\,\left[\frac{1}{\beta}\ln(1-\beta)+\frac{2\ln(\beta)-\frac{1}{2}\ln^2(\beta)}{1-\beta}\right]-\frac{\,w}{\beta\,(1-w\,\beta)}\Bigg\}\Bigg]\,,\label{eq:partSigqgdyn1finite}\\[0.3cm]
\hat{c}_{2}^{fg}(w,\beta) &=&\frac{\alpha_s}{2\pi}\frac{1}{\beta}\,\Bigg[-2\!\left(C_F\,(1-\beta)-\frac{N_c}{2}\right)\ln\!\left(\frac{\mu^2}{sw^2}\right)\nonumber\\
&& \hspace{-1.8cm}-\,2C_F\left(-w+\frac{w\beta}{2}+(1-\beta)(2-\ln(1-w))\right)+N_c\left(1-\ln(1-w)+\frac{1-w}{1-w\beta}\right)\Bigg],\label{eq:partSigqgdyn2finite}\\[0.3cm]
\hat{c}_{3}^{fg}(w,\beta) &=&\frac{\alpha_s}{2\pi}\Bigg[-2\pi\,\delta(1-w)\,\left(\frac{3}{2}\,C_F+(C_F-\tfrac{N_c}{2})\frac{\ln(\beta)}{1-\beta}\right)\Bigg],\label{eq:partSigqgdyn3finite}\\[0.3cm]
\hat{c}_{4}^{f\bar{f}}(w,\beta) &=&2\frac{\alpha_s} {2\pi}(C_F-\tfrac{N_c}{2})\Big[\frac{2(1-w)+w^3\beta(1-\beta)}{\beta(1-\beta)(1-w\beta)(1-w(1-\beta))}\ln\!\left(\frac{\mu^2}{sw^2}\right)\nonumber\\[0.15cm]
&&+\,\frac{4-5w+w^2+w^3\beta(1-\beta)-\left(2(1-w)+w^3\beta(1-\beta)\right)\,\ln(1-w)}{\beta(1-\beta)(1-w\beta)(1-w(1-\beta))}\Big]\,,\label{eq:partSigqqbdyn4finite}\\[0.3cm]
\hat{c}_{5}^{f\bar{f}}(w,\beta) &=&2\frac{\alpha_s} {2\pi}(C_F-\tfrac{N_c}{2})\Bigg[\frac{w^3(1-2\beta)}{(1-w\beta)(1-w(1-\beta))}\ln\!\left(\frac{\mu^2}{sw^2}\right)\nonumber\\[0.15cm]
&&\hspace{2cm}-\,\frac{w(1-2\beta)\left[1-w-w^2\beta(1-\beta)\,(1-\ln(1-w))\right]}{\beta(1-\beta)(1-w\beta)(1-w(1-\beta))}\Bigg],\label{eq:partSigqqbdyn5finite}\\[0.3cm]
\hat{c}^g_{D_T}(w)&=&4 \frac{C_F\,\alpha_s}{2\pi}\,\frac{1-w}{w}\Bigg[\ln\!\left(\frac{\mu^2}{sw^2(1-w)}\right)+3-w\Bigg]\,,\label{eq:DTgfinite}\\[0.3cm]
\hat{c}_{D_{1T}^{\perp (1)}}^g(w) &=&2 \frac{C_F\,\alpha_s}{2\pi}\Bigg[-\frac{1+(1-w)^2}{w}\ln\!\left(\frac{\mu^2}{sw^2(1-w)}\right)-6\frac{1-w}{w}\Bigg],\label{eq:D1Tperpgfinite}\\[0.3cm]
\hat{c}_{H_{1}}^g(w) &=&4 \frac{C_F\,\alpha_s}{2\pi}\Bigg[\frac{7-5w}{w}-2(1-w)\Bigg],\label{eq:H1g}\\[0.3cm]
\hat{c}^g_{1}(w,\beta)&=&
\frac{C_F\,\alpha_s}{2\pi}\Bigg[(1-w+2w\beta(1-\beta))\ln\!\left(\frac{\mu^2}{sw^2(1-w)}\right)+1-w-4w\beta(1-\beta)\Bigg]\,,\nonumber\\\label{eq:partSiggg1finite}\\
\hat{c}^g_{2}(w,\beta)
&=&\, \frac{C_F\,\alpha_s}{2\pi} (1-2\beta)\,\frac{1-w}{w}\Bigg[ \left(8-3w-4\beta(1-\beta)\right)\ln\!\left(\frac{\mu^2}{sw^2(1-w)}\right)\nonumber\\[0.1cm]
&&\hspace{4cm}+\,24-11w-4\beta(1-\beta)(3-w)\Bigg]\,,\label{eq:partSiggg2finite}\\[0.3cm]
\hat{c}^g_{3}(w,\beta)
&=&\frac{C_F\,\alpha_s}{2\pi}(1-2\beta)\frac{1-w}{w}\Bigg[(4-w-4\beta(1-\beta))\ln\!\left(\frac{\mu^2}{sw^2(1-w)}\right)\nonumber\\[0.1cm]
&&\hspace{4cm}+\,12-5w+4\beta(1-\beta)(2-w)\Big]\,.\label{eq:partSiggg3finite}
\eea
From this result, one can derive the LO evolution equation for $D^f_T(z)$ as\\

\begin{align}
\frac{\partial}{\partial \ln\mu^2}\!&\left(D_{T}^{f}(z;\mu)\right)=\frac{z} {2}\int_z^1\tfrac{\d w}{w^2}\int_0^1 \d\beta\,\Bigg[\,P_{1,f\to f}^{[1]}(w)\,D_{1T}^{\perp (1),f}(\tfrac{z}{w};\mu)+P_{1,f\to g}^{[1]}(w)\,D_{1T}^{\perp (1),g}(\tfrac{z}{w};\mu)\,\nonumber\\[0.3cm]
&+P_{2,f\to fg}^{[1]}(w,\beta)\,\frac{\Im[\hat{D}_{FT}^{fg}-\hat{G}_{FT}^{fg}](\tfrac{z}{w},\beta;\mu)}{1-\beta}+P_{3,f\to fg}^{[1]}(w,\beta)\,\frac{2\,\Im[\hat{D}_{FT}^{fg}](\tfrac{z}{w},\beta;\mu)}{(1-\beta)^2}\nonumber\\[0.3cm]
&+\sum_{f'=q^\prime,\bar{q}^\prime}P_{4,f\to f'\bar{f'}}^{[1]}(w,\beta)\,\Im[\hat{D}_{FT}^{f'\bar{f'}}(\tfrac{z}{w},\beta;\mu)]+\sum_{f'=q^\prime,\bar{q}^\prime}P_{5,f\to f'\bar{f'}}^{[1]}(w,\beta)\,\Im[\hat{G}_{FT}^{f'\bar{f'}}(\tfrac{z}{w},\beta;\mu)]\nonumber\\[0.3cm]
&+P_{6,f\to gg}^{[1]}(w,\beta)\,\frac{\Im[\hat{N}_{2}^{s}(\tfrac{z}{w},\beta;\mu)]}{\beta^2(1-\beta)^2}+P_{7,f\to gg}^{[1]}(w,\beta)\,\frac{\Im[\hat{N}_{2}^{a}(\tfrac{z}{w},\beta;\mu)]}{\beta^2(1-\beta)^2}\nonumber\\[0.3cm]
&+P_{8,f\to gg}^{[1]}(w,\beta)\,\frac{\Im[\hat{N}_{1}(\tfrac{z}{w},\beta;\mu)]}{\beta^2(1-\beta)^2}
\,\Bigg]\,,\label{eq:evoDT}
\end{align}
where
\bea
P_{1,f\to f}^{[1]}(w)&=&-2\frac{C_F \alpha_s}{2\pi}\left(\frac{1+w^2}{(1-w)_+}+\frac{3}{2}\,\delta(1-w)\right)\,,\label{eq:SplitFuncD}\\[0.3cm]
P_{1,f\to g}^{[1]}(w)&=&4\frac{C_F \alpha_s}{2\pi}\left(\frac{1+(1-w)^2}{w}\right)\,,\label{eq:SplitDT1qq}\\[0.3cm]
P_{2,f\to fg}^{[1]}(w,\beta)&=& -2\frac{\alpha_s}{2\pi}\left(\delta(1-w)\,\left[(C_F-\tfrac{N_c}{2})\frac{\ln(\beta)}{1-\beta}-\frac{1}{2}\delta_{\mathrm{UV}}\right]+C_F\frac{1-2w-w^2}{(1-w)_+}\right)\,,\nonumber\\ \\[0.3cm]
P_{3,f\to fg}^{[1]}(w,\beta)&=&2\frac{C_F\alpha_s}{2\pi}\left(C_F\,\frac{1-\beta}{\beta}-\frac{N_c}{2}\frac{1}{\beta}\right)\,,\\[0.3cm]
P_{4,f\to f'\bar{f'}}^{[1]}(w,\beta)&=&-2\frac{\alpha_s}{2\pi}\,\Big(\frac{\delta^{ff'}(C_F-\tfrac{N_c}{2})(2(1-w)+w^3\beta(1-\beta))}{\beta(1-\beta)(1-w\beta)(1-w(1-\beta))}-4\frac{1-w}{w}\Big)\,,\\[0.3cm]
P_{5,f\to f'\bar{f'}}^{[1]}(w,\beta)&=&-2\frac{\alpha_s}{2\pi}\,\frac{\delta^{ff'}(C_F-\tfrac{N_c}{2})\,w^3(1-2\beta)}{ (1-w\beta)(1-w(1-\beta))}\,,\\[0.3cm]
P_{6,f\to gg}^{[1]}(w,\beta)&=&-2\frac{C_F\,\alpha_s}{2\pi}\,\left(1-w+2w\beta(1-\beta)\right))\,\\[0.3cm]
P_{7,f\to gg}^{[1]}(w,\beta)&=&-2\frac{C_F\,\alpha_s}{2\pi}\,(1-2\beta)\frac{1-w}{w}\,\left(8-3w-4\beta(1-\beta)\right)\,,\\[0.3cm]
P_{8,f\to gg}^{[1]}(w,\beta)&=&-2\frac{C_F\,\alpha_s}{2\pi}\,(1-2\beta)\frac{1-w}{w}\,\left(4-w+4\beta(1-\beta)\right)\label{eq:SplitDT8gg}\,.
\eea
The expressions in eqs.~(\ref{eq:CSUTNLOfinite}), (\ref{eq:evoDT}) are new from this work and are our main results.

\section{Conclusions\label{Concl}}
In this paper we studied the production of polarized $\Lambda$-hyperons in electron-positron annihilation. We performed the perturbative QCD computations for the transverse-spin dependent differential cross section at both leading (LO) and next-to-leading order (NLO). Our leading-order result is given in eq.~\eqref{eq:CSUTLO}, which receives contributions from intrinsic, kinematic, and dynamic twist-3 fragmentation correlators. With the help of equation-of-motion relations, we find that the final result can be expressed in terms of a single intrinsic twist-3 fragmentation correlator $D_T^q(z)$, not the kinematical function $D_{1T}^{\perp\,q}(z)$ that one might have naively expected based on work in the Generalized Parton Model~\cite{Anselmino:2000vs,Anselmino:2001js}. Thus, the sizable transverse polarization measured in such a process indicates directly the size of $D_T^q(z)$. The next-to-leading order expression for the cross section involving hard partonic cross sections and interference terms is given in eq.~\eqref{eq:CSUTNLOfinite}. Assuming that collinear twist-3 factorization holds in this process, we derived the evolution equation in eq.~(\ref{eq:evoDT}) for the intrinsic twist-3 FF $D_T^q(z)$. The expressions in eqs.~(\ref{eq:CSUTNLOfinite}), (\ref{eq:evoDT}) are the main results of this work. As a cross-check of the collinear twist-3 factorization, an independent computation for the evolution equation of $D_T^q(z)$ is desirable. We will pursue such a study in a future publication, where we plan to derive such an evolution equation directly from the operator definition of $D_T^q(z)$. Another future research direction we are also pursuing at the moment is to study other related spin observables, such as the longitudinal lepton -- transverse hadron spin asymmetry. The techniques developed in our paper would be very useful in this regard. Last but not least, the phenomenology at NLO would be very interesting though it could be quite challenging. 

\acknowledgments 
The authors thank A.~Metz, T.~Rogers, N.~Sato, and W.~Vogelsang for fruitful discussions.
This work was supported by the U.S. Department of Energy, Office of Nuclear Physics under contract No.~DE-FG02-07ER41460 (L.~Gamberg),  
by the National Science Foundation under Grant No.~PHY-1720486 (Z.~Kang), by the U.S. Department of Energy, Office of Science under Contract No.~DE-AC52-06NA25396 and the LANL LDRD Program (S.~Yoshida),
and within the framework of the TMD Topical Collaboration.

\appendix
\section{Transverse hadron-spin partonic cross sections before subtraction}
In this Appendix we give the partonic cross sections for the transverse-hadron spin observable before the subtraction of collinear divergences (see the discussion in section \ref{Evol}).

\subsection{Quark-Quark \& Quark-Gluon-Quark}
The partonic cross sections in eq.~(\ref{eq:CSUTNLOqqqgq}) read
\bea
\hat{\sigma}_{D_{1T}^{\perp (1)}}^f(w) &=&-2\delta(1-w)+2\frac{C_F\alpha_s}{2\pi}S_\varepsilon \left(\frac{s}{\mu^2}\right)^{-\varepsilon}\Big[\frac{1}{\varepsilon}\left(\frac{1+w^2}{(1-w)_+}+\frac{3}{2}\,\delta(1-w)\right)\label{eq:partSigqD1Tp}\\
&&+\,\delta(1-w)\left(\frac{9}{2}-\frac{2\pi^2}{3}\right)-\left(\frac{\ln(1-w)}{1-w}\right)_+(1+w^2)+\frac{4+2w-3w^2}{2\,(1-w)_+}\Big]\,,\nonumber\\
\hat{\sigma}_{1}^{fg}(w,\beta) &=&2\,\delta(1-w)+2\frac{\alpha_s}{2\pi}S_\varepsilon \left(\frac{s}{\mu^2}\right)^{-\varepsilon}\label{eq:partSigqgdyn1}\\
&&\times\Bigg[\frac{1}{\varepsilon}\Bigg\{\delta(1-w)\,\Big[(C_F-\tfrac{N_c}{2})\,\frac{\ln(\beta)}{1-\beta}-\frac{1}{2}\left(\frac{s}{\mu^2}\right)^{\varepsilon}\delta_{\mathrm{UV}}\Big]+C_F\frac{1-2w-w^2}{(1-w)_+}\Bigg\}\nonumber\\
&&+\,C_F\,\Bigg\{\delta(1-w)\left[\frac{2\ln(\beta)-\frac{1}{2}\ln^2(\beta)}{1-\beta}-\tfrac{1}{2}\ln(1-\beta)-\tfrac{3}{2}+\frac{2\pi^2}{3}\right]\nonumber\\
&&-\,(1-2w-w^2)\left(\frac{\ln(1-w)}{1-w}\right)_++\frac{2-\frac{11}{2}w+2w^2}{(1-w)_+}-\frac{w}{\beta}\Bigg\}\nonumber\\
&&-\,\frac{N_c}{2}\Bigg\{\delta(1-w)\,\left[\frac{1}{\beta}\ln(1-\beta)+\frac{2\ln(\beta)-\frac{1}{2}\ln^2(\beta)}{1-\beta}\right]-\frac{\,w}{\beta\,(1-w\,\beta)}\Bigg\}\Bigg]\,,\nonumber
\eea
\bea
\hat{\sigma}_{2}^{fg}(w,\beta) &=&\frac{C_F\alpha_s}{2\pi}S_\varepsilon \left(\tfrac{s}{\mu^2}\right)^{-\varepsilon}\frac{1}{\beta}\,\Big[-\frac{2}{\varepsilon}\left(\,C_F\,(1-\beta)-\frac{N_c}{2}\right)\label{eq:partSigqgdyn2}\\
&& \hspace{-1cm}-2C_F\left(-w+\frac{w\beta}{2}+(1-\beta)(2-\ln(1-w))\right)+N_c\left(1-\ln(1-w)+\frac{1-w}{1-w\beta}\right)\Big]\,,\nonumber\\[0.3cm]
\hat{\sigma}_{3}^{fg}(w,\beta) &=&\frac{\alpha_s}{2\pi}S_\varepsilon \left(\frac{s}{\mu^2}\right)^{-\varepsilon}\Big[-2\pi\,\delta(1-w)\,\left(\frac{3}{2}\,C_F+(C_F-\tfrac{N_c}{2})\frac{\ln(\beta)}{1-\beta}\right)\Big]\,.\label{eq:partSigqgdyn3}
\eea
\\
\subsection{Quark-Anti-quark-Gluon}
The partonic cross sections in eq.~(\ref{eq:CSUTNLOqqbg}) read
\bea
\hat{\sigma}_{4}^{f\bar{f}}(w,\beta) &=&2 \frac{\alpha_s} {2\pi}(C_F-\tfrac{N_c}{2})S_{\varepsilon}\left(\tfrac{s}{\mu^2}\right)^{-\varepsilon}\Big[\frac{1}{\varepsilon}\frac{2(1-w)+w^3\beta(1-\beta)}{\beta(1-\beta)(1-w\beta)(1-w(1-\beta))}\nonumber\\
&&\hspace{-1cm}+\,\frac{4-5w+w^2+w^3\beta(1-\beta)-\left(2(1-w)+w^3\beta(1-\beta)\right)\,\ln(1-w)}{\beta(1-\beta)(1-w\beta)(1-w(1-\beta))}\Big]\,,\label{eq:partSigqqbdyn4}\\[0.3cm]
\hat{\sigma}_{5}^{f\bar{f}}(w,\beta) &=&2\frac{\alpha_s} {2\pi}(C_F-\tfrac{N_c}{2})S_{\varepsilon}\left(\tfrac{s}{\mu^2}\right)^{-\varepsilon}\Big[\frac{1}{\varepsilon}\frac{w^3(1-2\beta)}{(1-w\beta)(1-w(1-\beta))}\label{eq:partSigqqbdyn5}\\
&&-\frac{w(1-2\beta)\left[1-w-w^2\beta(1-\beta)\,(1-\ln(1-w))\right]}{\beta(1-\beta)(1-w\beta)(1-w(1-\beta))}\Big]\,.\nonumber
\eea
\\
\subsection{Gluon-Gluon \& Tri-Gluon}
The partonic cross sections in eq.~(\ref{eq:CSUTgNLO5g}) read
\bea
\hat{\sigma}^g_{D_T}(w)&=&4 \frac{C_F\,\alpha_s}{2\pi}S_\varepsilon\left(\frac{s}{\mu^2}\right)^{-\varepsilon}\,\frac{1-w}{w}\Big[\frac{1}{\varepsilon}+3-w-\ln(1-w)\Big]\,,\label{eq:DTg}\\
\hat{\sigma}_{D_{1T}^{\perp (1)}}^g(w) &=&2 \frac{C_F\,\alpha_s}{2\pi}S_\varepsilon\left(\frac{s}{\mu^2}\right)^{-\varepsilon}\Big[-\frac{1}{\varepsilon}\frac{1+(1-w)^2}{w}-6\frac{1-w}{w}+\frac{1+(1-w)^2}{w}\ln(1-w)\Big],\nonumber\\ \label{eq:D1Tperpg}\\
\hat{\sigma}_{H_{1}}^g(w) &=&4 \frac{C_F\,\alpha_s}{2\pi}S_\varepsilon\left(\frac{s}{\mu^2}\right)^{-\varepsilon}\Big[\frac{7-5w}{w}-2(1-w)\Big],\label{eq:H1ga}\\
\hat{\sigma}^g_{1}(w,\beta)&=&
\frac{C_F\,\alpha_s}{2\pi}S_\varepsilon \left(\frac{s}{\mu^2}\right)^{-\varepsilon}\Big[\frac{1-w+2w\beta(1-\beta)}{\varepsilon}\nonumber\\
&&\hspace{0cm}+1-w-4w\beta(1-\beta)-(1-w+2w\beta(1-\beta))\ln(1-w)\Big]\,,\label{eq:partSiggg1}\\
\hat{\sigma}^g_{2}(w,\beta)
&=&\, \frac{C_F\,\alpha_s}{2\pi}S_\varepsilon\left(\frac{s}{\mu^2}\right)^{-\varepsilon} (1-2\beta)\,\frac{1-w}{w}\Big[ \frac{1}{\varepsilon}\left(8-3w-4\beta(1-\beta)\right)\nonumber\\
&&\hspace{0cm}+24-11w-(8-3w)\ln(1-w)-4\beta(1-\beta)(3-w-\ln(1-w))\Big]\,,\label{eq:partSiggg2}\\
\hat{\sigma}^g_{3}(w,\beta)
&=&\frac{C_F\,\alpha_s}{2\pi}S_\varepsilon \left(\frac{s}{\mu^2}\right)^{-\varepsilon}(1-2\beta)\frac{1-w}{w}\Big[\frac{1}{\varepsilon}(4-w+4\beta(1-\beta))\nonumber\\
&&\hspace{0cm}+12-5w-(4-w)\ln(1-w)+4\beta(1-\beta)(2-w-\ln(1-w))\Big]\,.\label{eq:partSiggg3}
\eea

\section{UV counterterms for $\boldsymbol{D_{T}^{[1]}(z)}$}
The UV counterterms in eq.~(\ref{eq:renDT}) read
\bea
Z_{1,f\to f}^{[1]}(w)&=&2\frac{C_F \alpha_s}{2\pi}\frac{S_\varepsilon}{\varepsilon}\left(\frac{1+w^2}{(1-w)_+}+\frac{3}{2}\,\delta(1-w)\right)\,,\label{eq:ZDqq}\\[0.3cm]
Z_{1,f\to g}^{[1]}(w)&=&-4\frac{C_F \alpha_s}{2\pi}\frac{S_\varepsilon}{\varepsilon}\left(\frac{1+(1-w)^2}{w}\right)\,,\label{eq:ZDT1qq}\\[0.3cm]
Z_{2,f\to fg}^{[1]}(w,\beta)&=& 2\frac{\alpha_s}{2\pi}\frac{S_\varepsilon}{\varepsilon}\left(\delta(1-w)\,\left[(C_F-\tfrac{N_c}{2})\frac{\ln(\beta)}{1-\beta}-\frac{1}{2}\delta_{\mathrm{UV}}\right]+C_F\frac{1-2w-w^2}{(1-w)_+}\right)\,,\nonumber\\ \\[0.3cm]
Z_{3,f\to fg}^{[1]}(w,\beta)&=&-2\frac{ \alpha_s}{2\pi}\frac{S_\varepsilon}{\varepsilon}\left(C_F\,\frac{1-\beta}{\beta}-\frac{N_c}{2}\frac{1}{\beta}\right)\,,\\[0.3cm]
Z_{4,f\to f'\bar{f'}}^{[1]}(w,\beta)&=&2\frac{\alpha_s}{2\pi}\frac{S_\varepsilon}{\varepsilon}\,\Big(\frac{\delta^{ff'}(C_F-\tfrac{N_c}{2})(2(1-w)+w^3\beta(1-\beta))}{\beta(1-\beta)(1-w\beta)(1-w(1-\beta))}- 4\frac{1-w}{w}\Big)\,,\\[0.3cm]
Z_{5,f\to f'\bar{f'}}^{[1]}(w,\beta)&=&2\frac{\alpha_s}{2\pi}\frac{S_\varepsilon}{\varepsilon}\,\frac{\delta^{ff'}(C_F-\tfrac{N_c}{2})\,w^3(1-2\beta)}{(1-w\beta)(1-w(1-\beta))}\,,\\[0.3cm]
Z_{6,f\to gg}^{[1]}(w,\beta)&=& 2\frac{C_F\,\alpha_s}{2\pi}\frac{S_\varepsilon}{\varepsilon}\,\left(1-w+2w\beta(1-\beta)\right))\,\\[0.3cm]
Z_{7,f\to gg}^{[1]}(w,\beta)&=&2\frac{C_F\,\alpha_s}{2\pi}\frac{S_\varepsilon}{\varepsilon}\,(1-2\beta)\frac{1-w}{w}\,\left(8-3w-4\beta(1-\beta)\right)\,,\\[0.3cm]
Z_{8,f\to gg}^{[1]}(w,\beta)&=&2\frac{C_F\,\alpha_s}{2\pi}\frac{S_\varepsilon}{\varepsilon}\,(1-2\beta)\frac{1-w}{w}\,\left(4-w+4\beta(1-\beta)\right)\label{eq:ZDT8gg}\,.
\eea

\bibliography{Referenzen}{}

\providecommand{\href}[2]{#2}\begingroup\raggedright\begin{thebibliography}{10}

\bibitem{Bunce:1976yb}
G.~Bunce et~al., \emph{{Lambda0 Hyperon Polarization in inclusive production by
  300-GeV protons on beryllium}}, {\emph{Phys. Rev. Lett.} {\bfseries 36}
  (1976) 1113--1116}.

\bibitem{Schachinger:1978qs}
L.~Schachinger et~al., \emph{{A Precise Measurement of the $\Lambda^0$ Magnetic
  Moment}}, \href{https://doi.org/10.1103/PhysRevLett.41.1348}{\emph{Phys. Rev.
  Lett.} {\bfseries 41} (1978) 1348}.

\bibitem{Heller:1983ia}
K.~J. Heller et~al., \emph{{Polarization of XI0 and Lambda Hyperons produced bz
  400-GeV/c protons}},
  \href{https://doi.org/10.1103/PhysRevLett.51.2025}{\emph{Phys. Rev. Lett.}
  {\bfseries 51} (1983) 2025--2028}.

\bibitem{Lundberg:1989hw}
B.~Lundberg et~al., \emph{{Polarization in Inclusive $\Lambda$ and
  $\bar{\Lambda}$ Production at Large $p_T$}},
  \href{https://doi.org/10.1103/PhysRevD.40.3557}{\emph{Phys. Rev.} {\bfseries
  D40} (1989) 3557--3567}.

\bibitem{Yuldashev:1990az}
B.~S. Yuldashev et~al., \emph{{Neutral strange particle production in p Ne-20
  and p N interactions at 300-GeV/c}},
  \href{https://doi.org/10.1103/PhysRevD.43.2792}{\emph{Phys. Rev.} {\bfseries
  D43} (1991) 2792--2802}.

\bibitem{Ramberg:1994tk}
E.~J. Ramberg et~al., \emph{{Polarization of Lambda and anti-Lambda produced by
  800-GeV protons}},
  \href{https://doi.org/10.1016/0370-2693(94)91397-8}{\emph{Phys. Lett.}
  {\bfseries B338} (1994) 403--408}.

\bibitem{Fanti:1998px}
V.~Fanti et~al., \emph{{A Measurement of the transverse polarization of Lambda
  hyperons produced in inelastic p N reactions at 450-GeV proton energy}},
  \href{https://doi.org/10.1007/s100520050337}{\emph{Eur. Phys. J.} {\bfseries
  C6} (1999) 265--269}.

\bibitem{Abt:2006da}
{\scshape HERA-B} collaboration, I.~Abt et~al., \emph{{Polarization of Lambda
  and anti-Lambda in 920-GeV fixed-target proton-nucleus collisions}},
  \href{https://doi.org/10.1016/j.physletb.2006.05.040}{\emph{Phys. Lett.}
  {\bfseries B638} (2006) 415--421},
  [\href{https://arxiv.org/abs/hep-ex/0603047}{{\ttfamily hep-ex/0603047}}].

\bibitem{Erhan:1979xm}
S.~Erhan et~al., \emph{{$\Lambda^0$ Polarization in Proton Proton Interactions
  at $\sqrt{s}=53$-{GeV} and 62-{GeV}}},
  \href{https://doi.org/10.1016/0370-2693(79)90761-5}{\emph{Phys. Lett.}
  {\bfseries 82B} (1979) 301--304}.

\bibitem{ATLAS:2014ona}
{\scshape ATLAS} collaboration, G.~Aad et~al., \emph{{Measurement of the
  transverse polarization of $\Lambda$ and $\bar{\Lambda}$ hyperons produced in
  proton-proton collisions at $\sqrt{s}=7$ TeV using the ATLAS detector}},
  \href{https://doi.org/10.1103/PhysRevD.91.032004}{\emph{Phys. Rev.}
  {\bfseries D91} (2015) 032004},
  [\href{https://arxiv.org/abs/1412.1692}{{\ttfamily 1412.1692}}].

\bibitem{Airapetian:2006ee}
{\scshape HERMES} collaboration, A.~Airapetian et~al., \emph{{Longitudinal Spin
  Transfer to the Lambda Hyperon in Semi-Inclusive Deep-Inelastic Scattering}},
  \href{https://doi.org/10.1103/PhysRevD.74.072004}{\emph{Phys. Rev.}
  {\bfseries D74} (2006) 072004},
  [\href{https://arxiv.org/abs/hep-ex/0607004}{{\ttfamily hep-ex/0607004}}].

\bibitem{Airapetian:2007mx}
{\scshape HERMES} collaboration, A.~Airapetian et~al., \emph{{Transverse
  Polarization of Lambda and anti-Lambda Hyperons in Quasireal
  Photoproduction}},
  \href{https://doi.org/10.1103/PhysRevD.76.092008}{\emph{Phys. Rev.}
  {\bfseries D76} (2007) 092008},
  [\href{https://arxiv.org/abs/0704.3133}{{\ttfamily 0704.3133}}].

\bibitem{Airapetian:2014tyc}
{\scshape HERMES} collaboration, A.~Airapetian et~al., \emph{{Transverse
  polarization of $\Lambda$ hyperons from quasireal photoproduction on
  nuclei}}, \href{https://doi.org/10.1103/PhysRevD.90.072007}{\emph{Phys. Rev.}
  {\bfseries D90} (2014) 072007},
  [\href{https://arxiv.org/abs/1406.3236}{{\ttfamily 1406.3236}}].

\bibitem{Astier:2000ax}
{\scshape NOMAD} collaboration, P.~Astier et~al., \emph{{Measurement of the
  Lambda polarization in nu/mu charged current interactions in the NOMAD
  experiment}},
  \href{https://doi.org/10.1016/S0550-3213(00)00503-4}{\emph{Nucl. Phys.}
  {\bfseries B588} (2000) 3--36}.

\bibitem{Astier:2001ve}
{\scshape NOMAD} collaboration, P.~Astier et~al., \emph{{Measurement of the
  anti-Lambda polarization in muon-neutrino charged current interactions in the
  NOMAD experiment}},
  \href{https://doi.org/10.1016/S0550-3213(01)00181-X}{\emph{Nucl. Phys.}
  {\bfseries B605} (2001) 3--14},
  [\href{https://arxiv.org/abs/hep-ex/0103047}{{\ttfamily hep-ex/0103047}}].

\bibitem{Anselmino:2000vs}
M.~Anselmino, D.~Boer, U.~D'Alesio and F.~Murgia, \emph{{Lambda polarization
  from unpolarized quark fragmentation}},
  \href{https://doi.org/10.1103/PhysRevD.63.054029}{\emph{Phys. Rev.}
  {\bfseries D63} (2001) 054029},
  [\href{https://arxiv.org/abs/hep-ph/0008186}{{\ttfamily hep-ph/0008186}}].

\bibitem{Anselmino:2001js}
M.~Anselmino, D.~Boer, U.~D'Alesio and F.~Murgia, \emph{{Transverse lambda
  polarization in semiinclusive DIS}},
  \href{https://doi.org/10.1103/PhysRevD.65.114014}{\emph{Phys. Rev.}
  {\bfseries D65} (2002) 114014},
  [\href{https://arxiv.org/abs/hep-ph/0109186}{{\ttfamily hep-ph/0109186}}].

\bibitem{Boer:2010ya}
D.~Boer, Z.-B. Kang, W.~Vogelsang and F.~Yuan, \emph{{Test of the Universality
  of Naive-time-reversal-odd Fragmentation Functions}},
  \href{https://doi.org/10.1103/PhysRevLett.105.202001}{\emph{Phys. Rev. Lett.}
  {\bfseries 105} (2010) 202001},
  [\href{https://arxiv.org/abs/1008.3543}{{\ttfamily 1008.3543}}].

\bibitem{Metz:2016swz}
A.~Metz and A.~Vossen, \emph{{Parton Fragmentation Functions}},
  \href{https://doi.org/10.1016/j.ppnp.2016.08.003}{\emph{Prog. Part. Nucl.
  Phys.} {\bfseries 91} (2016) 136--202},
  [\href{https://arxiv.org/abs/1607.02521}{{\ttfamily 1607.02521}}].

\bibitem{Ackerstaff:1997nh}
{\scshape OPAL} collaboration, K.~Ackerstaff et~al., \emph{{Polarization and
  forward - backward asymmetry of Lambda baryons in hadronic Z0 decays}},
  \href{https://doi.org/10.1007/s100520050123}{\emph{Eur. Phys. J.} {\bfseries
  C2} (1998) 49--59}, [\href{https://arxiv.org/abs/hep-ex/9708027}{{\ttfamily
  hep-ex/9708027}}].

\bibitem{Niiyama:2017wpp}
{\scshape Belle} collaboration, M.~Niiyama et~al., \emph{{Production cross
  sections of hyperons and charmed baryons from $e^+e^-$ annihilation near
  $\sqrt{s} = 10.52$~GeV}},  \href{https://arxiv.org/abs/1706.06791}{{\ttfamily
  1706.06791}}.

\bibitem{Abdesselam:2016nym}
{\scshape Belle} collaboration, A.~Abdesselam et~al., \emph{{Observation of
  Transverse $\Lambda/\bar{\Lambda}$ Hyperon Polarization in $e^+e^-$
  Annihilation at Belle}},  \href{https://arxiv.org/abs/1611.06648}{{\ttfamily
  1611.06648}}.

\bibitem{Guan:2018ckx}
{\scshape Belle} collaboration, Y.~Guan et~al., \emph{{Observation of
  Transverse $\Lambda/\bar{\Lambda}$ Hyperon Polarization in $e^+e^-$
  Annihilation at Belle}},  \href{https://arxiv.org/abs/1808.05000}{{\ttfamily
  1808.05000}}.

\bibitem{Qiu:1991wg}
J.-w. Qiu and G.~Sterman, \emph{Single transverse spin asymmetries in direct
  photon production}, {\emph{Nucl. Phys.} {\bfseries B378} (1992) 52--78}.

\bibitem{Qiu:1998ia}
J.-w. Qiu and G.~Sterman, \emph{Single transverse-spin asymmetries in hadronic
  pion production}, {\emph{Phys. Rev.} {\bfseries D59} (1999) 014004},
  [\href{https://arxiv.org/abs/hep-ph/9806356}{{\ttfamily hep-ph/9806356}}].

\bibitem{Kouvaris:2006zy}
C.~Kouvaris, J.-W. Qiu, W.~Vogelsang and F.~Yuan, \emph{{Single transverse-spin
  asymmetry in high transverse momentum pion production in pp collisions}},
  \href{https://doi.org/10.1103/PhysRevD.74.114013}{\emph{Phys. Rev.}
  {\bfseries D74} (2006) 114013},
  [\href{https://arxiv.org/abs/hep-ph/0609238}{{\ttfamily hep-ph/0609238}}].

\bibitem{Eguchi:2006qz}
H.~Eguchi, Y.~Koike and K.~Tanaka, \emph{{Single Transverse Spin Asymmetry for
  Large-p(T) Pion Production in Semi-Inclusive Deep Inelastic Scattering}},
  \href{https://doi.org/10.1016/j.nuclphysb.2006.05.036}{\emph{Nucl. Phys.}
  {\bfseries B752} (2006) 1--17},
  [\href{https://arxiv.org/abs/hep-ph/0604003}{{\ttfamily hep-ph/0604003}}].

\bibitem{Eguchi:2006mc}
H.~Eguchi, Y.~Koike and K.~Tanaka, \emph{{Twist-3 Formalism for Single
  Transverse Spin Asymmetry Reexamined: Semi-Inclusive Deep Inelastic
  Scattering}},
  \href{https://doi.org/10.1016/j.nuclphysb.2006.11.016}{\emph{Nucl. Phys.}
  {\bfseries B763} (2007) 198--227},
  [\href{https://arxiv.org/abs/hep-ph/0610314}{{\ttfamily hep-ph/0610314}}].

\bibitem{Koike:2009ge}
Y.~Koike and T.~Tomita, \emph{{Soft-fermion-pole contribution to single-spin
  asymmetry for pion production in pp collisions}},
  \href{https://doi.org/10.1016/j.physletb.2009.04.017}{\emph{Phys. Lett.}
  {\bfseries B675} (2009) 181--189},
  [\href{https://arxiv.org/abs/0903.1923}{{\ttfamily 0903.1923}}].

\bibitem{Beppu:2010qn}
H.~Beppu, Y.~Koike, K.~Tanaka and S.~Yoshida, \emph{{Contribution of Twist-3
  Multi-Gluon Correlation Functions to Single Spin Asymmetry in Semi-Inclusive
  Deep Inelastic Scattering}},
  \href{https://doi.org/10.1103/PhysRevD.82.054005}{\emph{Phys. Rev.}
  {\bfseries D82} (2010) 054005},
  [\href{https://arxiv.org/abs/1007.2034}{{\ttfamily 1007.2034}}].

\bibitem{Koike:2011nx}
Y.~Koike and S.~Yoshida, \emph{{Three-gluon contribution to the single spin
  asymmetry in Drell-Yan and direct-photon processes}},
  \href{https://doi.org/10.1103/PhysRevD.85.034030}{\emph{Phys. Rev.}
  {\bfseries D85} (2012) 034030},
  [\href{https://arxiv.org/abs/1112.1161}{{\ttfamily 1112.1161}}].

\bibitem{Kang:2010zzb}
Z.-B. Kang, F.~Yuan and J.~Zhou, \emph{{Twist-three fragmentation function
  contribution to the single spin asymmetry in p p collisions}},
  \href{https://doi.org/10.1016/j.physletb.2010.07.003}{\emph{Phys. Lett.}
  {\bfseries B691} (2010) 243--248},
  [\href{https://arxiv.org/abs/1002.0399}{{\ttfamily 1002.0399}}].

\bibitem{Metz:2012ct}
A.~Metz and D.~Pitonyak, \emph{{Fragmentation contribution to the transverse
  single-spin asymmetry in proton-proton collisions}},
  \href{https://doi.org/10.1016/j.physletb.2013.05.043,
  10.1016/j.physletb.2016.10.011}{\emph{Phys. Lett.} {\bfseries B723} (2013)
  365--370}, [\href{https://arxiv.org/abs/1212.5037}{{\ttfamily 1212.5037}}].

\bibitem{Kanazawa:2013uia}
K.~Kanazawa and Y.~Koike, \emph{{Contribution of twist-3 fragmentation function
  to single transverse-spin asymmetry in semi-inclusive deep inelastic
  scattering}}, \href{https://doi.org/10.1103/PhysRevD.88.074022}{\emph{Phys.
  Rev.} {\bfseries D88} (2013) 074022},
  [\href{https://arxiv.org/abs/1309.1215}{{\ttfamily 1309.1215}}].

\bibitem{Beppu:2013uda}
H.~Beppu, K.~Kanazawa, Y.~Koike and S.~Yoshida, \emph{{Three-gluon contribution
  to the single spin asymmetry for light hadron production in pp collision}},
  \href{https://doi.org/10.1103/PhysRevD.89.034029}{\emph{Phys. Rev.}
  {\bfseries D89} (2014) 034029},
  [\href{https://arxiv.org/abs/1312.6862}{{\ttfamily 1312.6862}}].

\bibitem{Pitonyak:2016hqh}
D.~Pitonyak, \emph{{Transverse spin observables in hard-scattering hadronic
  processes within collinear factorization}},
  \href{https://doi.org/10.1142/S0217751X16300490}{\emph{Int. J. Mod. Phys.}
  {\bfseries A31} (2016) 1630049},
  [\href{https://arxiv.org/abs/1608.05353}{{\ttfamily 1608.05353}}].

\bibitem{Kanazawa:2000cx}
Y.~Kanazawa and Y.~Koike, \emph{{Polarization in hadronic Lambda hyperon
  production and chiral odd twist - three quark distribution}},
  \href{https://doi.org/10.1103/PhysRevD.64.034019}{\emph{Phys. Rev.}
  {\bfseries D64} (2001) 034019},
  [\href{https://arxiv.org/abs/hep-ph/0012225}{{\ttfamily hep-ph/0012225}}].

\bibitem{Zhou:2008fb}
J.~Zhou, F.~Yuan and Z.-T. Liang, \emph{{Hyperon Polarization in Unpolarized
  Scattering Processes}},
  \href{https://doi.org/10.1103/PhysRevD.78.114008}{\emph{Phys. Rev.}
  {\bfseries D78} (2008) 114008},
  [\href{https://arxiv.org/abs/0808.3629}{{\ttfamily 0808.3629}}].

\bibitem{Kanazawa:2015jxa}
K.~Kanazawa, A.~Metz, D.~Pitonyak and M.~Schlegel, \emph{{Single-spin
  asymmetries in the leptoproduction of transversely polarized Lambda
  hyperons}},
  \href{https://doi.org/10.1016/j.physletb.2015.04.011}{\emph{Phys.Lett.}
  {\bfseries B744} (2015) 385--390},
  [\href{https://arxiv.org/abs/1503.02003}{{\ttfamily 1503.02003}}].

\bibitem{Koike:2015zya}
Y.~Koike, K.~Yabe and S.~Yoshida, \emph{{Hyperon polarization from the twist-3
  distribution in unpolarized proton-proton collision}},
  \href{https://doi.org/10.1103/PhysRevD.92.094011}{\emph{Phys. Rev.}
  {\bfseries D92} (2015) 094011},
  [\href{https://arxiv.org/abs/1509.06830}{{\ttfamily 1509.06830}}].

\bibitem{Koike:2017fxr}
Y.~Koike, A.~Metz, D.~Pitonyak, K.~Yabe and S.~Yoshida, \emph{{Twist-3
  fragmentation contribution to polarized hyperon production in unpolarized
  hadronic collisions}},
  \href{https://doi.org/10.1103/PhysRevD.95.114013}{\emph{Phys. Rev.}
  {\bfseries D95} (2017) 114013},
  [\href{https://arxiv.org/abs/1703.09399}{{\ttfamily 1703.09399}}].

\bibitem{Vogelsang:2009pj}
W.~Vogelsang and F.~Yuan, \emph{{Next-to-leading Order Calculation of the
  Single Transverse Spin Asymmetry in the Drell-Yan Process}},
  \href{https://doi.org/10.1103/PhysRevD.79.094010}{\emph{Phys. Rev.}
  {\bfseries D79} (2009) 094010},
  [\href{https://arxiv.org/abs/0904.0410}{{\ttfamily 0904.0410}}].

\bibitem{Kang:2012ns}
Z.-B. Kang, I.~Vitev and H.~Xing, \emph{{Transverse momentum-weighted Sivers
  asymmetry in semi-inclusive deep inelastic scattering at next-to-leading
  order}}, \href{https://doi.org/10.1103/PhysRevD.87.034024}{\emph{Phys. Rev.}
  {\bfseries D87} (2013) 034024},
  [\href{https://arxiv.org/abs/1212.1221}{{\ttfamily 1212.1221}}].

\bibitem{Dai:2014ala}
L.-Y. Dai, Z.-B. Kang, A.~Prokudin and I.~Vitev, \emph{{Next-to-leading order
  transverse momentum-weighted Sivers asymmetry in semi-inclusive deep
  inelastic scattering: the role of the three-gluon correlator}},
  \href{https://doi.org/10.1103/PhysRevD.92.114024}{\emph{Phys. Rev.}
  {\bfseries D92} (2015) 114024},
  [\href{https://arxiv.org/abs/1409.5851}{{\ttfamily 1409.5851}}].

\bibitem{Yoshida:2016tfh}
S.~Yoshida, \emph{{New pole contribution to $P_{h\perp}$-weighted
  single-transverse spin asymmetry in semi-inclusive deep inelastic
  scattering}}, \href{https://doi.org/10.1103/PhysRevD.93.054048}{\emph{Phys.
  Rev.} {\bfseries D93} (2016) 054048},
  [\href{https://arxiv.org/abs/1601.07737}{{\ttfamily 1601.07737}}].

\bibitem{Chen:2016dnp}
A.~P. Chen, J.~P. Ma and G.~P. Zhang, \emph{{One-loop corrections to single
  spin asymmetries at twist-3 in Drell-Yan processes}},
  \href{https://doi.org/10.1103/PhysRevD.95.074005}{\emph{Phys. Rev.}
  {\bfseries D95} (2017) 074005},
  [\href{https://arxiv.org/abs/1607.08676}{{\ttfamily 1607.08676}}].

\bibitem{Chen:2017lvx}
A.~P. Chen, J.~P. Ma and G.~P. Zhang, \emph{{One-Loop Corrections of Single
  Spin Asymmetries in Semi-Inclusive DIS}},
  \href{https://arxiv.org/abs/1708.09091}{{\ttfamily 1708.09091}}.

\bibitem{Meissner:2008yf}
S.~Meissner and A.~Metz, \emph{{Partonic pole matrix elements for
  fragmentation}},
  \href{https://doi.org/10.1103/PhysRevLett.102.172003}{\emph{Phys. Rev. Lett.}
  {\bfseries 102} (2009) 172003},
  [\href{https://arxiv.org/abs/0812.3783}{{\ttfamily 0812.3783}}].

\bibitem{Gamberg:2010uw}
L.~P. Gamberg, A.~Mukherjee and P.~J. Mulders, \emph{{A model independent
  analysis of gluonic pole matrix elements and universality of TMD
  fragmentation functions}},
  \href{https://doi.org/10.1103/PhysRevD.83.071503}{\emph{Phys. Rev.}
  {\bfseries D83} (2011) 071503},
  [\href{https://arxiv.org/abs/1010.4556}{{\ttfamily 1010.4556}}].

\bibitem{Koike:2008du}
Y.~Koike, K.~Tanaka and S.~Yoshida, \emph{{Drell-Yan double-spin asymmetry
  A(LT) in polarized p anti-p collisions: Wandzura-Wilczek contribution}},
  \href{https://doi.org/10.1016/j.physletb.2008.08.049}{\emph{Phys. Lett.}
  {\bfseries B668} (2008) 286--292},
  [\href{https://arxiv.org/abs/0805.2289}{{\ttfamily 0805.2289}}].

\bibitem{Zhou:2009jm}
J.~Zhou, F.~Yuan and Z.-T. Liang, \emph{{Transverse momentum dependent quark
  distributions and polarized Drell-Yan processes}},
  \href{https://doi.org/10.1103/PhysRevD.81.054008}{\emph{Phys. Rev.}
  {\bfseries D81} (2010) 054008},
  [\href{https://arxiv.org/abs/0909.2238}{{\ttfamily 0909.2238}}].

\bibitem{Liang:2012rb}
Z.-T. Liang, A.~Metz, D.~Pitonyak, A.~Schaefer, Y.-K. Song and J.~Zhou,
  \emph{{Double spin asymmetry A$_{LT}$ in direct photon production}},
  \href{https://doi.org/10.1016/j.physletb.2012.04.072}{\emph{Phys. Lett.}
  {\bfseries B712} (2012) 235--239},
  [\href{https://arxiv.org/abs/1203.3956}{{\ttfamily 1203.3956}}].

\bibitem{Metz:2012fq}
A.~Metz, D.~Pitonyak, A.~Schaefer and J.~Zhou, \emph{{Analysis of the
  double-spin asymmetry $A_LT$ in inelastic nucleon-nucleon collisions}},
  \href{https://doi.org/10.1103/PhysRevD.86.114020}{\emph{Phys. Rev.}
  {\bfseries D86} (2012) 114020},
  [\href{https://arxiv.org/abs/1210.6555}{{\ttfamily 1210.6555}}].

\bibitem{Kanazawa:2014tda}
K.~Kanazawa, A.~Metz, D.~Pitonyak and M.~Schlegel,
  \emph{{Longitudinal-transverse double-spin asymmetries in single-inclusive
  leptoproduction of hadrons}},
  \href{https://doi.org/10.1016/j.physletb.2015.02.005}{\emph{Phys.Lett.}
  {\bfseries B742} (2015) 340--346},
  [\href{https://arxiv.org/abs/1411.6459}{{\ttfamily 1411.6459}}].

\bibitem{Koike:2016ura}
Y.~Koike, D.~Pitonyak and S.~Yoshida, \emph{{Twist-3 effect from the
  longitudinally polarized proton for $A_{LT}$ in hadron production from $pp$
  collisions}},
  \href{https://doi.org/10.1016/j.physletb.2016.05.043}{\emph{Phys. Lett.}
  {\bfseries B759} (2016) 75--81},
  [\href{https://arxiv.org/abs/1603.07908}{{\ttfamily 1603.07908}}].

\bibitem{Kanazawa:2015ajw}
K.~Kanazawa, Y.~Koike, A.~Metz, D.~Pitonyak and M.~Schlegel, \emph{{Operator
  Constraints for Twist-3 Functions and Lorentz Invariance Properties of
  Twist-3 Observables}},
  \href{https://doi.org/10.1103/PhysRevD.93.054024}{\emph{Phys. Rev.}
  {\bfseries D93} (2016) 054024},
  [\href{https://arxiv.org/abs/1512.07233}{{\ttfamily 1512.07233}}].

\bibitem{Ji:1993vw}
X.-D. Ji, \emph{{Chiral odd and spin dependent quark fragmentation functions
  and their applications}},
  \href{https://doi.org/10.1103/PhysRevD.49.114}{\emph{Phys. Rev.} {\bfseries
  D49} (1994) 114--124},
  [\href{https://arxiv.org/abs/hep-ph/9307235}{{\ttfamily hep-ph/9307235}}].

\bibitem{Chen:1994ar}
K.~Chen, G.~R. Goldstein, R.~L. Jaffe and X.-D. Ji, \emph{{Probing quark
  fragmentation functions for spin 1/2 baryon production in unpolarized e+ e-
  annihilation}},
  \href{https://doi.org/10.1016/0550-3213(95)00193-V}{\emph{Nucl. Phys.}
  {\bfseries B445} (1995) 380--398},
  [\href{https://arxiv.org/abs/hep-ph/9410337}{{\ttfamily hep-ph/9410337}}].

\bibitem{Mulders:1995dh}
P.~J. Mulders and R.~D. Tangerman, \emph{The complete tree-level result up to
  order 1/q for polarized deep-inelastic leptoproduction}, {\emph{Nucl. Phys.}
  {\bfseries B461} (1996) 197--237},
  [\href{https://arxiv.org/abs/hep-ph/9510301}{{\ttfamily hep-ph/9510301}}].

\bibitem{Boer:1997mf}
D.~Boer, R.~Jakob and P.~J. Mulders, \emph{{Asymmetries in polarized hadron
  production in e+ e- annihilation up to order 1/Q}},
  \href{https://doi.org/10.1016/S0550-3213(97)00456-2}{\emph{Nucl. Phys.}
  {\bfseries B504} (1997) 345--380},
  [\href{https://arxiv.org/abs/hep-ph/9702281}{{\ttfamily hep-ph/9702281}}].

\bibitem{Collins:1981uw}
J.~C. Collins and D.~E. Soper, \emph{Parton distribution and decay functions},
  {\emph{Nucl. Phys.} {\bfseries B194} (1982) 445}.

\bibitem{Collins:2011zzd}
J.~Collins, \emph{{Foundations of perturbative QCD}}, {\emph{Camb. Monogr.
  Part. Phys. Nucl. Phys. Cosmol.} {\bfseries 32} (2011) 1--624}.

\bibitem{Collins:2002kn}
J.~C. Collins, \emph{Leading-twist single-transverse-spin asymmetries:
  Drell-yan and deep-inelastic scattering}, {\emph{Phys. Lett.} {\bfseries
  B536} (2002) 43--48}, [\href{https://arxiv.org/abs/hep-ph/0204004}{{\ttfamily
  hep-ph/0204004}}].

\bibitem{Ji:2002aa}
X.-d. Ji and F.~Yuan, \emph{Parton distributions in light-cone gauge: Where are
  the final-state interactions?}, {\emph{Phys. Lett.} {\bfseries B543} (2002)
  66--72}, [\href{https://arxiv.org/abs/hep-ph/0206057}{{\ttfamily
  hep-ph/0206057}}].

\bibitem{Belitsky:2002sm}
A.~V. Belitsky, X.~Ji and F.~Yuan, \emph{Final state interactions and gauge
  invariant parton distributions}, {\emph{Nucl. Phys.} {\bfseries B656} (2003)
  165--198}, [\href{https://arxiv.org/abs/hep-ph/0208038}{{\ttfamily
  hep-ph/0208038}}].

\bibitem{Boer:2003cm}
D.~Boer, P.~J. Mulders and F.~Pijlman, \emph{Universality of t-odd effects in
  single spin and azimuthal asymmetries}, {\emph{Nucl. Phys.} {\bfseries B667}
  (2003) 201--241}, [\href{https://arxiv.org/abs/hep-ph/0303034}{{\ttfamily
  hep-ph/0303034}}].

\bibitem{Collins:2004nx}
J.~C. Collins and A.~Metz, \emph{Universality of soft and collinear factors in
  hard- scattering factorization}, {\emph{Phys. Rev. Lett.} {\bfseries 93}
  (2004) 252001}, [\href{https://arxiv.org/abs/hep-ph/0408249}{{\ttfamily
  hep-ph/0408249}}].

\bibitem{Bacchetta:2006tn}
A.~Bacchetta, M.~Diehl, K.~Goeke, A.~Metz, P.~Mulders and M.~Schlegel,
  \emph{Semi-inclusive deep inelastic scattering at small transverse momentum},
  {\emph{JHEP} {\bfseries 02} (2007) 093},
  [\href{https://arxiv.org/abs/hep-ph/0611265}{{\ttfamily hep-ph/0611265}}].

\bibitem{Koike:2011mb}
Y.~Koike and S.~Yoshida, \emph{{Probing the three-gluon correlation functions
  by the single spin asymmetry in $p^\uparrow p\to DX$}},
  \href{https://doi.org/10.1103/PhysRevD.84.014026}{\emph{Phys. Rev.}
  {\bfseries D84} (2011) 014026},
  [\href{https://arxiv.org/abs/1104.3943}{{\ttfamily 1104.3943}}].

\bibitem{Jaffe:1983hp}
R.~Jaffe, \emph{{Parton Distribution Functions for Twist Four}},
  \href{https://doi.org/10.1016/0550-3213(83)90361-9}{\emph{Nucl.Phys.}
  {\bfseries B229} (1983) 205}.

\bibitem{Christ:1966zz}
N.~Christ and T.~D. Lee, \emph{{Possible Tests of Cst and Tst Invariances in l
  + /- + N $\to$ l + /- + Gamma and A $\to$ B+e++e-}},
  \href{https://doi.org/10.1103/PhysRev.143.1310}{\emph{Phys. Rev.} {\bfseries
  143} (1966) 1310--1321}.

\bibitem{Metz:2006pe}
A.~Metz, M.~Schlegel and K.~Goeke, \emph{Transverse single spin asymmetries in
  inclusive deep- inelastic scattering}, {\emph{Phys. Lett.} {\bfseries B643}
  (2006) 319--324}, [\href{https://arxiv.org/abs/hep-ph/0610112}{{\ttfamily
  hep-ph/0610112}}].

\bibitem{Afanasev:2007ii}
A.~Afanasev, M.~Strikman and C.~Weiss, \emph{{Transverse target spin asymmetry
  in inclusive DIS with two-photon exchange}},
  \href{https://doi.org/10.1103/PhysRevD.77.014028}{\emph{Phys. Rev.}
  {\bfseries D77} (2008) 014028},
  [\href{https://arxiv.org/abs/0709.0901}{{\ttfamily 0709.0901}}].

\bibitem{Metz:2012ui}
A.~Metz, D.~Pitonyak, A.~Schafer, M.~Schlegel, W.~Vogelsang et~al.,
  \emph{{Single-spin asymmetries in inclusive deep inelastic scattering and
  multiparton correlations in the nucleon}},
  \href{https://doi.org/10.1103/PhysRevD.86.094039}{\emph{Phys.Rev.} {\bfseries
  D86} (2012) 094039}, [\href{https://arxiv.org/abs/1209.3138}{{\ttfamily
  1209.3138}}].

\bibitem{Schlegel:2012ve}
M.~Schlegel, \emph{{Partonic description of the transverse target single-spin
  asymmetry in inclusive deep-inelastic scattering}},
  \href{https://doi.org/10.1103/PhysRevD.87.034006}{\emph{Phys.Rev.} {\bfseries
  D87} (2013) 034006}, [\href{https://arxiv.org/abs/1211.3579}{{\ttfamily
  1211.3579}}].

\bibitem{Goeke:2005hb}
K.~Goeke, A.~Metz and M.~Schlegel, \emph{Parameterization of the quark-quark
  correlator of a spin- 1/2 hadron}, {\emph{Phys. Lett.} {\bfseries B618}
  (2005) 90--96}, [\href{https://arxiv.org/abs/hep-ph/0504130}{{\ttfamily
  hep-ph/0504130}}].

\bibitem{Leibbrandt:1994wj}
G.~Leibbrandt, \emph{{Noncovariant gauges: Quantization of Yang-Mills and
  Chern-Simons theory in axial type gauges}}, {\emph{Singapore, Singapore:
  World Scientific (1994) 212 p} (1994) }.

\bibitem{deFlorian:1997zj}
D.~de~Florian, M.~Stratmann and W.~Vogelsang, \emph{{QCD analysis of
  unpolarized and polarized Lambda baryon production in leading and
  next-to-leading order}},
  \href{https://doi.org/10.1103/PhysRevD.57.5811}{\emph{Phys. Rev.} {\bfseries
  D57} (1998) 5811--5824},
  [\href{https://arxiv.org/abs/hep-ph/9711387}{{\ttfamily hep-ph/9711387}}].

\bibitem{deFlorian:1998ba}
D.~de~Florian, M.~Stratmann and W.~Vogelsang, \emph{{Polarized Lambda baryon
  production in p p collisions}},
  \href{https://doi.org/10.1103/PhysRevLett.81.530}{\emph{Phys. Rev. Lett.}
  {\bfseries 81} (1998) 530--533},
  [\href{https://arxiv.org/abs/hep-ph/9802432}{{\ttfamily hep-ph/9802432}}].

\bibitem{Hooft:1972fi}
G.~t'~Hooft and M.~J.~G. Veltman, \emph{Regularization and renormalization of
  gauge fields}, {\emph{Nucl. Phys.} {\bfseries B44} (1972) 189--213}.

\bibitem{Breitenlohner:1977hr}
P.~Breitenlohner and D.~Maison, \emph{{Dimensional Renormalization and the
  Action Principle}}, \href{https://doi.org/10.1007/BF01609069}{\emph{Commun.
  Math. Phys.} {\bfseries 52} (1977) 11--38}.

\bibitem{Collins:1988wj}
J.~C. Collins and J.-w. Qiu, \emph{{A New Derivation of the Altarelli-parisi
  Equations}}, \href{https://doi.org/10.1103/PhysRevD.39.1398}{\emph{Phys.
  Rev.} {\bfseries D39} (1989) 1398}.

\bibitem{Belitsky:1996hg}
A.~V. Belitsky and E.~A. Kuraev, \emph{{Evolution of chiral odd spin
  independent fracture functions in quantum chromodynamics}},
  \href{https://doi.org/10.1016/S0550-3213(97)00306-4}{\emph{Nucl. Phys.}
  {\bfseries B499} (1997) 301--318},
  [\href{https://arxiv.org/abs/hep-ph/9612256}{{\ttfamily hep-ph/9612256}}].

\bibitem{Kang:2010xv}
Z.-B. Kang, \emph{{QCD evolution of naive-time-reversal-odd fragmentation
  functions}}, \href{https://doi.org/10.1103/PhysRevD.83.036006}{\emph{Phys.
  Rev.} {\bfseries D83} (2011) 036006},
  [\href{https://arxiv.org/abs/1012.3419}{{\ttfamily 1012.3419}}].

\bibitem{Ma:2017upj}
J.~P. Ma and G.~P. Zhang, \emph{{Evolution of Chirality-odd Twist-3
  Fragmentation Functions}},
  \href{https://doi.org/10.1016/j.physletb.2017.07.025}{\emph{Phys. Lett.}
  {\bfseries B772} (2017) 559--566},
  [\href{https://arxiv.org/abs/1701.04141}{{\ttfamily 1701.04141}}].

\end{thebibliography}\endgroup
\bibliographystyle{JHEP}

\end{document}